\documentclass{jaa}
\usepackage{natbib}
\usepackage{pdflscape}
\usepackage{multirow}
\usepackage{hyperref}
\usepackage{graphicx}
\usepackage{xcolor}
\usepackage{placeins}

\begin{document}\sloppy

\title{Anomalous scattering of pulsars towards the Gum Nebula}


\author{M.~A. Krishnakumar\textsuperscript{1,*}, Bhal Chandra Joshi\textsuperscript{2} and P.~K. Manoharan\textsuperscript{3,4}}
\affilOne{\textsuperscript{1}Radio Astronomy Centre, NCRA-TIFR, Ooty, 643 001, Tamil Nadu, India.\\}
\affilTwo{\textsuperscript{2}National Centre for Radio Astrophysics, Tata Institute of Fundamental Research, Pune, 411007, Maharashtra, India\\}
\affilThree{\textsuperscript{3}Heliophysics Science Division, NASA Goddard Space Flight Center, Greenbelt, MD 20771, USA\\}
\affilFour{\textsuperscript{4}The Catholic University of America, Washington, DC 20064, USA\\}


\twocolumn[{

\maketitle

\corres{kkma@ncra.tifr.res.in}


\msinfo{8 January 2026}{-}

\begin{abstract}
We report wideband scatter-broadening estimates of 14 pulsars towards the 
Gum nebula region using the Band-3 of the upgraded GMRT. This work increases
the measurements of frequency scaling index of scatter-broadening ($\alpha$)  
across the nebula by more than 3 times. A strong correlation between the 
distance and the scattering strength is observed for pulsars behind the nebula.
It is also observed that for distant pulsars ($> 2 kpc$), the effect of the Gum 
nebula in DM and scattering strength is not substantial. We also report a much
flatter $\alpha$ for the Vela pulsar and argue that its scattering is not
caused by the Gum nebula, but the Vela supernova remnant.
\end{abstract}


}]


\doinum{12.3456/s78910-011-012-3}
\artcitid{\#\#\#\#}
\volnum{000}
\year{0000}
\pgrange{1--}
\setcounter{page}{1}
\lp{1}

\section{Introduction}
\label{sec:intro}

Pulsars are fast spinning neutron stars discovered serendipitously in 1967 \citep{1968Natur.217..709H}. As they spin at fast rate, the radio emission coming from their polar regions appears as periodic pulses. As these pulses travel through the ionised interstellar medium (IISM), they get dispersed, resulting in a frequency dependent delay. IISM is not uniform throughout, but clumpy with localised dense structures like supernova remnants, H{\small II} regions, etc. When the pulsed signals from pulsars pass through such dense regions in the IISM, they get scattered. At lower radio frequencies, these otherwise narrow pulses get broadened in time exhibiting an exponential decay tail. This scatter-broadening scales as a function of frequency, which provides an insight about the turbulence in the medium. For example, a medium with Kolmogorov type turbulence leads to scatter-broadening, which varies as a function of observing frequency ($\nu$) as  $\nu^{-4.4}$. But this does not hold true for all the lines of sight (LoS) due to the presence of different structures along them and the scatter-broadening can vary between $\nu^{-3}$ -- $\nu^{-5}$. Measuring this scaling relation can help in studying the different IISM structures causing the scattering as shown by \citet{rickett2009}. If the LoS of several pulsars pass through different regions of a nebula or an H{\small II} region, the density turbulence in different regions of it can be modelled \citep{ocker2024}.

The Gum nebula is one of the largest optical emission nebulae discovered in the H$\alpha$ observations of the southern sky by \citet{gum52}. The shape of the nebula is almost circular with a diameter of $\sim$36$^{\circ}$ centered at the Galactic longitude of 260$^{\circ}$ and latitude of $-$2$^{\circ}$. The latest distance estimate to the nebula by \citet{eggen90} and \citet{woermann01} shows that it is $\sim$450 pc away from the Sun with a radius of 125 pc and thickness of $\sim$20 pc \citep{sun08,jansson12,purcell15}. The origin and evolution of the nebula was under scrutiny for many years after its discovery, with no model being able to explain all the observed characteristics.

According to \citet{purcell15}, the origin of the nebula due to an old supernova model is unlikely, and the model of H{\small II} region around a wind blown bubble sounds as a better explanation. \citet{sahu93} found that the Vela OB2 star is making a smaller bubble within the Gum nebula called the IRAS Vela shell. It has a diameter of 7.5$^{\circ}$ and is interpreted as a wind blown bubble, driven by the Vela OB2. Recently, \citet{gao25} revisited the morphology and origin of the IRAS Vela shell using a parsec-scale 3D dust tomography and reported that stellar winds are not the primary driver of the shell's expansion, but an H{\small II} region and supernova feedback. The electron density in the nebula varies a lot (0.1 $-$ 100 cm$^{-3}$), as evident from the H$\alpha$ images of the region.

In the pulsar distance model of \citet{tc93} and \citet{ne2001}, (and in a recent update of it by \citep{ne2025}), Gum nebula was considered as a separate component with an angular diameter of 30$^{\circ}$ and an  assumed free electron density of 0.2 cm$^{-3}$ everywhere. The fluctuation parameter, which determines the scattering in each LoS, was assumed to be zero in the early models due to the lack of observations available at that time. After \citet{mr01} measured the scatter-broadening towards Gum nebula at 327~MHz using the Ooty Radio Telescope, \citet{ne2001} improved the model distances and the scattering towards Gum nebula. Recently, \cite{ne2025} revisited the Gum nebula and updated the model of electron densities and DMs using the new measurements towards this region. Though there were 25 measurements of scatter-broadening by \citet{mr01}, most of them had large uncertainties. Their fitting procedure considered the pulse width also as a free parameter. Hence, some of the measurements were probably unreliable. Moreover, only 13 pulsars in this sample  covered the Gum nebula out of the 25 they reported. The discovery of new pulsars 
towards this region have prompted us to carry out a new study to understand the region in much 
better detail, particularly as the new pulsar discoveries since 2001 afford a better sampling 
of the spatial extent of the Gum nebula and its inner bubble using multi-frequency observations.

The varying electron content across the spatial extent of the Gum nebula has 
a direct bearing on the distance estimates using the dispersion measure (DM) for pulsars that 
lie behind the Gum nebula. The distances to pulsars were modelled by \citet{tc93, ne2001, ymw16} 
and recently by \citet{ne2025} by considering the spiral arm structure of the Galaxy,  
average electron density, radio continuum emission associated with 
the Galaxy and H{\small II} regions towards their individual LoS. The distance 
estimates of pulsars that lie behind the Gum nebula are unreliable due to assumptions 
about the electron content of Gum nebula in these models. Moreover, 
the DM independent distance estimates (i.e., either using parallax or VLBI measurements) 
of these pulsars are very few. In our list, only seven of the 20 pulsars 
have an independent distance estimate.  Hence the results from this study will motivate further modelling of the distances to pulsars.

The main goal of this study is to understand the distribution of the scattering strength in the
Gum nebula. With the help of this, we will be able to understand the turbulence characteristics 
in different regions of the nebula. The results of this study will also constrain 
better the distance estimates of the pulsars that lie behind the Gum nebula.
In addition, along with the rotation measure (RM) and DM of the pulsars, we will be able to 
understand the orientation and strength of the magnetic field towards the Gum nebula region. 
In this work, measurements derived from the observations of 20 pulsars in 
a pilot program, conducted with the uGMRT, are presented. Detailed interpretation 
combining these measurements with those available in the literature and future observations is planned in a subsequent 
follow-up work. Observations and analysis procedure are described in Section \ref{sec:obs} followed 
by a presentation of our measurements and discussion of these results in Section \ref{sec:result}. We 
conclude with a summary in Section \ref{sec:sum}.

\begin{figure*}[t]
\centering
\includegraphics[scale=0.4]{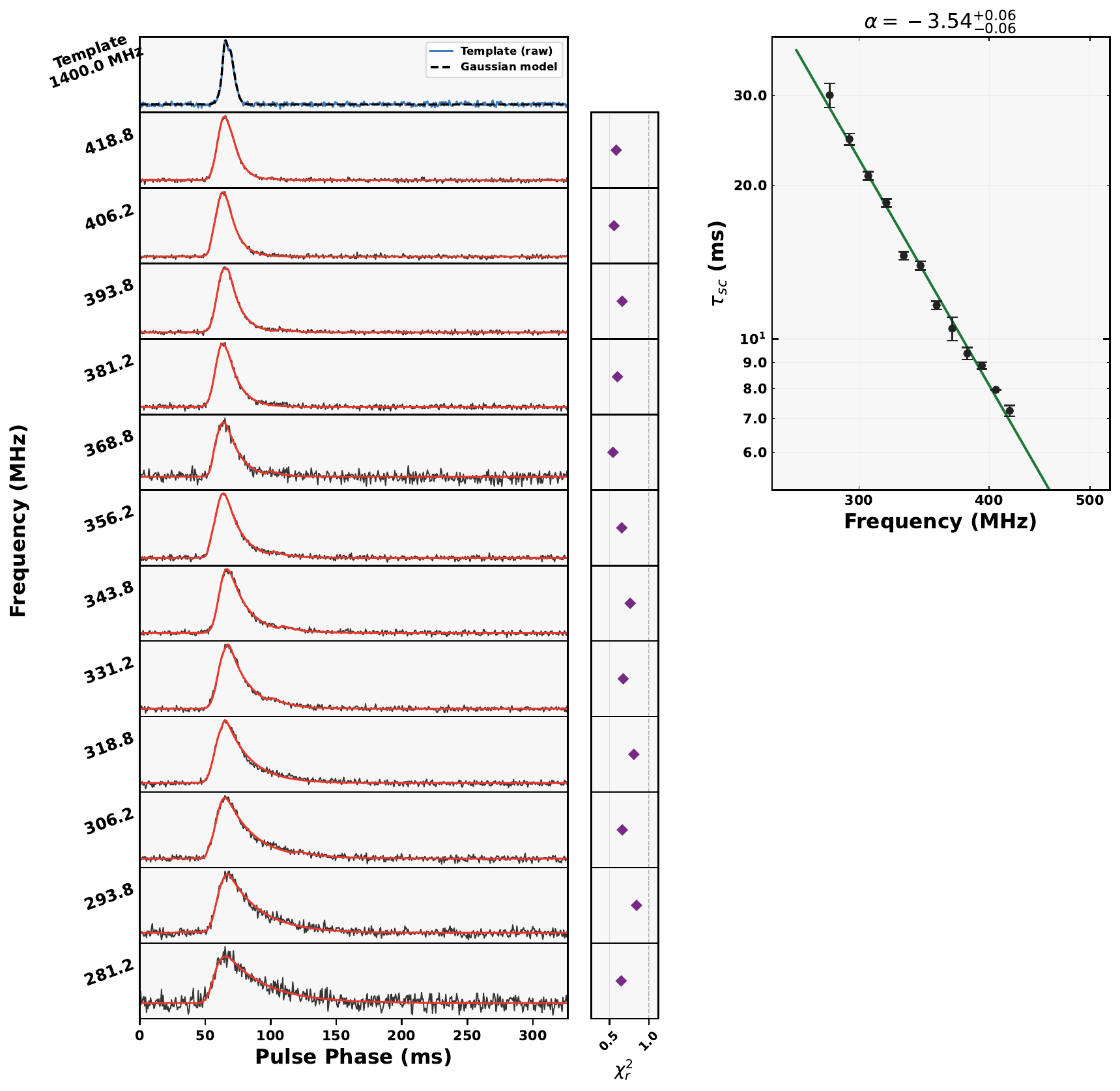}
\caption{The left panel shows the profile at each sub-band of the uGMRT band3 for PSR J0857$-$4424. The black curves show the observed profile and the red curves show the best fit model for PBF1. The template (intrinsic) profile used for fitting the band-3 profiles is shown at the top of the panel where blue curve shows the observed profile and the black dashed curve shows the template extracted for fitting purpose. The reduced $\chi^2$ of each profile fit is also shown in the right side panel near the profiles. The right side panel shows the measured $\tau_{sc}$ in black filled circles as a function of observing frequency in log-log plot. The green continuous line shows the best fit model to the data set. The best fit $\alpha$ measurement is given at the top of the panel.} 
\label{fig1}
\end{figure*}

\section{Observation and Analysis}\label{sec:obs}

There are about 100 pulsars discovered in the direction towards the Gum nebula. We have selected a list of 20 pulsars, which lie in and around the Gum nebula region for this study. The sample size is smaller compared to the available pulsars towards the Gum nebula as we removed several faint pulsars from it as well as applied a rough cut-off based on the DM--$\tau_{sc}$ relation \citep{kmnjm15}. These pulsars sample different regions of the nebula with possibly varying scatter-broadening times, so that we could get a holistic picture of the distribution of the scattering strength across the nebula. The observations were made using the uGMRT during December 2016 -- January 2017. We have used the Band--3 (250 -- 500 MHz) employing the wide-band backend of the uGMRT for our observations.  We used the 200 MHz bandwidth available with uGMRT between the frequency range of 275 -- 475 MHz. The data were recorded in a digital filterbank mode with 8192 channels ($\sim24.4$~kHz wide) across the bandwidth at a time resolution of 327.68 $\mu$s. We observed 10 pulsars in two different session using a 16 antenna phased array setup. The phase calibration of the array was performed every 45 minutes to keep the sensitivity at the expected levels.

The uGMRT filterbank data were converted to SIGPROC filterbank format for further processing. Dedispersion was performed using the nominal DM of the pulsar available from the ATNF pulsar catalogue\footnote{\href{https://www.atnf.csiro.au/research/pulsar/psrcat/}{ATNF Pulsar Catalogue}}. The dedispersed data were collapsed to 16 sub-bands across the whole band, giving a profile in every 12.5 MHz band. Since the channel resolution was coarse, the dispersion smear in each channel is likely to be considerable. For the highest DM pulsar in our sample, PSR J0831$-$4406, the dispersion smear in the lowest sub-band profile will be $\sim$1.2~ms. This is taken care during the fitting procedure to estimate the scattering time, $\tau_{sc}$.

We assume the observed profile to be the convolution of the intrinsic pulse profile, $P_{i}(t)$,  with the IISM transfer function $s(t)$ and the dispersion smear $D(t)$ as shown below:

\begin{equation}
    P(t) = P_{i}(t) * s(t) * D(t)
\end{equation}

\noindent where $P_{i}(t)$ is the intrinsic unscattered pulse profile, $s(t)$ is the IISM transfer function for scattering and $D(t)$ is a rectangular function of temporal width corresponding to the dispersion smear in a particular frequency channel.

Given that the scattering will be caused majorly by the turbulence within the nebula for most of the pulsars, a thin screen approximation may not be the ideal model for fitting the scattering tail. Because of this, we choose three different IISM transfer functions for fitting the profiles to measure scattering time. These three pulse broadening functions are:

\begin{equation}
    \mathrm{PBF1}\quad s(t) = e^{-1/\tau}\quad U(t)
\end{equation}

\begin{equation}
    \mathrm{PBF2} \qquad
s(t) =
\sqrt{\frac{\pi \tau_{\rm sc}}{4t^3}}
e^{(-\frac{\pi^2 \tau_{\rm sc}}{16t})}\quad U(t)
\end{equation}

\begin{equation}
    \mathrm{PBF3} \qquad
s(t) = \frac{1}{\sqrt{\pi \tau_{\rm sc} t}}
e^{(-\frac{t}{\tau_{\rm sc}})}\quad U(t)
\end{equation}

where $U(t)$ is a unit step function. PBF1 represents the commonly used thin screen approximation. The scattering material is concentrated on a phase changing thin screen placed close to the center of the pulsar and the observer. For most of the pulsars, this model alone is sufficient for fitting the observed profile. PBF2 represents a thick screen approximation as formulated by \cite{williamson72}. This will result in a slow rise which will decay fast into an exponential tail. PBF3 is a truncated exponential decay function. The assumption is that the the line of sight is highly anisotropic where the scattering is limited to one dimension \citep{gkk17}.

For every pulsar, we found a high frequency profile for fitting purposes where scatter-broadening is expected to be low. For some of them, we found that the scattering is only detected at the lower part of our observing band and hence used the highest frequency profile (470 MHz) as template. The template profiles are fit with multiple Gaussians to create a noise-free template. The fitted values of Gaussian components, namely, their phase, amplitude and width were stored. These analytic templates were used for fitting the profiles to measure scatter-broadening using all the three PBFs.

All the profiles were fitted using the three PBFs by minimising the $\chi^2$ between the profile and the convolved template as performed in \citet{kmnjm15}. The profiles where the fitting did not converge were flagged as bad measurements and removed. The amplitude and phase of the Gaussians are given freedom to fit (maximum of 5\% from its original value) while the width of them were fixed so that it will not affect the obtained $\tau_{sc}$ values. This helped obtain reliable post-fit residuals for multi-component profiles which evolved from a higher frequency template to the band-3 frequency ranges.

After obtaining $\tau_{sc}$ for all the available frequencies for a given pulsar, a MCMC method was used to obtain the scattering scaling index, $\alpha$ and its uncertainty. Each of the $\tau_{sc}$ values were randomised within their uncertainty range and were given unity error bars and a straight line fit was performed in the log--log plane ($log \tau_{sc}$ = $-\alpha log\nu + b$) so that $\alpha$ can be obtained easily as the slope of the fit. This process was repeated for 10,000 iterations. From the distribution of $\alpha$, the median was taken as the value of $\alpha$ and the uncertainties are quoted from 5 and 95 percentiles. Figure~\ref{fig1} displays the results of the above analysis on PSR J0857$-$4424 for PBF1. For this particular pulsar, the measured $\alpha$ is somewhat lower than expected for a medium with Kolmogorov turbulence. The fit plots for all the pulsars are provided in the same format in \ref{app1}

To identify which PBF is the most preferred model for a given pulsar, we used the multi-model comparison framework using the Akaike Information Criterion \citealp[(AIC)]{Akaike1974} and Bayesian Information Criterion \citealp[(BIC)]{Schwarz1978}. By definition, AIC = $2k - 2ln(L)$ and BIC = $kln(n)-2ln(L)$, where $k$ is the number of constraints in a fit, $n$ is the total number of samples and $L$ is maximum value of the likelihood function for a model. In the case where the samples are normally distributed, which is our case for post-fit residuals, $-2ln(L)$ can be considered as $\chi^2$ of the fit. This simplifies the formulation into AIC = $2k + \chi^2_{total}$ and BIC = $kln(n) + \chi^2_{total}$. A set of channels where all the three PBFs could fit was taken as the data set for model comparison to avoid any bias. When the profile fit did not converge for many channels, only two models were considered for comparison. The total values of $\chi^2$, total number of constraints, $k$ and the total number of samples, $n$ were applied in the above formula for obtaining AIC/BIC. The differences $\Delta$BIC and $\Delta$AIC relative to the best model are interpreted on the Jeffreys scale, with $\Delta$BIC $>$ 10 constituting decisive evidence against the inferior model, $\Delta$BIC $>$ 6 strong evidence, and $\Delta$BIC $>$ 2 positive evidence \citep{KassRaftery1995}. The lower the value of $\Delta$AIC/$\Delta$BIC, the better the model is. We used this approach to obtain the best PBF for a given pulsar.

\begin{figure*}[t]
\centering
\includegraphics[scale=0.6]{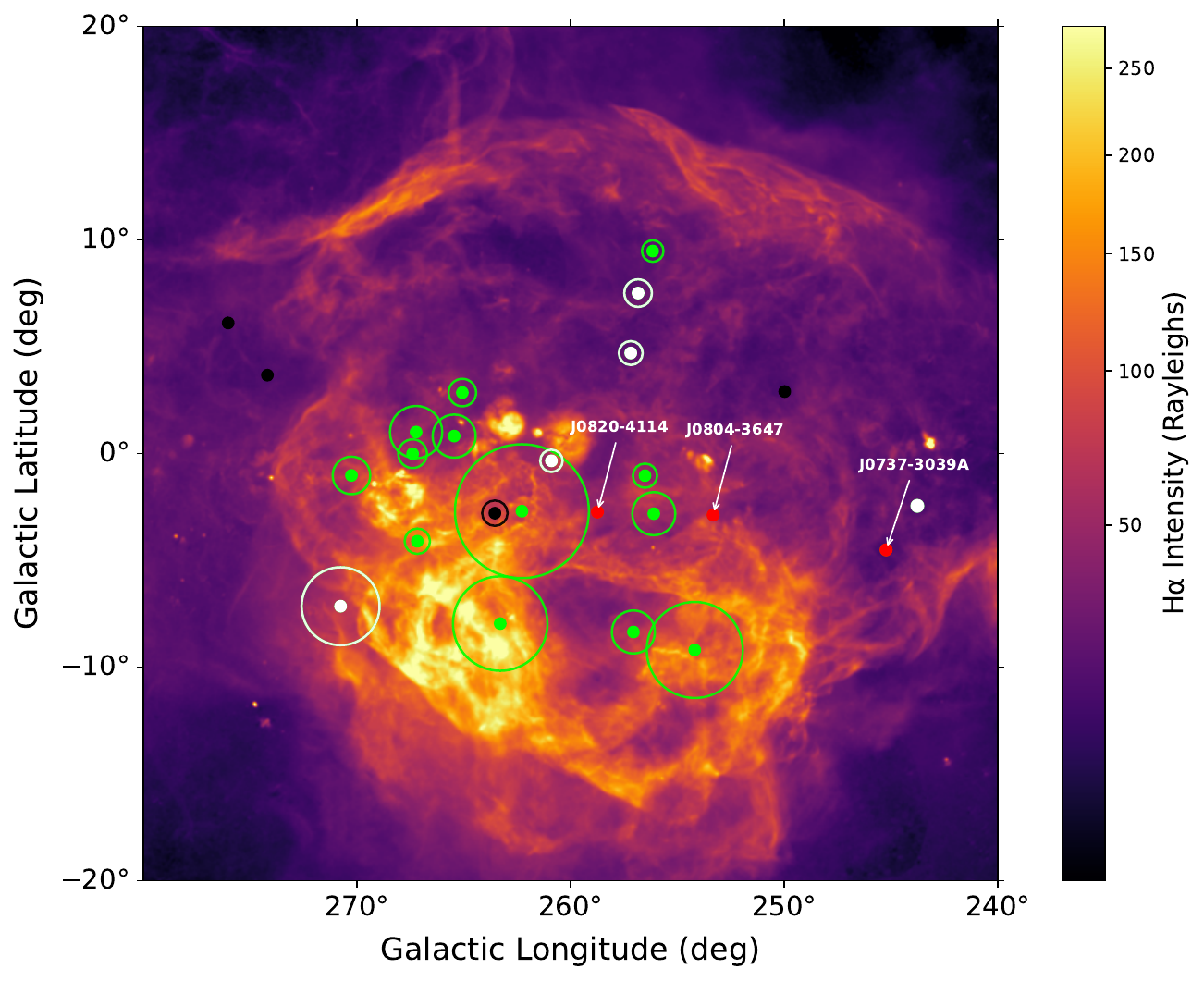}
\caption[The distribution of $\tau_{sc}$ estimates of pulsars across the Gum nebula at 325 MHz overlaid on the H{$\alpha$} image of the region]{The distribution of the $\tau_{sc}$ of pulsars across the Gum nebula is shown.
The black filled circles show the pulsars in front of the nebula (by distance), the green filled circles show the pulsar positions that are behind the nebula. The red filled circles represent the pulsar behind the nebula that did not show any scatter-broadening. The white circles are the $\tau_{sc}$ estimates from literature scaled to 325 MHz and \citep{lw2,kmnjm15}. 
The diameter of the circles indicate the $\tau_{sc}$ of each pulsar at 325 MHz. The background image
is taken from the H$\alpha$ images produced by \citet{fink03}.}
\label{fig2}
\end{figure*}

\section{Results and Discussions}\label{sec:result}

The aim of this work is to understand the turbulence characteristics of the Gum nebula and its
dependence on parameters like DM, RM and the position in the nebula. We have observed 20
pulsars with the uGMRT band-3 with 200 MHz bandwidth in the frequency range of 275 -- 475 MHz. 
All the pulsars were detected with good 
signal to noise ratio profiles. A set of six pulsars did not show any measurable scatter-broadening 
even at the lowest of frequencies, namely, PSRs J0737$-$3039A, J0804$-$3647, J0818$-$3049, 
J0820$-$4114, J0945$-$4833 and J1003$-$4747. For the rest of the 14 pulsars, we measured 
$\tau_{sc}$ at several sub-bands across the observing band and estimated the $\alpha$. Using the model selection approach described above with $\Delta$AIC/$\Delta$BIC, nine pulsars strongly chose PBF1, four pulsars chose PBF3 and only one pulsar chose PBF2 (PSR J0908$-$4913). Plots of all the three PBF $\alpha$ fits and $\Delta$AIC/$\Delta$BIC values of all the 14 pulsars along with a table listing the values of $\alpha$, $\tau_{sc}$ at 325~MHz and $\Delta$BIC are given in ~\ref{app2} It was observed that the evidence for PBF2 for PSR J0908$-$4913 is not very high compared to PBF1. Given that most of the pulsars in the sample chose PBF1 as the preferred model, we report $\tau_{sc}$ and $\alpha$ values from that fit for consistency. Nevertheless, the profile fit and results for all the three models are given in ~\ref{app3}

\begin{figure*}[t]
\centering
\includegraphics[scale=0.6]{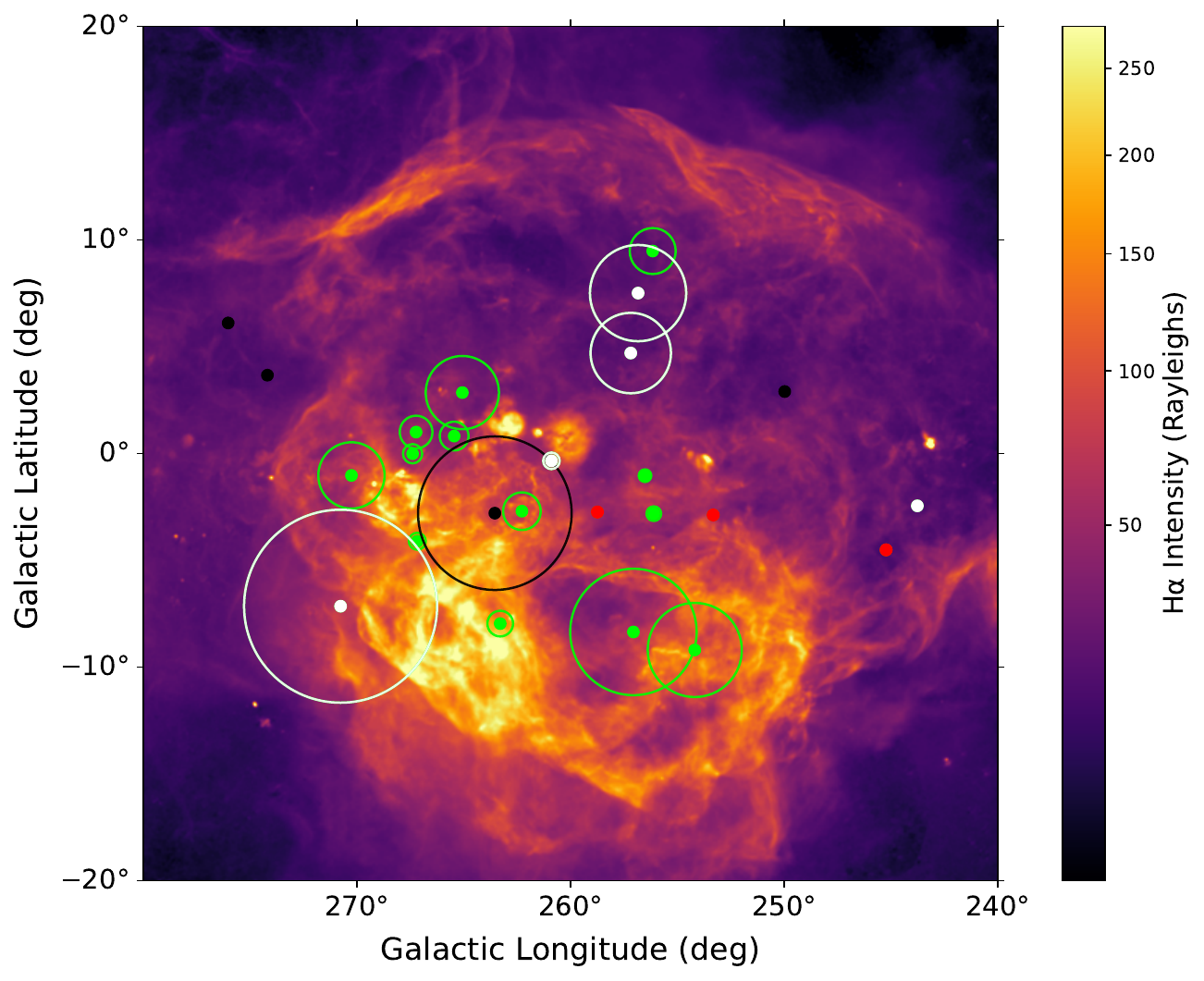}
\caption[Overlay of $C_{n_e}^2$ of pulsars on the H{$\alpha$} image of the Gum nebula]{Same as Figure~\ref{fig2}. The diameter of the circles indicate the scattering strength
in that line of sight, instead of $\tau_{sc}$ as in Figure~\ref{fig2}. The colour code is also the same.}
\label{fig3}
\end{figure*}

We also found 
four $\alpha$ estimates from the literature: PSR J0742$-$2822 from \citet{gkk17}, PSR J0835$-$4510 
from \citet{jnk98}, PSRs J0837$-$4135 and J0840-5332 from \citet{lw3}. 
The results of the fit of PBF1 model, 
including $\tau_{sc}$ at 325~MHz, scattering strength, log $C_{n_{e}}^{2}$, catalogue RM and the magnetic field along the 
line of sight, B$_{\|}$, etc. are given in Table~\ref{c6tab1}. The measurements of $\alpha$ across 
the Gum nebula have increased from a mere four to 17 with our study. Although this is still a small
number, we have started seeing trends in the observed characteristics. A distribution of 
$\tau_{sc}$ and scattering strength across the Gum nebula are shown in Figures~\ref{fig2} and 
\ref{fig3} respectively.

With the assumed distance of 450 pc to the Gum nebula, four pulsars from our sample are not behind the nebula. 
They are PSRs J0818$-$3049, J0835$-$4510, J0945$-$4833 and J1003$-$4747. The following discussion
has to be considered with caution since only PSR J0835$-$4510 has a DM independent distance estimate, 
and for the rest three, distance estimates are from the YMW16 model. Out of these, only for PSR 
J0835$-$4510 we were able to measure $\tau_{sc}$ and estimate $\alpha$, whereas the other three did 
not show any measurable scatter-broadening even at the lowest frequencies. The position of all these four pulsars are plotted in black filled circles in Figures \ref{fig2} and \ref{fig3}. J0835$-$4510 is the pulsar 
in the Vela supernova remnant and scatter-broadening was reported in this pulsar from the earliest observations. It was 
shown by \citet{jnk98} by combining $\tau_{sc}$ and $\delta\nu_d$ measurements across a large 
range of frequencies that the $\alpha$ is 3.93, somewhat lower than the Kolmogorov value. They argued
that this might be due to scattering from discrete sources, like the supernova remnant itself.
We find, with our observations that the scattering spectral index is much flatter ( $\alpha$ = 2.5) than what 
was reported by \citet{jnk98}. This indicates three possibilities: (1) the scattering is a 
frequency dependent phenomena so that the scintillation and scatter-broadening happens at very different scales, (2) combining $\tau_{sc}$ and $\delta\nu_d$ measurements without
the knowledge of the real value of the conversion factor C$_1$ ($2\pi\tau_{sc}\delta\nu_{d} = C_1$) will lead to uncertain $\alpha$ or (3) 
there is a variation in scatter-broadening as a function of observing epochs. We do not have any 
arguments in support of the first case, since our observations are limited to the 275 -- 475 MHz
frequency range only. \citet{lw2} indicated that, the conversion factor between $\tau_{sc}$
and $\delta\nu_d$ might be higher than 1 for this pulsar. This suggests that the scaling of all 
these estimates should be done with caution. Moreover, this will lead to a systematic error, 
which gets propagated to $\alpha$ estimates. Since this pulsar is in a supernova remnant it may show  
variation in $\tau_{sc}$ as a function of observing epochs. If this is true, then combining 
$\tau_{sc}$ measurements from different epochs will produce incorrect $\alpha$ estimates, which should
be treated with caution.
In our study, the observations were performed with the wideband receiver that helped avoid the influence of these time dependent variations. Hence, the estimate of $\alpha$ presented in here is robust
and it is possible that the scatter-broadening of the Vela pulsar is dominated by the materials within the supernova remnant (pulsar wind nebula), but not due to the Gum nebula. The other three pulsars which are located in 
front of the nebula are not showing any measurable scattering evolution in the observed band, indicating 
that they are not being affected by the Gum nebula, although their DMs are moderately high ($\sim$ 100 pc 
cm$^{-3}$). One requires observations below 200~MHz for measuring their scatter-broadening evolution properly and hence to estimate scattering strength.

\begin{table*}[ht]
\caption[The parameters of the pulsars in the Gum nebula region]{The parameters and results of the 23 pulsars studied in this paper is given. For each pulsar, table lists its position in the Galactic longitude and latitude, period, DM, distance, RM, B$_{\|}$, $\tau_{sc}$ at 325 MHz, pulse width W50, $\alpha$ and  log $C_{n_{e}}^{2}$. Pulsars in bold letters indicate the $\alpha$ estimates from literature as discussed in Section~\ref{sec:result}.}
\label{c6tab1}
\centering
\footnotesize
\begin{tabular}{|llcccc|cc|cccc|}
\hline
PSR J      &   Gl    &   Gb   &  Period  &  DM & Distance & RM & B$_{\|}$ & $\tau_{sc}^{325MHz}$ & W50$^{325 MHz}$ & $\alpha$ & log $C_{n_{e}}^{2}$\\
        & (degrees) & (degrees) & (s) & pc cm$^{-3}$ & (kpc) & rad m$^2$ & $\mu$G &  (ms)   & (ms) &    &  (m$^{-20/3}$) \\ 
\hline
J0737$-$3039A & 245.236 & $-$4.505 & 0.022699 & 48.92 & 1.10 & 112.30 &  2.83 & --- & 1.8$\dagger$ &    ---              &  ---  \\
J0738$-$4042 & 254.194 & $-$9.192 & 0.374920 & 160.80 & 1.60 &  12.10 &  0.09 & 82.9 & 89.1 &   3.62$_{-0.01}^{+0.01}$ &  $-$0.10 \\
{\bf J0742$-$2822} & 243.773 & $-$2.444 & 0.166762 &  73.78 & 2.00 & 149.95  & 2.50  & --- & --- & 2.52$_{-0.3}^{+0.3}$ & $-$2.6 \\
J0749$-$4247 & 257.066 & $-$8.349 & 1.095452 & 104.59 & 0.56 & 125.00 &  1.47  & 16.7 &  28.7 & 4.3$_{-0.1}^{+0.1}$ & 0.16 \\
J0804$-$3647 & 253.331 & $-$2.872 & 2.191987 & 196.00 & 3.35 &  ---   &  ---    & --- & 39.3 &    ---              & ---   \\
J0809$-$4753 & 263.301 & $-$7.957 & 0.547199 & 228.30 & 6.49 & 105.00 &  0.57   & 80.2 & 64.3 & 4.07$_{-0.03}^{+0.03}$ & $-$1.23 \\
J0812$-$3905 & 256.111 & $-$2.811 & 0.482596 & 327.00 & 5.83 &  ---   &  ---    & 16.8 & 54.7 & 3.7$_{-0.4}^{+0.4}$ & $-$1.71 \\
J0818$-$3049 & 249.983 &  2.908 & 0.763742 & 133.70 & 0.43 &  ---   &  ---    & --- & 60.1 &    ---              & ---   \\
J0820$-$3826 & 256.523 & $-$1.032 & 0.124836 & 195.56 & 4.09 &  ---   &  ---    & 5.3 & 10.9 & 3.9$_{-0.3}^{+0.3}$ & $-$1.84 \\
J0820$-$4114 & 258.749 & $-$2.735 & 0.545446 & 113.40 & 0.57 &  57.70 &  0.63   & --- & 159.9 &    ---              & ---   \\
J0831$-$4406 & 262.285 & $-$2.694 & 0.311674 & 254.00 & 5.88 & 509.00 &  2.47   & 161.0 & 131.7* & 4.2$_{-0.8}^{+0.8}$ & $-$0.90 \\
J0835$-$4510 & 263.552 & $-$2.787 & 0.089328 &  67.99 & 0.28 &  31.38 &  0.57   & 5.8 & 5.8 & 2.53$_{-0.01}^{+0.01}$ &  0.33 \\
{\bf J0837$-$4135} & 260.904 & $-$0.336 & 0.751624 & 147.29 & 1.50 & 145.00 &  1.21   & --- & --- & 2.4$_{-0.1}^{+0.1}$ & $-$1.61 \\
{\bf J0840$-$5332} & 270.774 & $-$7.143 & 0.720612 & 156.50 & 0.57 &  81.00 &  0.64   & --- & --- & 4.6$_{-0.1}^{+0.1}$ &  0.53 \\
J0842$-$4851 & 267.182 & $-$4.101 & 0.644354 & 196.85 & 3.10 &  ---   &  ---    & 5.8 & 13.1 & 3.99$_{-0.05}^{+0.05}$ & $-$1.59 \\
J0857$-$4424 & 265.455 &  0.820 & 0.326774 & 184.43 & 2.83 & $-$75.00 & -0.50   & 16.7 & 22.0 & 3.5$_{-0.1}^{+0.1}$ & $-$1.13 \\
J0900$-$3144 & 256.162 &  9.486 & 0.011110 &  75.71 & 0.89 &  ---   &  ---    & 4.2 & 4.2 & 3.9$_{-0.7}^{+0.7}$ & $-$0.72 \\
J0901$-$4624 & 267.404 & $-$0.004 & 0.441995 & 198.80 & 3.00 & 289.00 &  1.79   & 7.4 & 12.8 & 2.7$_{-0.4}^{+0.4}$ & $-$1.48 \\
J0904$-$4246 & 265.075 &  2.859 & 0.965171 & 145.80 & 0.68 & 284.00 &  2.40   & 6.8 & 29.7 & 5.1$_{-0.3}^{+0.3}$ &  $-$0.32 \\
J0905$-$4536 & 267.239 &  1.011 & 0.988281 & 179.70 & 1.96 & 153.00 &  1.05   & 24.6 & 198.0 & 4.7$_{-0.5}^{+0.5}$ &  $-$1.02 \\
J0908$-$4913 & 270.266 & $-$1.019 & 0.106755 & 180.37 & 1.00 &  10.00 &  0.07   & 12.7 & 11.6 & 3.3$_{-0.1}^{+0.1}$ &  $-$0.40 \\
J0945$-$4833 & 274.199 &  3.674 & 0.331586 &  98.10 & 0.35 &  ---   &  ---    & --- & 5.7 &    ---              & ---   \\
J1003$-$4747 & 276.037 &  6.117 & 0.307072 &  98.10 & 0.37 &  18.00 &  0.23   & --- & 11.9 &    ---              & ---   \\
\hline
\end{tabular}
\end{table*}

Three more pulsars did not show any scatter-broadening in our observations up to the lowest frequency of 275 MHz, namely, PSRs J0737$-$3039A, 
J0804$-$3647, J0820$-$4114. As can be seen in Figure~\ref{fig2}, the LoS to these pulsars pass 
through regions of low H$\alpha$ intensity, which may be the reason for the low scatter-broadening, although their 
DMs are moderately high ($>$100 pc cm$^{-3}$) except for PSR J0737$-$3039A. The distance to PSR 
J0737$-$3039A is also very well known (1.1$_{-0.1}^{+0.2}$ kpc) and is well beyond the nebula. 
The low scatter-broadening of this pulsar even at the lowest of our observed frequencies is puzzling. 
It is also inferred that the B$_{\|}$ in the LoS towards J0737$-$3039A is comparatively higher than the 
other pulsars in the nebula. More observations below the currently observed frequency range would provide some insight into these pulsars.

Figures~\ref{fig2} and \ref{fig3} show the distribution of the pulsar in the Galactic plane towards
the Gum nebula in terms of $\tau_{sc}$ and $C_{n_e}^2$ respectively. The value of $C_{n_e}^2$ is estimated using the relation $C_{n_e}^2$ = 0.002$\nu^{11/3}$ D$^{-11/6}$ $\delta\nu_{d}^{-5/6}$, where $\nu$ is the observing frequency in GHz, $D$ is the distance in kpc and $\delta\nu_d$ is the scintillation bandwidth converted from $\tau_{sc}$ using the relation $2\pi\tau_{sc}\delta\nu_d=1$ \citep{cordes86}. The variation in the density of the
nebula can be seen in the figures. For pulsars passing through high H$\alpha$ intensity regions in the nebula, the scatter-broadening 
shows enhancements. This indicates the fact that an increase in the density increases the turbulence 
(n$_e$ $\propto$ $\delta n_e$). This enhancement in the density turbulence did not show any effect in flattening $\alpha$. The density turbulence can also affect the magnetic field strength of the nebula or vice versa. 
We did not see any evidence for this also, indicating that the effects of magnetic field on density turbulence 
for the nebula is negligible, based on the current sample. It should be noted that the above measurements of B$_{\|}$ used integrated values of DM and RM which will include contributions from magnetic field and density outside the nebula as well. A Spearman rank correlation between $C_{n_e}^2$ and B$_{\|}$ was found to be 0.1 with a significance of 0.67 and between $C_{n_e}^2$ and $\alpha$ was 0.24 with a significance of 0.36 confirming the above conclusion.

\begin{figure*}[t]
\centering
\includegraphics[scale=0.5]{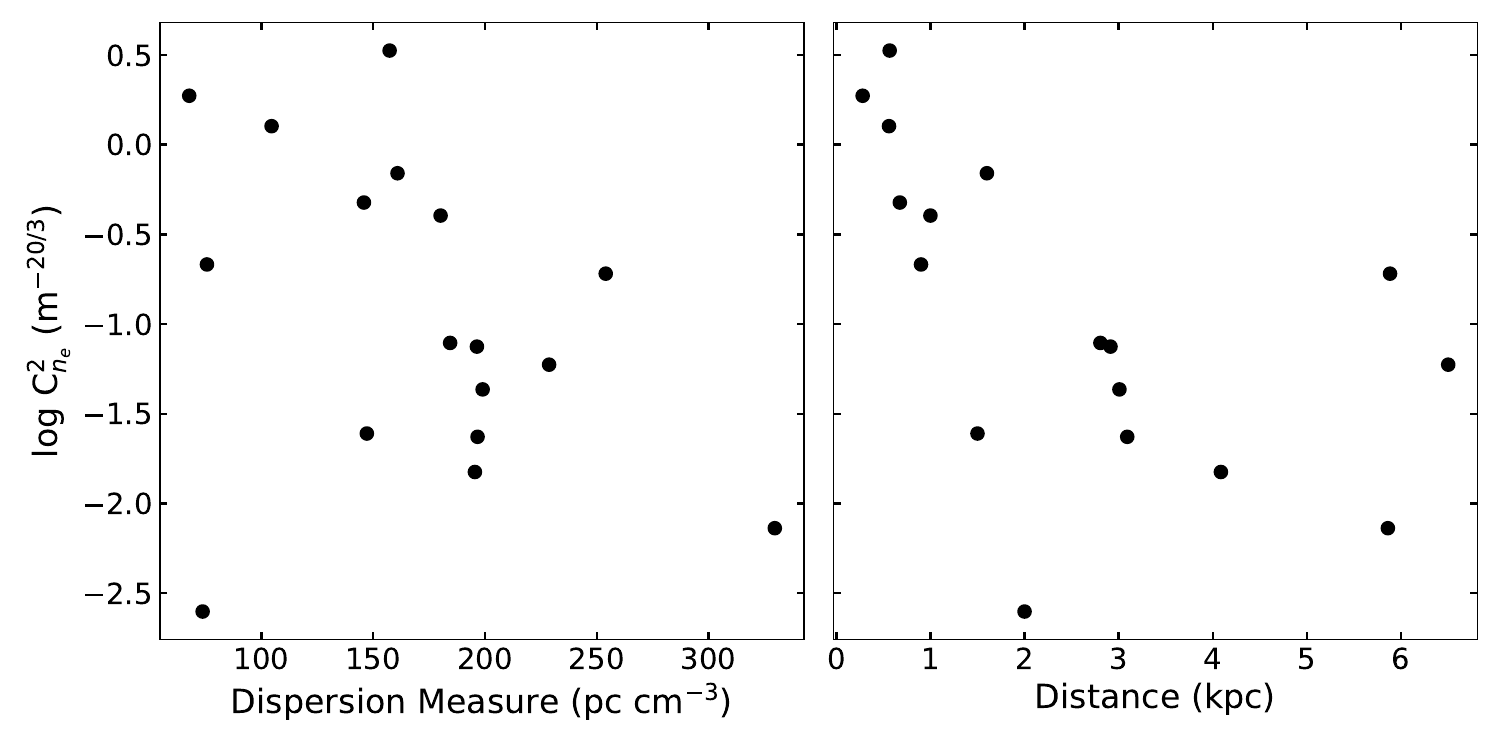}
\caption[Distribution of log $C_{n_e}^2$ as a function of DM and distance]{Distribution of the scattering strength of the 17 pulsars under study is plotted as a function of 
DM and distance in left and right hand side panels. A trend of decrease in scattering strength as a 
function of DM and distance is clearly visible.}
\label{fig4}
\end{figure*}

It was noticed in \cite{kjm17} that the scattering strength, in general show an increase as a 
function of DM and distance due to the fact that the inhomogeneities in the LoS increases as a function of 
both these parameters. Using the sample of 17 $\alpha$ estimates from this study, we examined the distribution 
of scattering strength or $\alpha$ in different parameter spaces. Although the current sample is small, we 
did see a dependence of scattering strength on DM and distance as shown in Figure~\ref{fig4}.
This is contrary to what is expected and seen till date. This indicates that, low DM
pulsars, in general show enhanced scattering instead of the high DM pulsars in the LoS towards Gum nebula. 
A Spearman's $\rho$ rank correlation analysis between $C_{n_e}^2$ and distance/DM wass performed. 
The Spearman rank correlation between $C_{n_e}^2$ and distance was found to be --0.75 
with a significance of 0.0002 and that between $C_{n_e}^2$ and DM was found to be 
-0.46 with a significance of 0.05. 
This indicates the scattering strength and DM/distance
to pulsars are highly anti-correlated, i.e., as the distance/DM increases, $C_{n_e}^{2}$ decreases. A clear trend is seen in the figure.
In addition, we did not see any clear dependence of $\alpha$ on any of these parameters. With this result, 
one could say that the scatter-broadening of pulsars close to the Gum nebula (D $< \sim2kpc$) are affected strongly by the density 
turbulence in the nebula, whereas for the distant pulsars, such a strong dependence on the density of the 
nebula is not seen. This conclusion has to be considered with caution, since the sampling of the nebula with this study 
is sparse. One need to increase the sample to at least 50 pulsars across the nebula to make any strong 
conclusion about the dependence of scattering strength and $\alpha$ on various other LoS parameters.

\section{Summary}\label{sec:sum}
Scatter-broadening of pulsars show anomalous behaviour in some LoS. In this work, we studied one such region in the Galaxy, 
the Gum nebula, spread over a diameter of about 35 degrees at a distance of 450~kpc, and its effects on 
scatter-broadening of pulsars. We have observed a set of 20 pulsars in total using the uGMRT Band--3 for 
this purpose. We were able to measure $\tau_{sc}$ and $\alpha$ only for a set of 14 pulsars while the rest of them did not show any measurable scatter-broadening. In addition, 
for four pulsars (one is common to our study) in this region, $\alpha$ estimates are available in literature. We performed a multi-model comparison using the $\Delta$AIC/$\Delta$BIC approach and found that for most of the pulsars in the sample ($\sim$65\%), a thin screen approximation model (PBF1) was the best choice.
A clear dependence of scatter-broadening and $C_{n_e}^2$ on the density variations across the 
nebula is seen. A clear anti-correlation between the $C_{n_e}^2$ and distance/DM is seen, showing that
the effect of the density turbulence of the nebula is more seen on the pulsars close to the nebula than the distant ones. 
We did not find any direct dependence of scattering strength on RM or magnetic field strength, indicating
that these two are not related, at least with the sample of pulsars that we have. 

To understand the nebula characteristics better, we would require simultaneous wide frequency range measurements of $\tau_{sc}$, $\delta\nu_d$ and $\alpha$ estimates. This requires a concerted observing campaign at 
much lower and higher frequencies apart from the band which we have used in this study. An observing campaign to expand this sample to understand the turbulence characteristics better, is being planned.

\section{Acknowledgements}

We thank the staff of the GMRT who have made these observations possible. The GMRT is run by the National Centre for Radio Astrophysics of the Tata Institute of Fundamental Research. We also thank the referee, whose comments have helped improve the manuscript. KMA acknowledges support from the Department of Atomic Energy project ``Next Generation Instrumentation for Radio Astronomy'' with UID RTI4017. PKM acknowledges support from NASA GSFC through the Cooperative Agreement to the Catholic University of America in support of the Partnership for Heliophysics and Space Environment Research (PhaSER) under the grant 80NSSC21M0180.
\vspace{-1em}

\bibliography{references}

\clearpage
\onecolumn

\appendix
\section{Profile scatter-broadening fit plots}
\label{app1}
\renewcommand{\thefigure}{A\arabic{figure}}
\setcounter{figure}{0}
The scatter-broadening fits for all the 14 pulsars with PBF1 is shown in this section. In each figure, the left side panel shows the observed profile in each sub-band (black curves) and the best fit model (red curves). The template used to fit the profile is shown at the top in black dashed curves and the original profile in blue curves. The center frequency of each sub-band is given on the left side of each profile. The middle panel shows the reduced $\chi^{2}_{r}$ of each profile fit. The right side panel shows the plot of $\tau_{sc}$ as a function of frequency in log-log scale. The best fit power law model is shown as a continuous curve in green colour. The final value of $\alpha$ is printed at the top of the panel. All the figures in this appendix are in the same format.
\begin{figure}[H]
\centering
\includegraphics[scale=0.5]{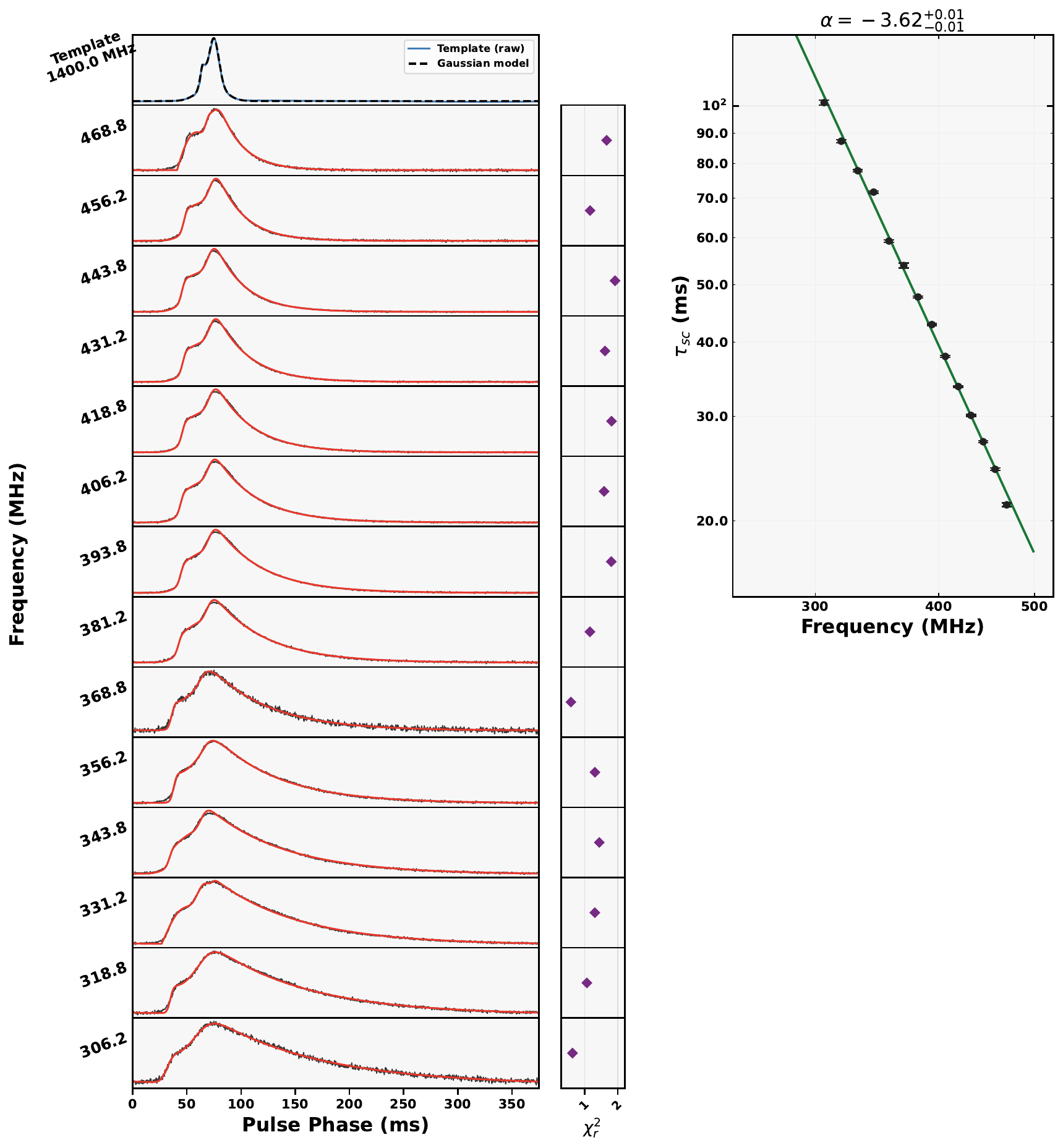}
\caption{PSR J0738$-$4042}
\label{appfig1}
\end{figure}


\begin{figure}[H]
\centering
\includegraphics[scale=0.5]{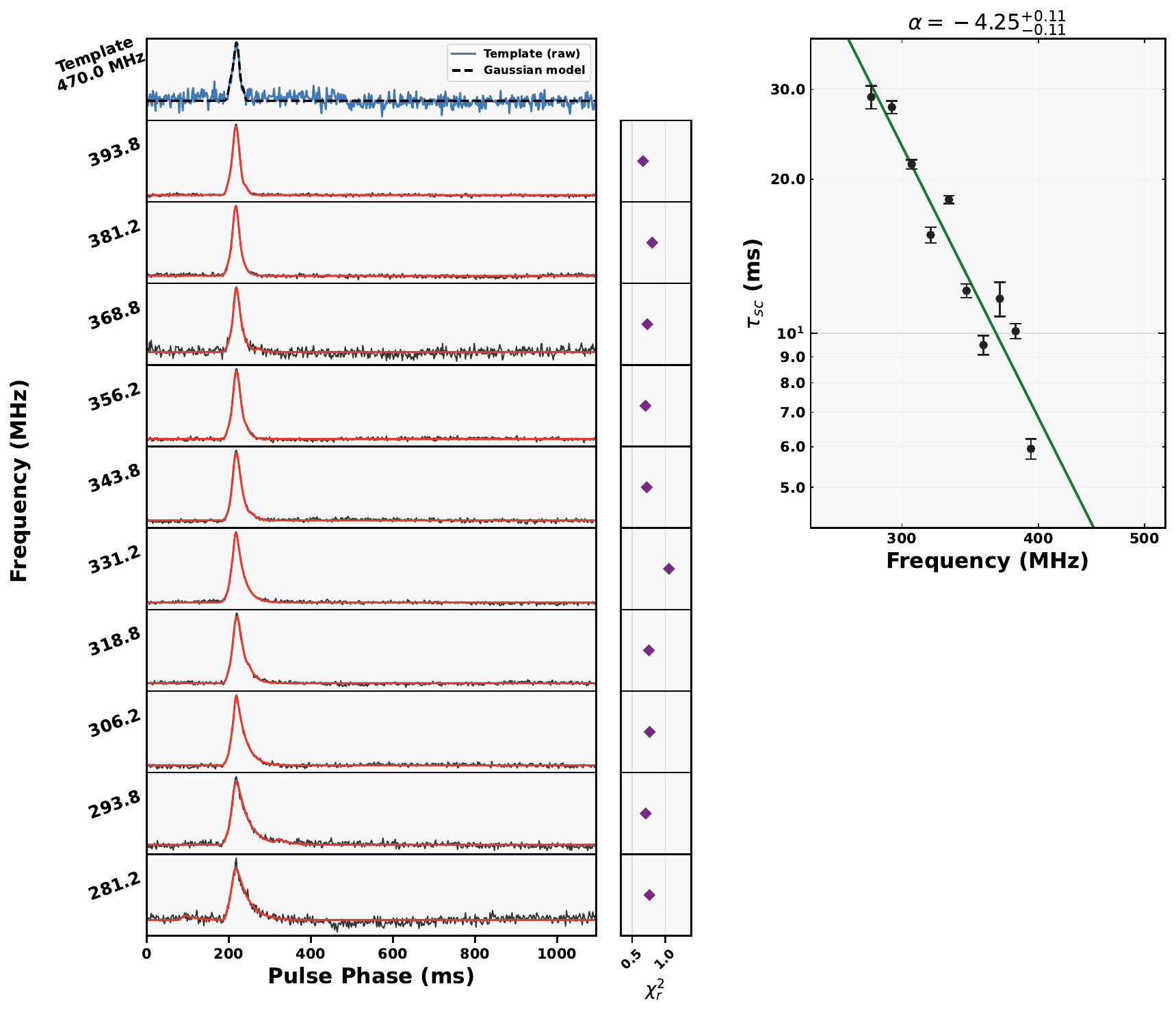}
\caption{PSR J0749$-$4247} 
\label{appfig2}
\end{figure}

\begin{figure}[H]
\centering
\includegraphics[scale=0.5]{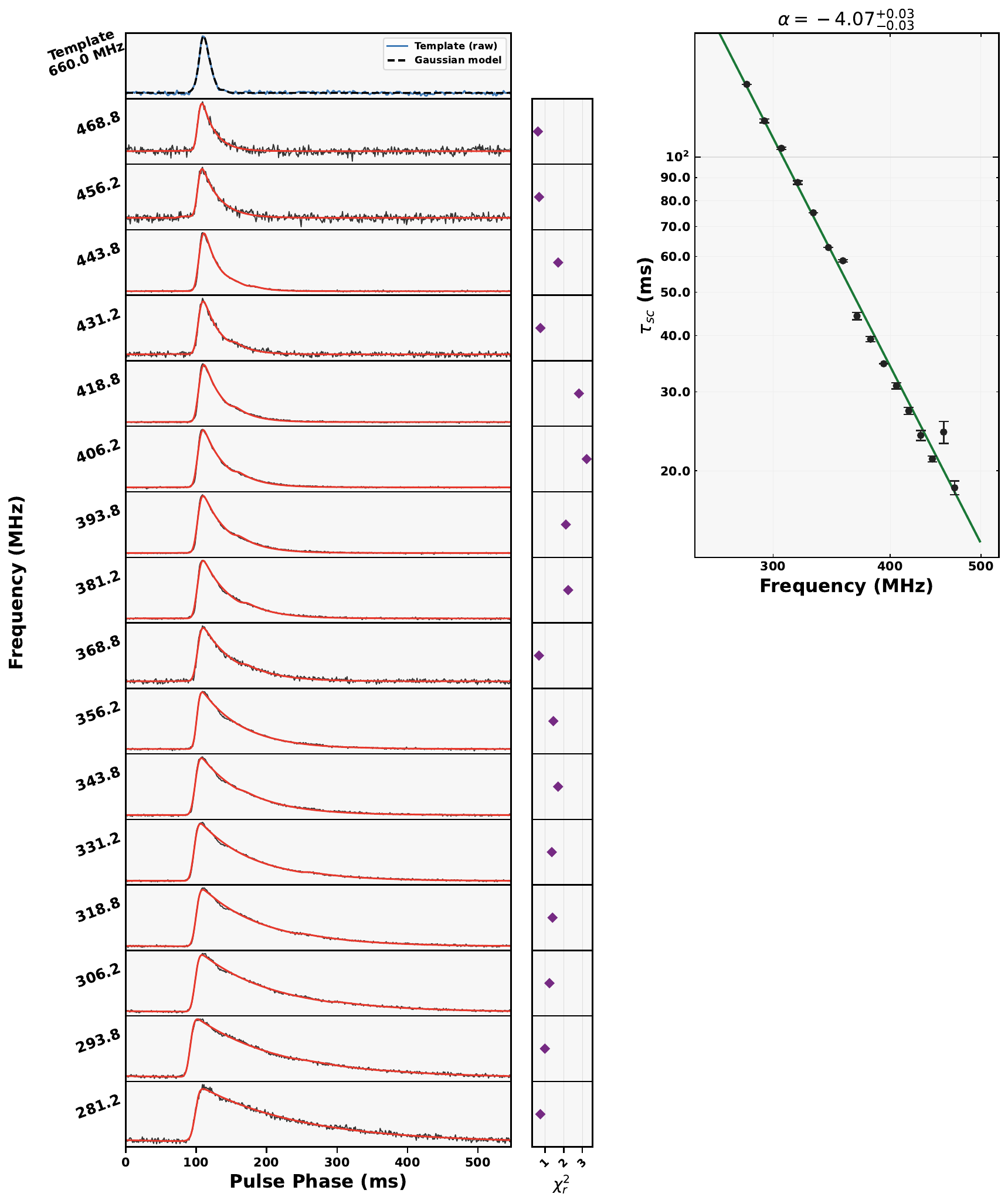}
\caption{PSR J0809$-$4753} 
\label{appfig3}
\end{figure}

\begin{figure}[H]
\centering
\includegraphics[scale=0.5]{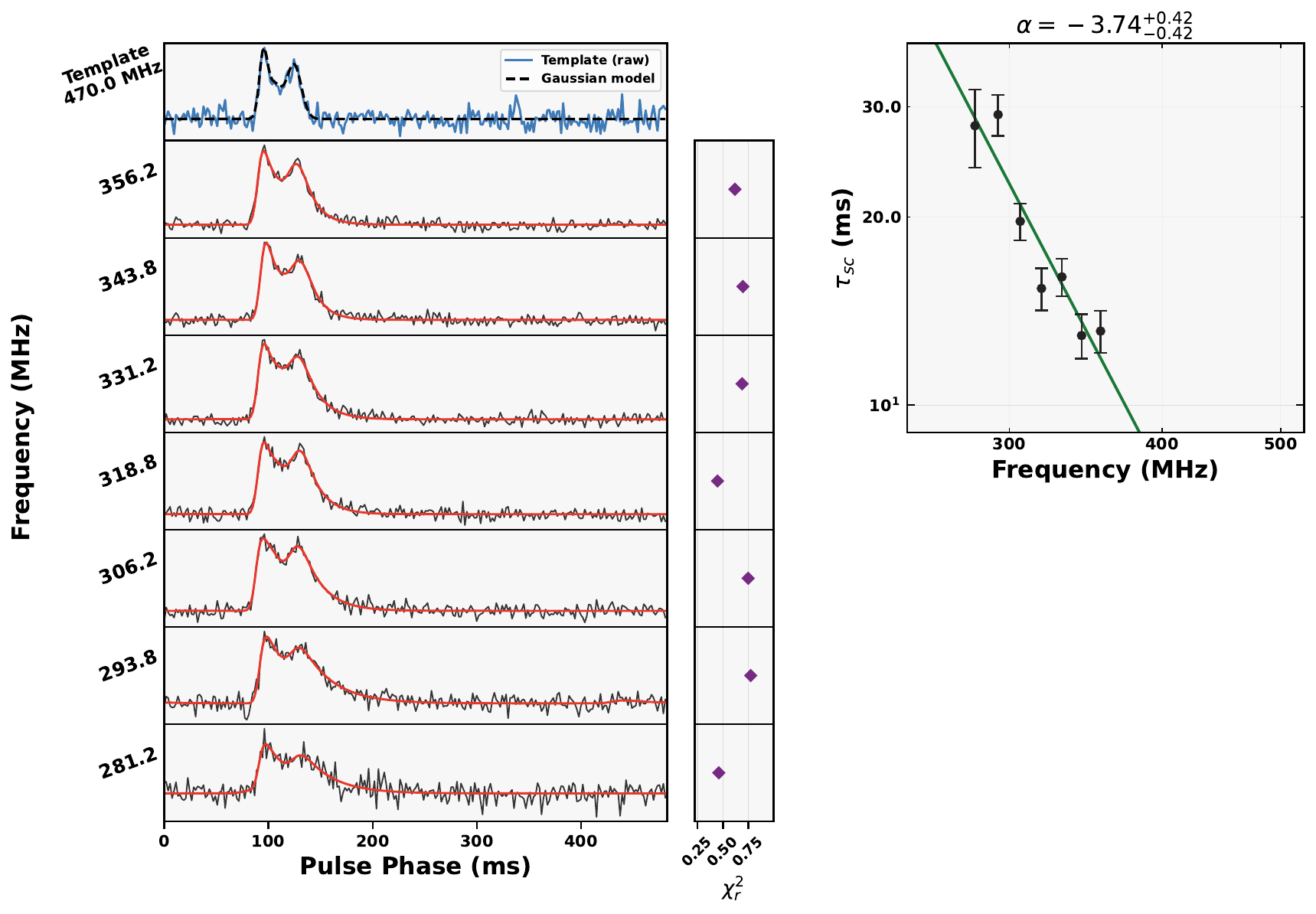}
\caption{PSR J0812$-$3905} 
\label{appfig4}
\end{figure}

\begin{figure}[H]
\centering
\includegraphics[scale=0.5]{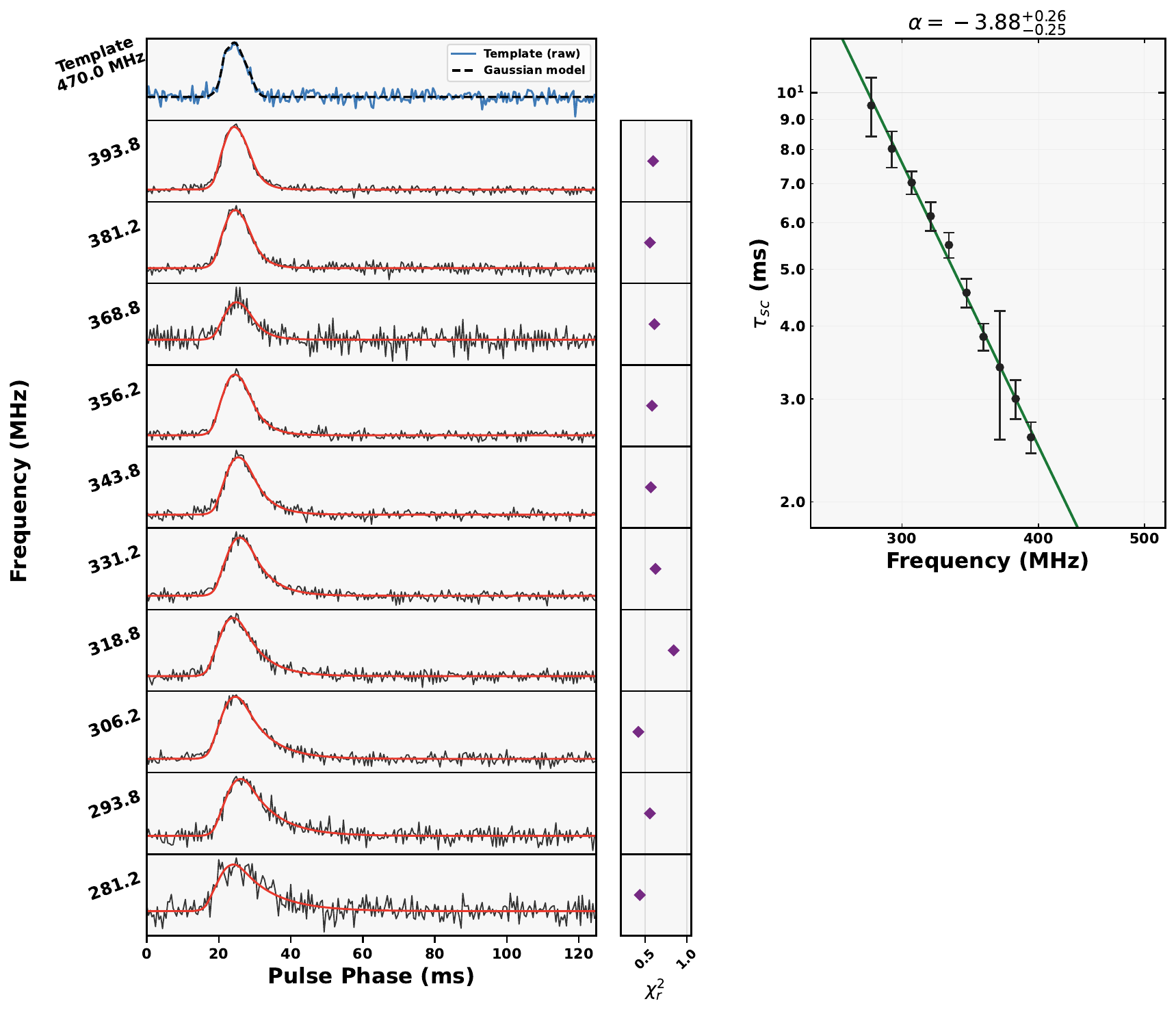}
\caption{PSR J0820$-$3826} 
\label{appfig5}
\end{figure}

\begin{figure}[H]
\centering
\includegraphics[scale=0.5]{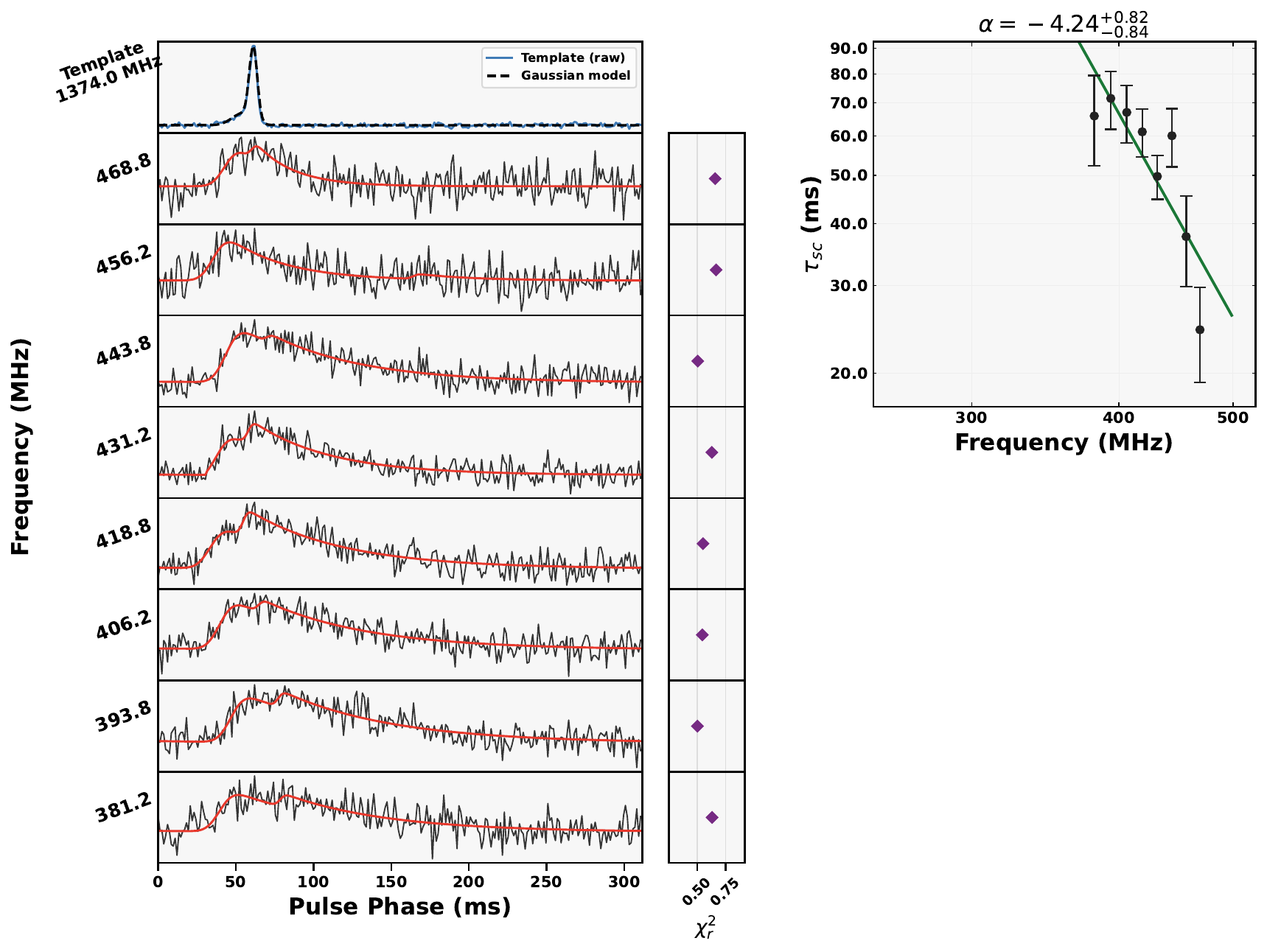}
\caption{PSR J0831$-$4406} 
\label{appfig6}
\end{figure}

\begin{figure}[H]
\centering
\includegraphics[scale=0.5]{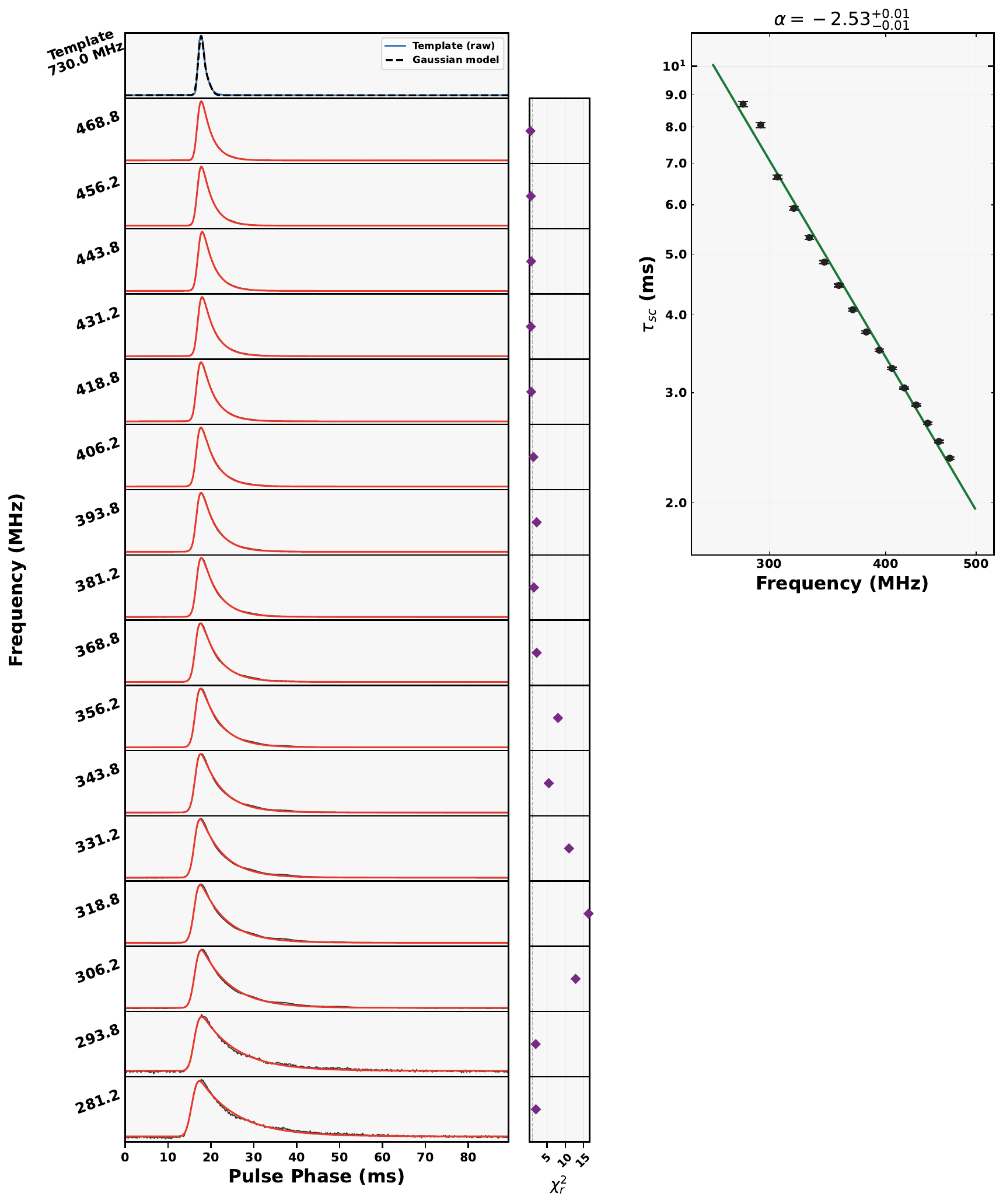}
\caption{PSR J0835$-$4510} 
\label{appfig7}
\end{figure}

\begin{figure}[H]
\centering
\includegraphics[scale=0.5]{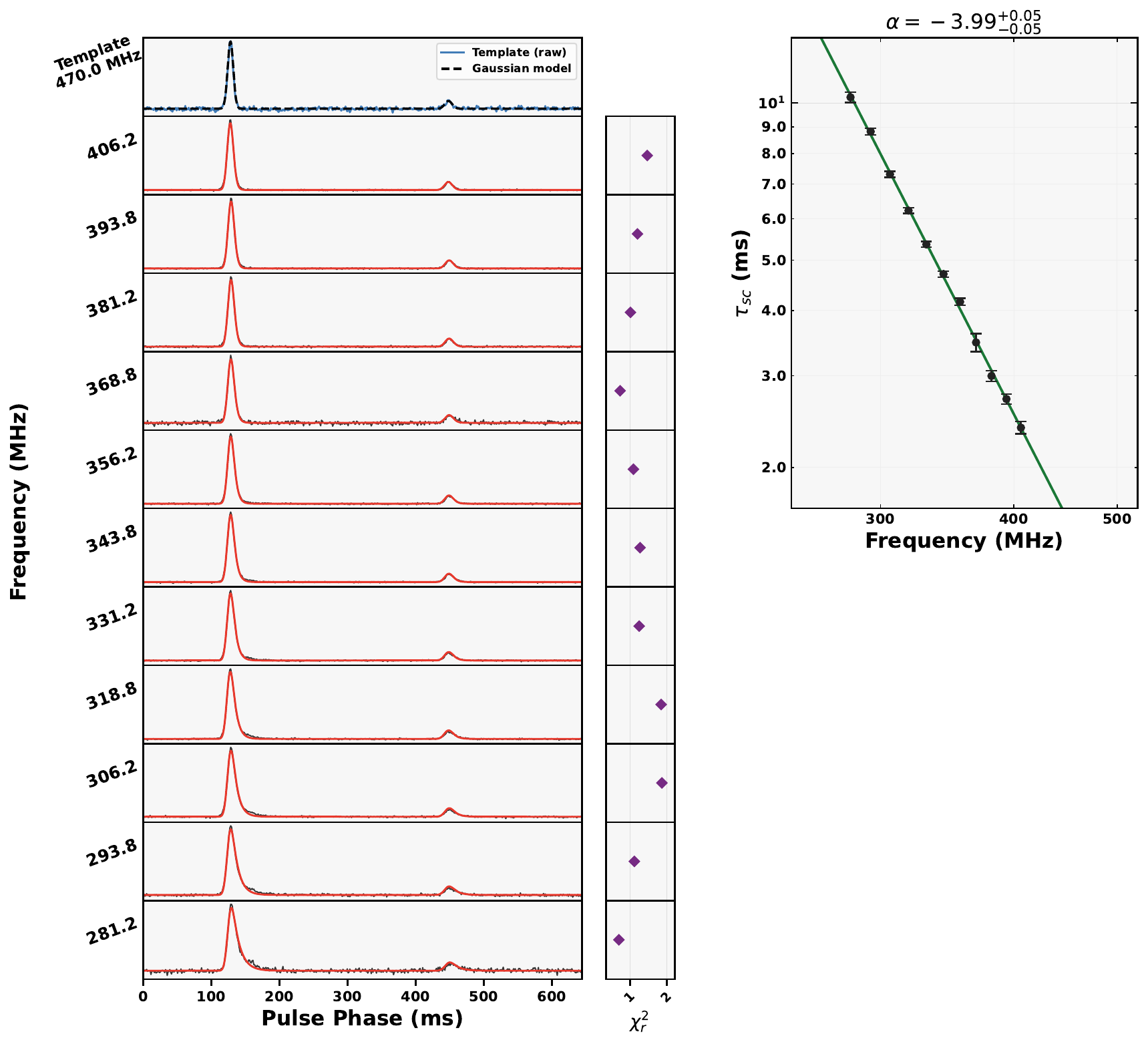}
\caption{PSR J0842$-$4851} 
\label{appfig8}
\end{figure}

\begin{figure}[H]
\centering
\includegraphics[scale=0.5]{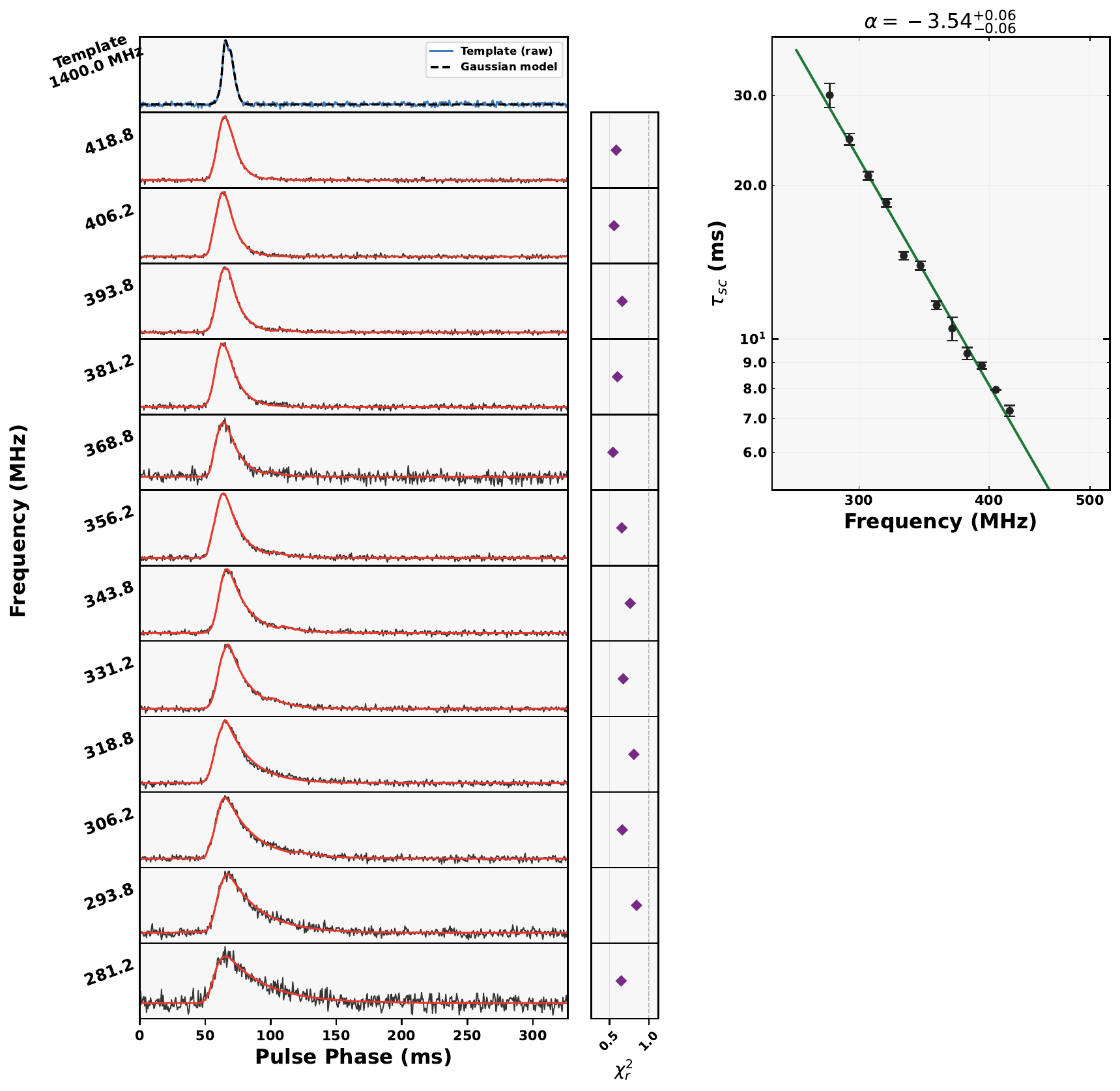}
\caption{PSR J0857$-$4424} 
\label{appfig9}
\end{figure}

\begin{figure}[H]
\centering
\includegraphics[scale=0.5]{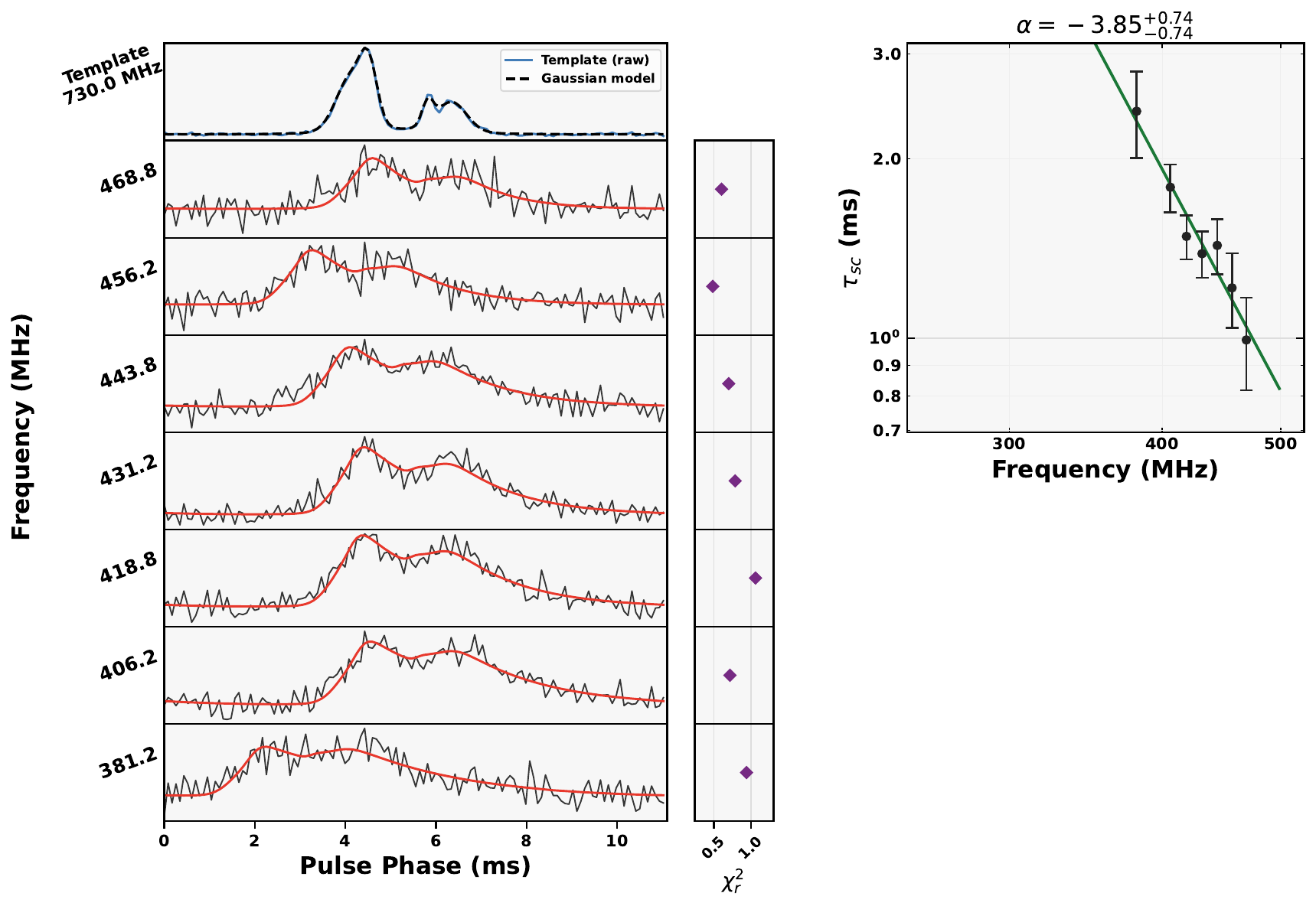}
\caption{PSR J0900$-$3144} 
\label{appfig10}
\end{figure}

\begin{figure}[H]
\centering
\includegraphics[scale=0.5]{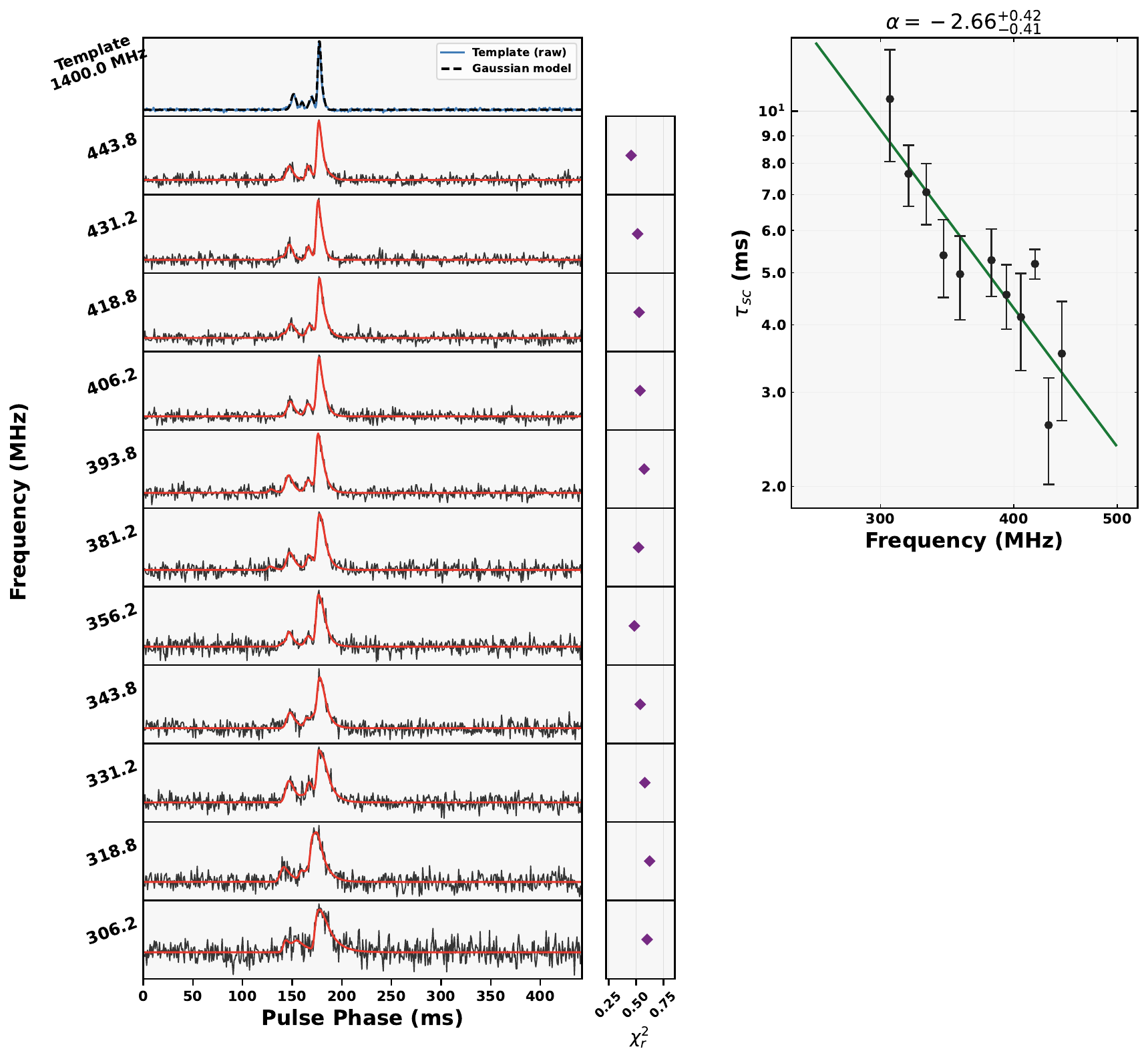}
\caption{PSR J0901$-$4624} 
\label{appfig11}
\end{figure}

\begin{figure}[H]
\centering
\includegraphics[scale=0.5]{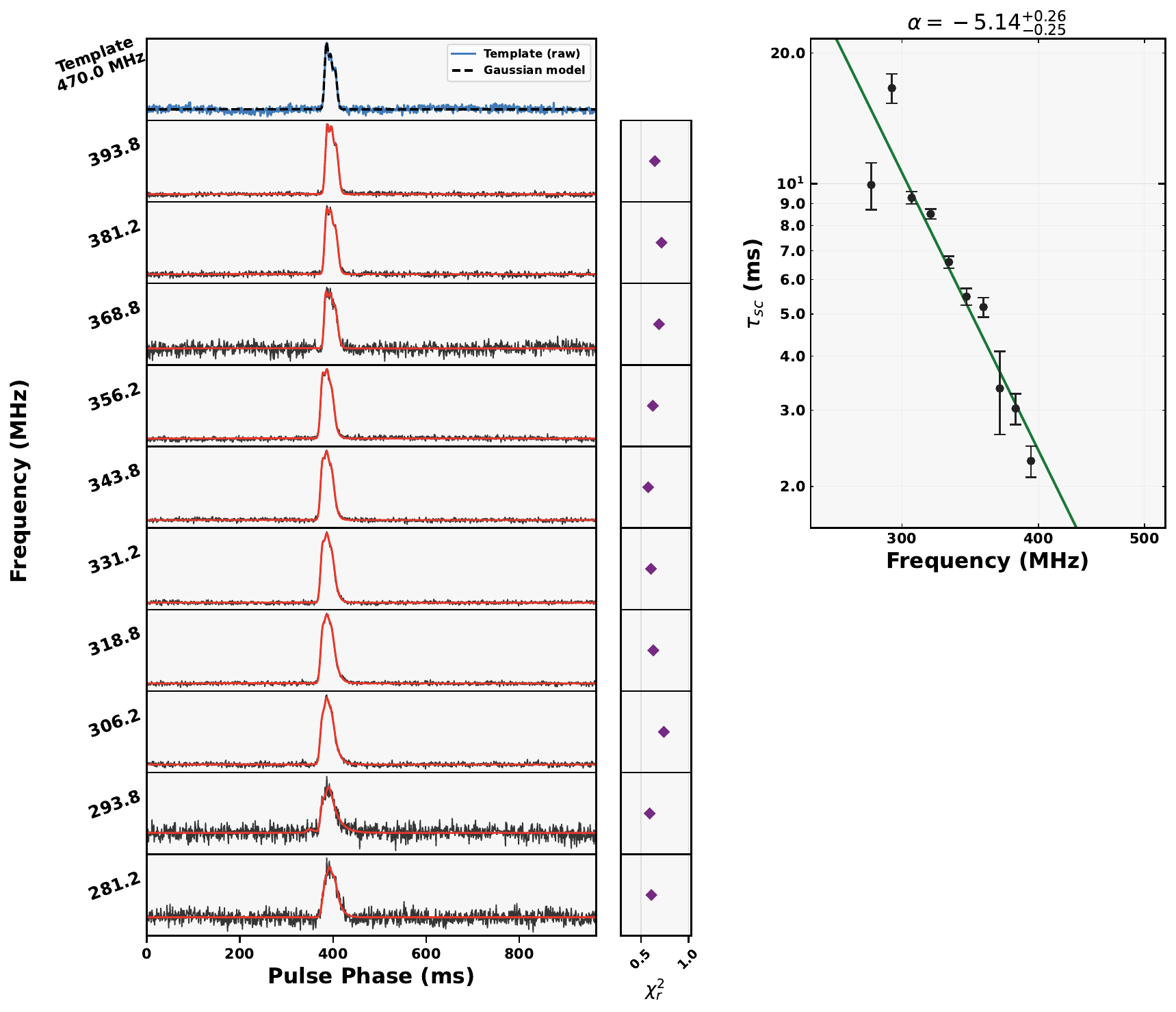}
\caption{PSR J0904$-$4246} 
\label{appfig12}
\end{figure}

\begin{figure}[H]
\centering
\includegraphics[scale=0.5]{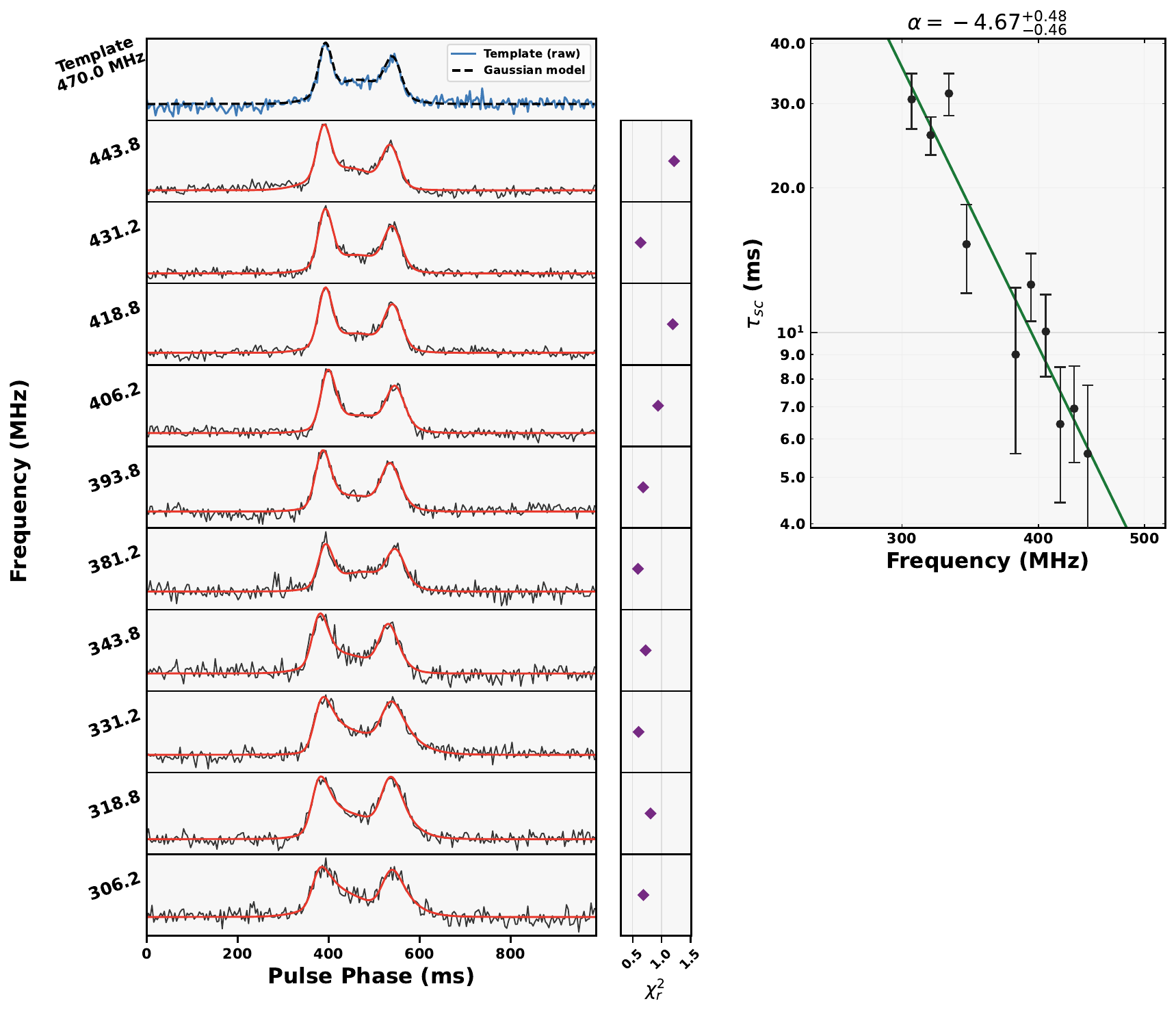}
\caption{PSR J0905$-$4536} 
\label{appfig13}
\end{figure}

\begin{figure}[H]
\centering
\includegraphics[scale=0.5]{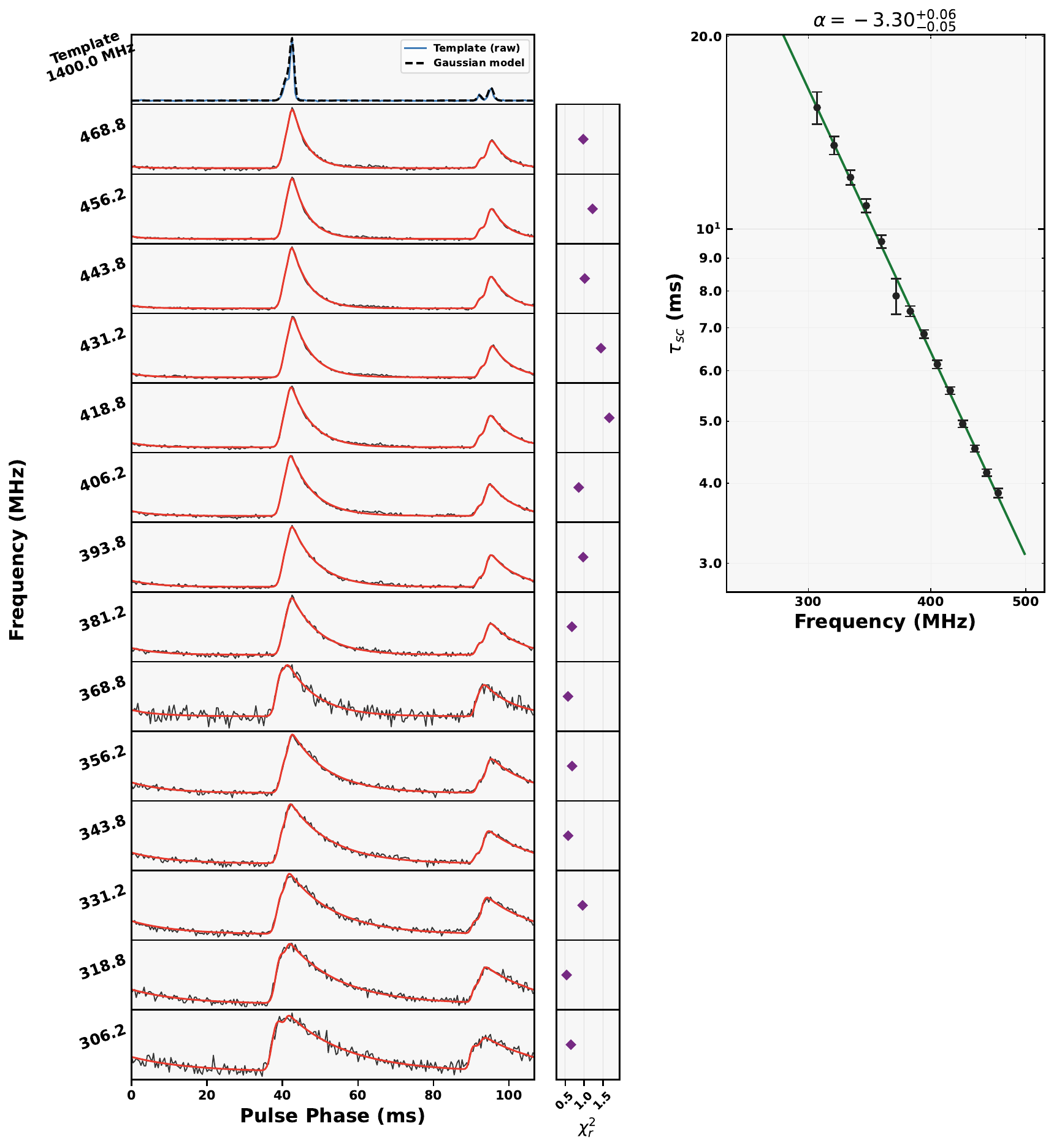}
\caption{PSR J0908$-$4913} 
\label{appfig14}
\end{figure}

\clearpage
\FloatBarrier
\section{PBF Model Selection and AIC/BIC Analysis}
In this appendix, we show the results of the analysis for detecting the best pulse broadening function for a specific pulsar. For each pulsar, the left panel shows the scattering measurements as a function of frequency and respective fits of their frequency scaling index, $\alpha$. Orange filled circles and curves  represent the profiles fitted with PBF1 model, blue filled squares and curves represent PBF2 and green filled triangles and curves represent PBF3 respectively. The right panel shows the model comparison where evidence against each model relative to the best is shown. Values that has zero (or close to zero) $\Delta$AIC/$\Delta$BIC is selected as the best fit model. The table below lists th edifferent $\alpha$ values recovered by each models and the $\Delta$AIC/$\Delta$BIC values.
\label{app2}
\renewcommand{\thefigure}{B\arabic{figure}}
\setcounter{figure}{0}
\renewcommand{\thetable}{B\arabic{table}}
\setcounter{table}{0}

\begin{table}[ht]
\centering
\caption{Fit results for all PBFs. The table lists, for each pulsar, the values of $\alpha$, $\tau_{sc}$ at 325 MHz and $\Delta$BIC values of all the three PBFs. The PBF having $\Delta$BIC value set at 0.0 is the chosen PBF by the $\Delta$BIC method for each of the pulsars.}
\label{tab:placeholder}

\begin{tabular}{|c|ccc|ccc|ccc|}
\hline
\multirow{2}{*}{PSR}
& \multicolumn{3}{c|}{$\alpha$}
& \multicolumn{3}{c|}{$\tau_{sc}^{325MHz}$}
& \multicolumn{3}{c|}{$\Delta$BIC} \\
\cline{2-10}
& PBF1 & PBF2 & PBF3
& PBF1 & PBF2 & PBF3
& PBF1 & PBF2 & PBF3 \\
\hline
J0738-4042 & 3.6 & 6.7 & 6.5 & 82.9 & 78.4 & 353.2 & \textbf{0.0} & 21709 & 99041 \\
J0749-4247 & 4.3 & -- & 4.4 & 16.5  & -- & 30.6 & 40.0 & -- & \textbf{0.0} \\
J0809-4753 & 4.1 & 6.7 & 6.3 & 80.2  & 44.0 & 275.6 & \textbf{0.0} & 2944 & 11187 \\
J0812-3905 & 3.7 & 5.9 & 3.6 & 16.8  & 4.7 & 33.2 & 34.3 & 2.0 & \textbf{0.0} \\
J0820-3826 & 3.9 & -- & 4.9 & 5.3   & -- & 11.5 & \textbf{0.0} & -- & 48.6 \\
J0831-4406 & 4.2 & 6.5 & 5.6 & 161.0 & 269.0 & 847.5 & \textbf{0.0} & 1.2 & 205 \\
J0835-4510 & 2.5 & 4.7 & 4.0 & 5.8   & 3.4 & 16.1 & \textbf{0.0} & 17587 & 20306 \\
J0842-4851 & 4.0 & -- & 5.1 & 5.8   & -- & 11.5 & 1390 & -- & \textbf{0.0} \\
J0857-4424 & 3.5 & 6.2 & 4.6 & 16.7  & 8.6 & 38.9 & \textbf{0.0} & 359.9 & 252.6 \\
J0900-3144 & 3.8 & 4.4 & 5.9 & 4.2   & 3.6 & 22.0 & \textbf{0.0} & 19.2 & 96.0 \\
J0901-4624 & 2.7 & -- & 2.9 & 7.4   & -- & 13.3 & \textbf{0.0} & -- & 6.6 \\
J0904-4246 & 5.1 & -- & 4.8 & 6.8   & -- & 11.8 & 73.2 & -- & \textbf{0.0} \\
J0905-4536 & 4.6 & -- & 5.2 & 24.6  & -- & 43.6 & \textbf{0.0} & -- & 41.4 \\
J0908-4913 & 3.3 & 4.3 & 4.8 & 12.7  & 11.3 & 61.8 & 26.4 & \textbf{0.0} & 7617 \\
\hline
\end{tabular}

\end{table}

\begin{figure}[H]
\centering
\includegraphics[scale=0.5]{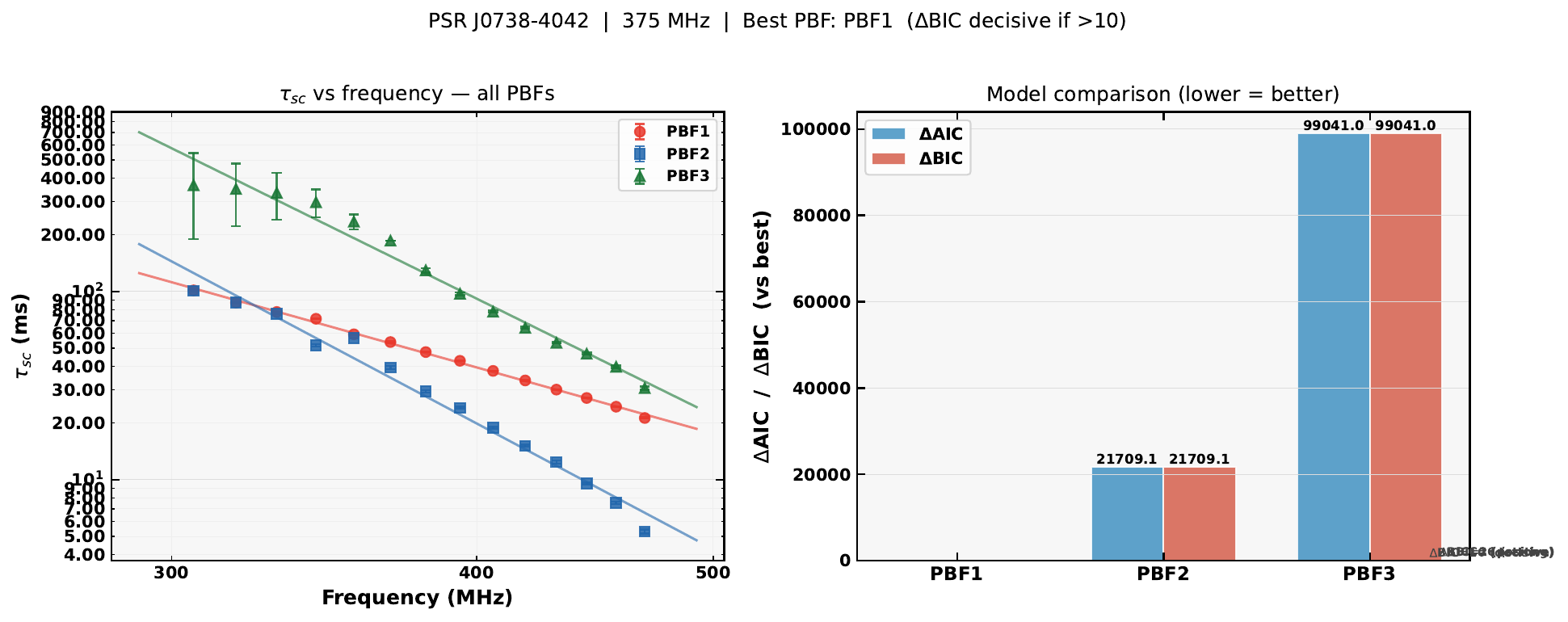}
\caption{PSR J0738$-$4042} 
\label{appfig15}
\end{figure}

\begin{figure}[H]
\centering
\includegraphics[scale=0.5]{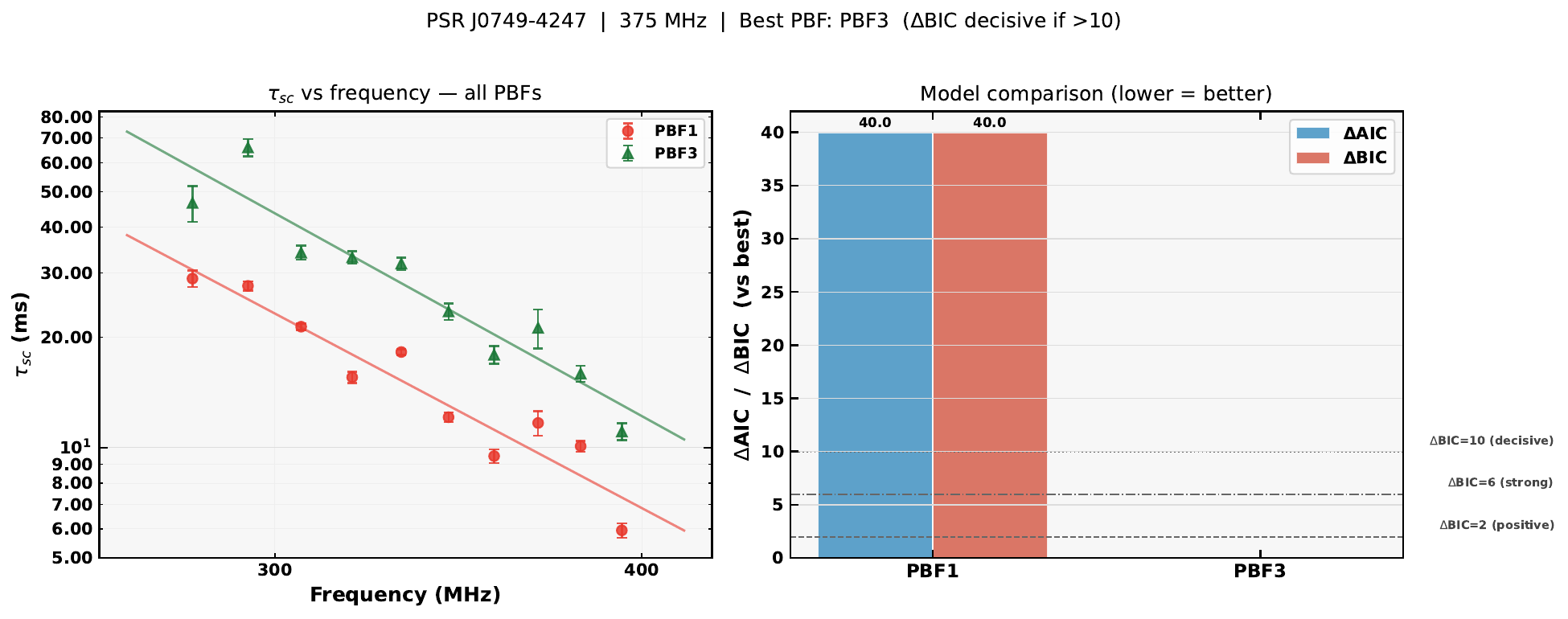}
\caption{PSR J0749$-$4247} 
\label{appfig16}
\end{figure}

\begin{figure}[H]
\centering
\includegraphics[scale=0.5]{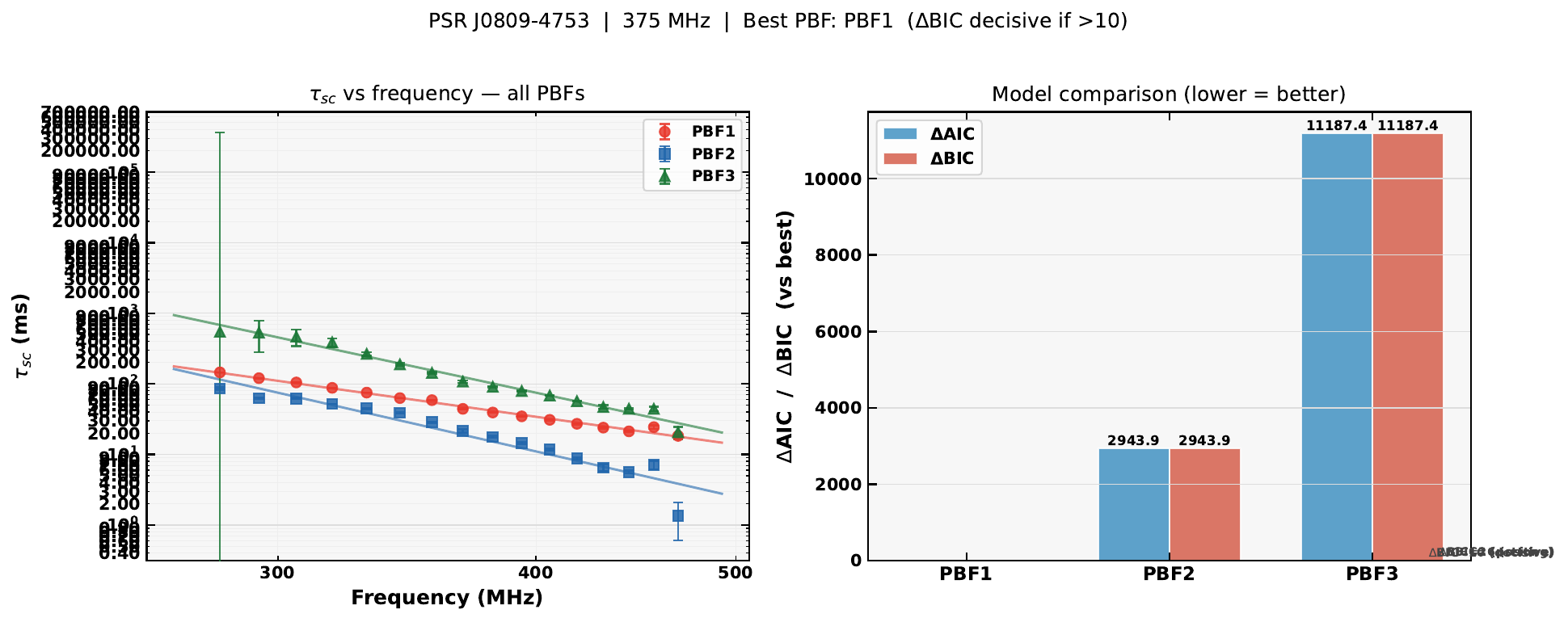}
\caption{PSR J0809$-$4753} 
\label{appfig17}
\end{figure}

\begin{figure}[H]
\centering
\includegraphics[scale=0.5]{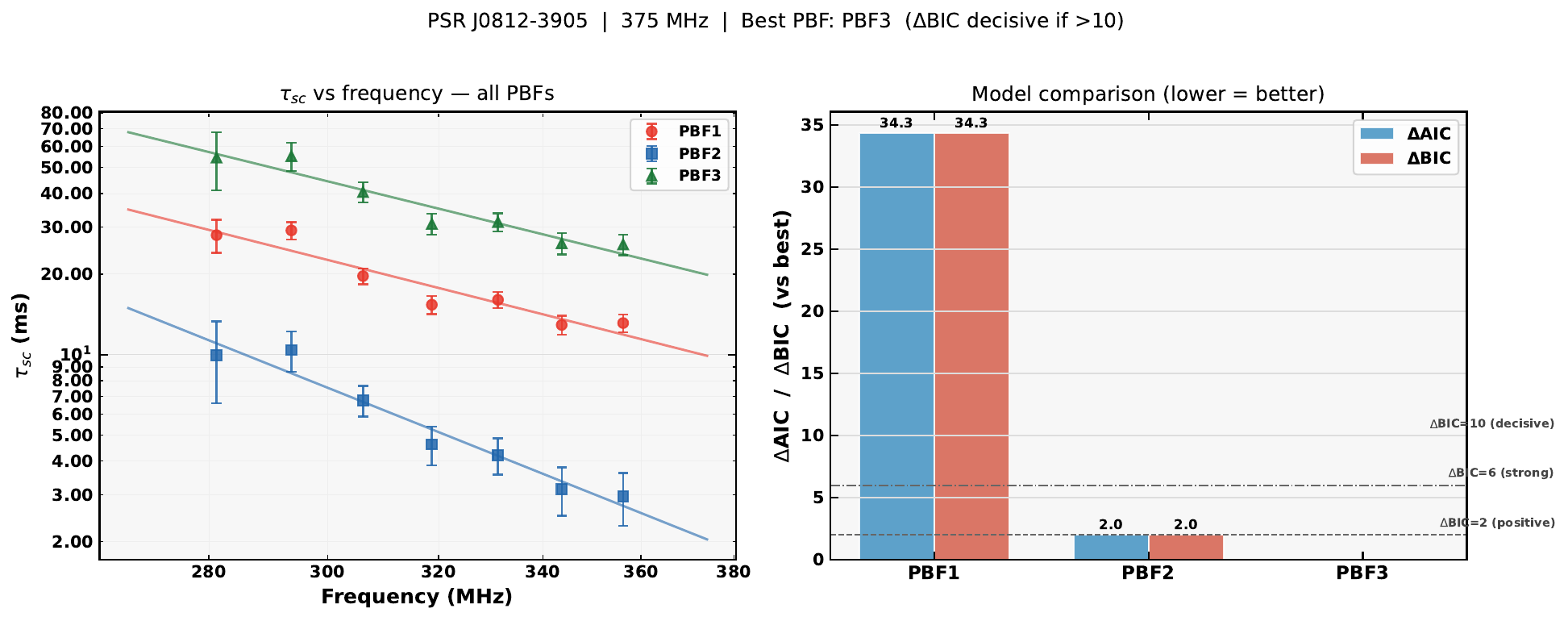}
\caption{PSR J0812$-$3905} 
\label{appfig18}
\end{figure}

\begin{figure}[H]
\centering
\includegraphics[scale=0.5]{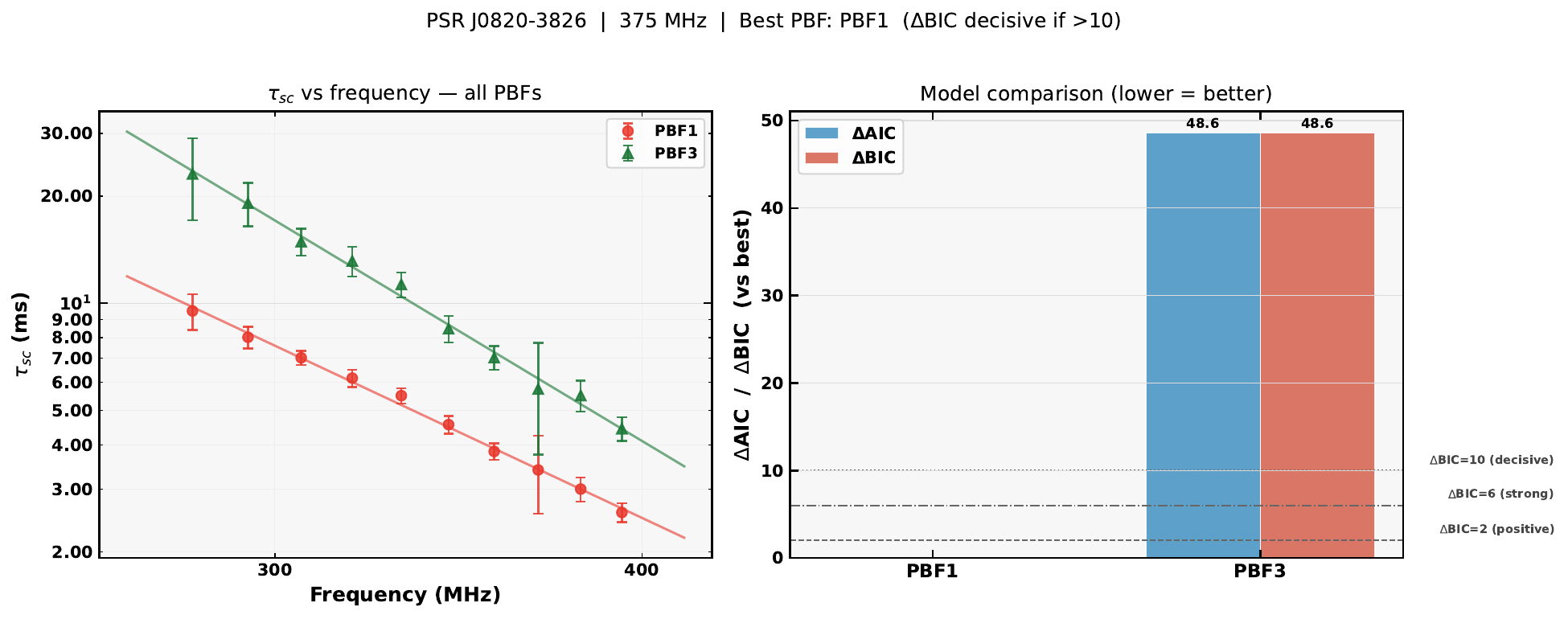}
\caption{PSR J0820$-$3826} 
\label{appfig19}
\end{figure}

\begin{figure}[H]
\centering
\includegraphics[scale=0.5]{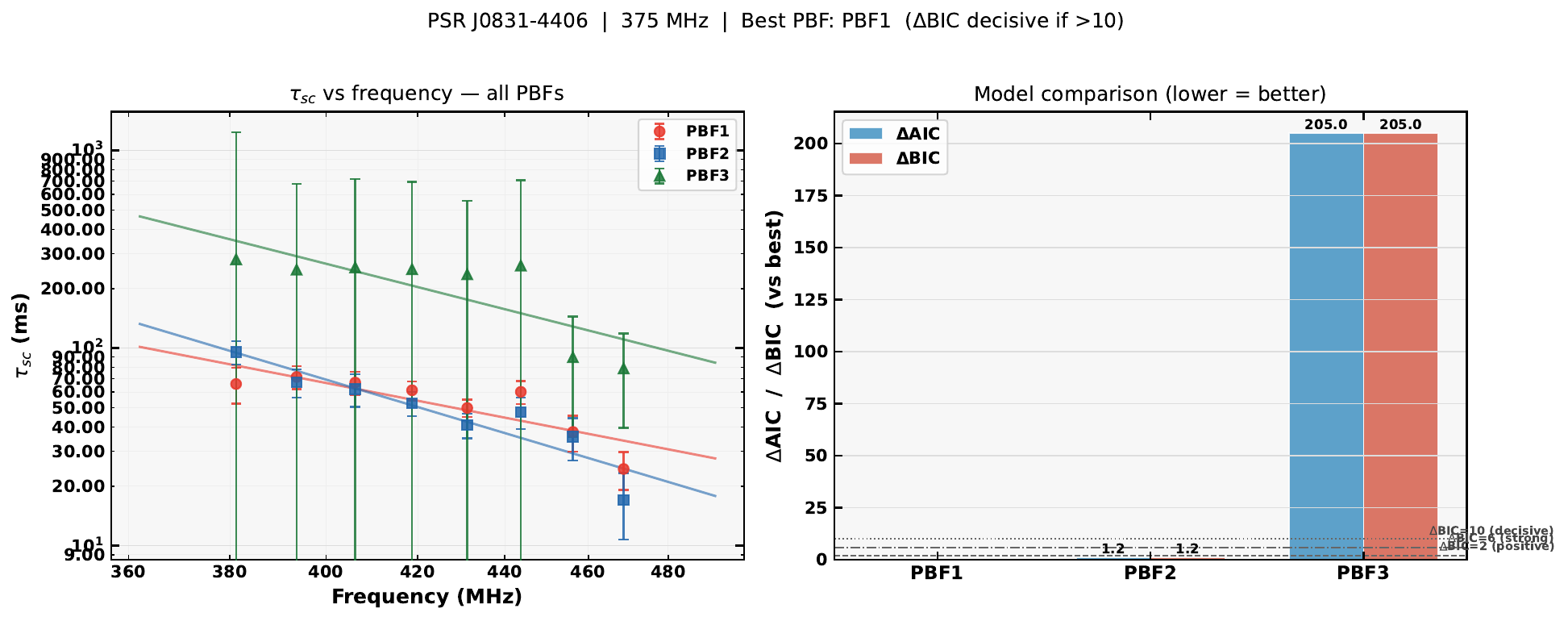}
\caption{PSR J0831$-$4406} 
\label{appfig20}
\end{figure}

\begin{figure}[H]
\centering
\includegraphics[scale=0.5]{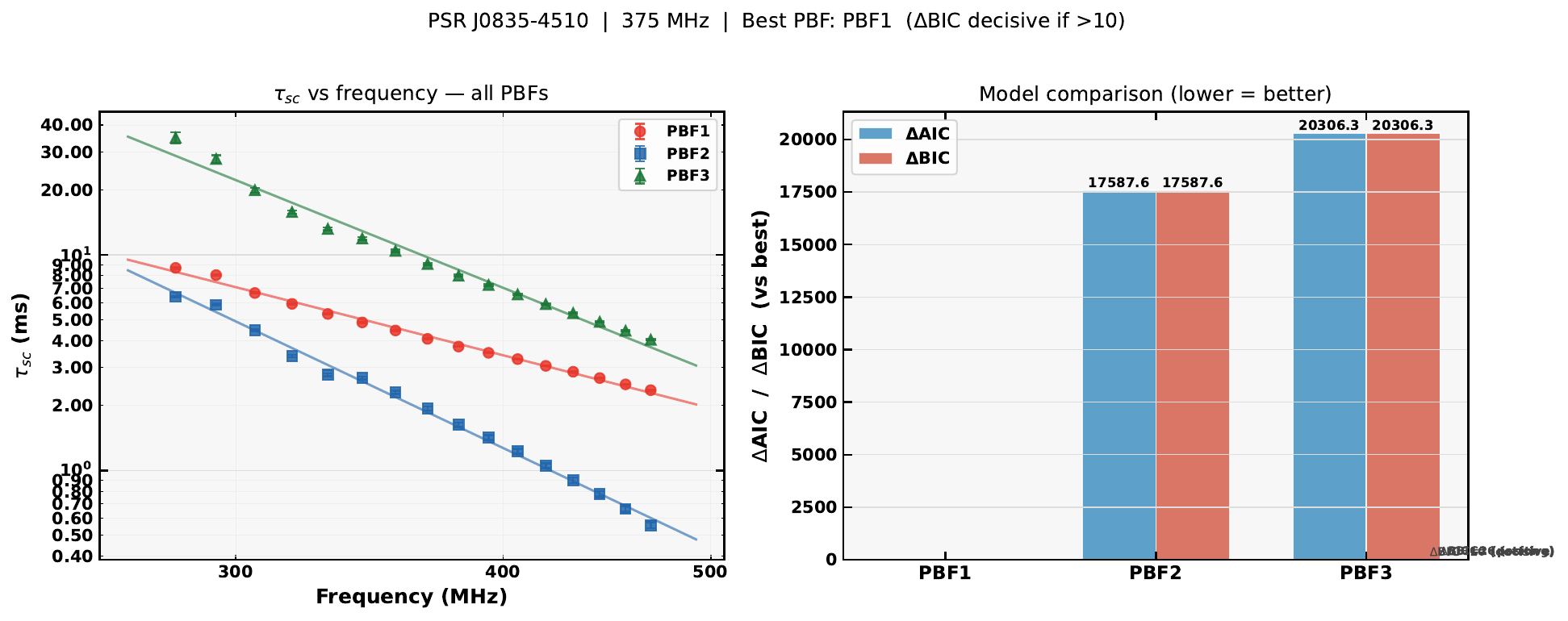}
\caption{PSR J0835$-$4510} 
\label{appfig21}
\end{figure}

\begin{figure}[H]
\centering
\includegraphics[scale=0.5]{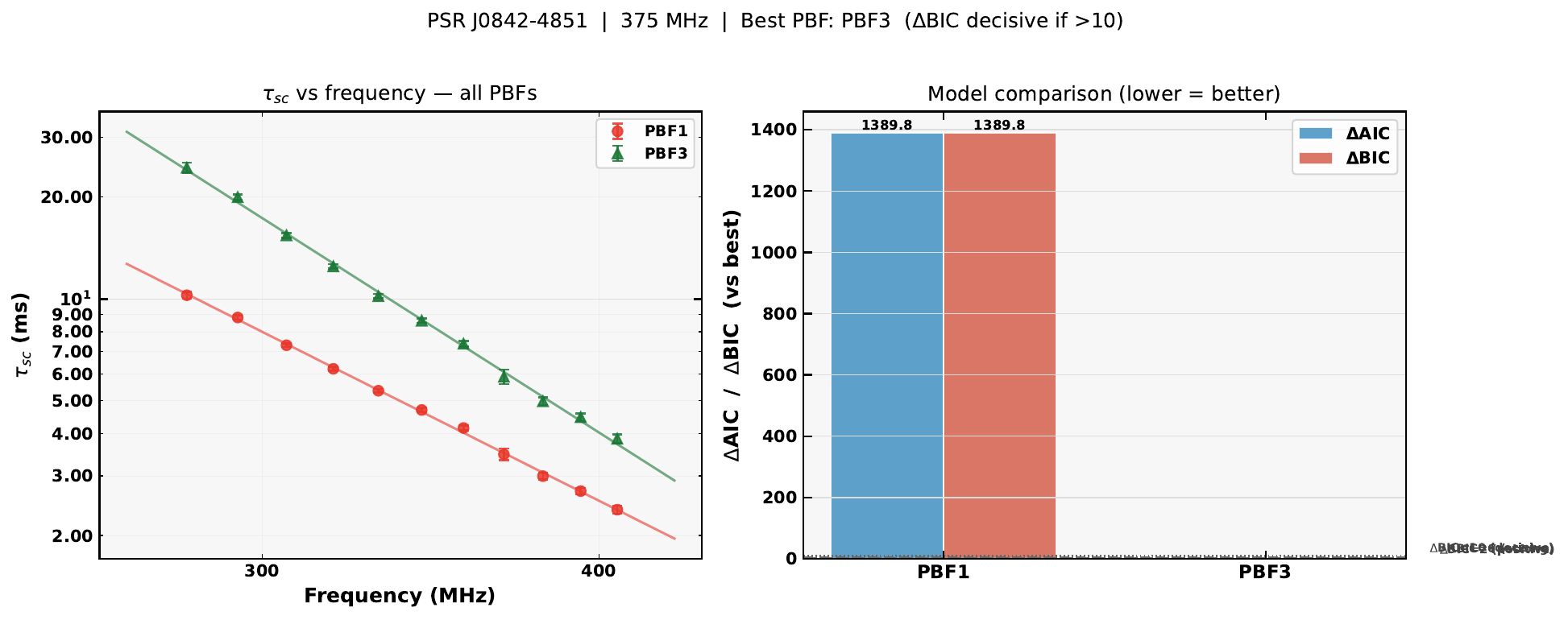}
\caption{PSR J0842$-$4851} 
\label{appfig22}
\end{figure}

\begin{figure}[H]
\centering
\includegraphics[scale=0.5]{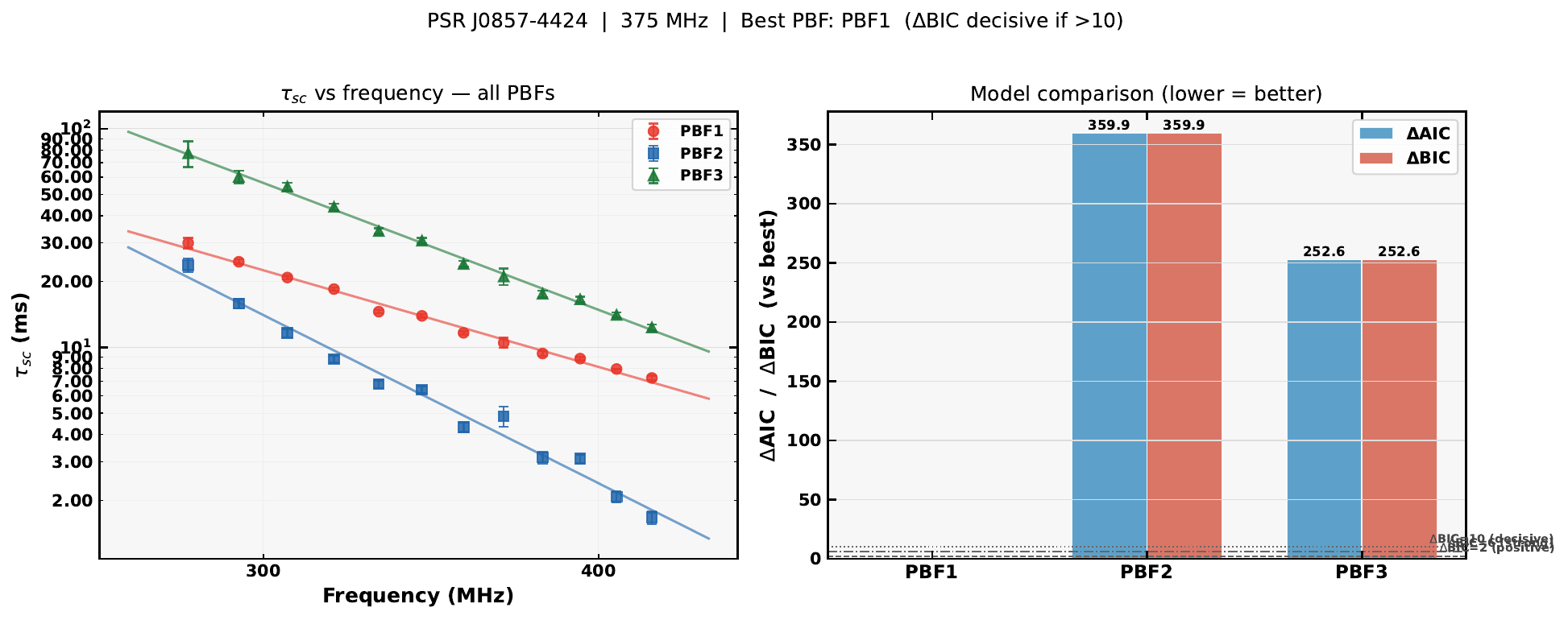}
\caption{PSR J0857$-$4424} 
\label{appfig23}
\end{figure}

\begin{figure}[H]
\centering
\includegraphics[scale=0.5]{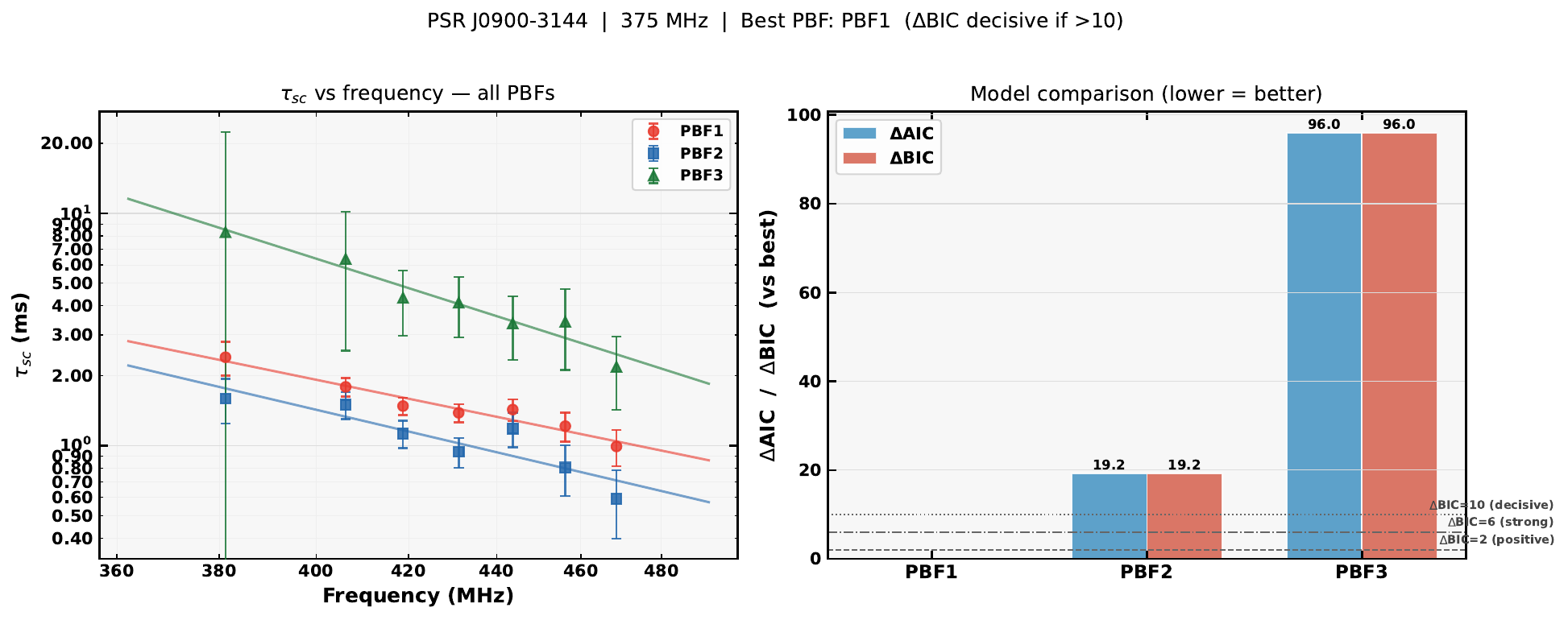}
\caption{PSR J0900$-$3144} 
\label{appfig24}
\end{figure}

\begin{figure}[H]
\centering
\includegraphics[scale=0.5]{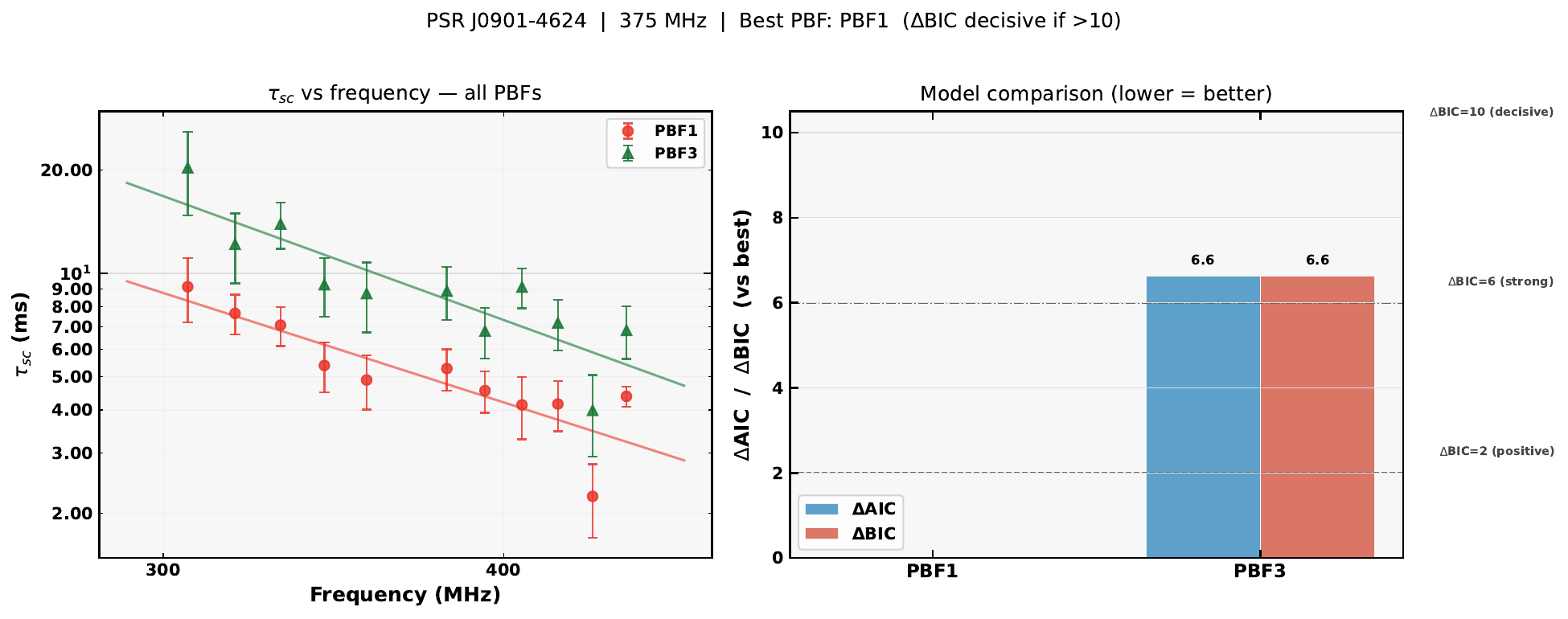}
\caption{PSR J0901$-$4624} 
\label{appfig25}
\end{figure}

\begin{figure}[H]
\centering
\includegraphics[scale=0.5]{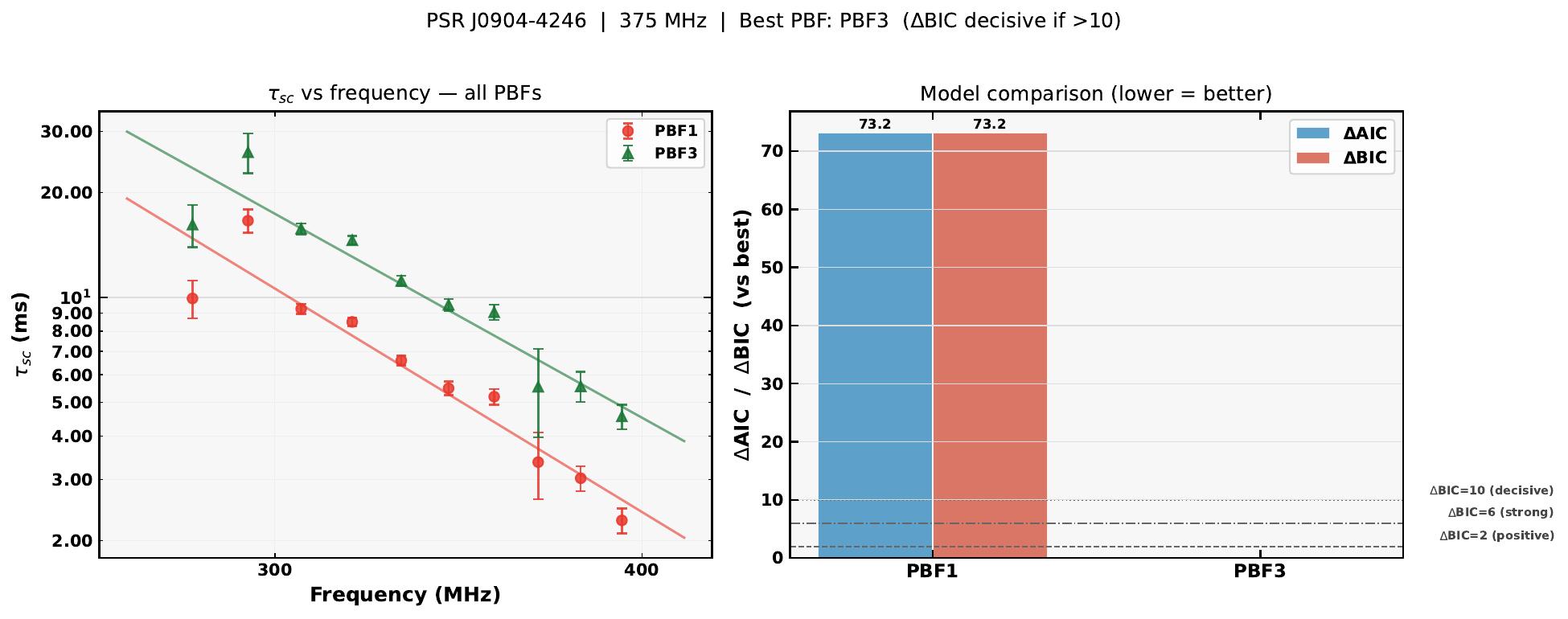}
\caption{PSR J0904$-$4246} 
\label{appfig26}
\end{figure}

\begin{figure}[H]
\centering
\includegraphics[scale=0.5]{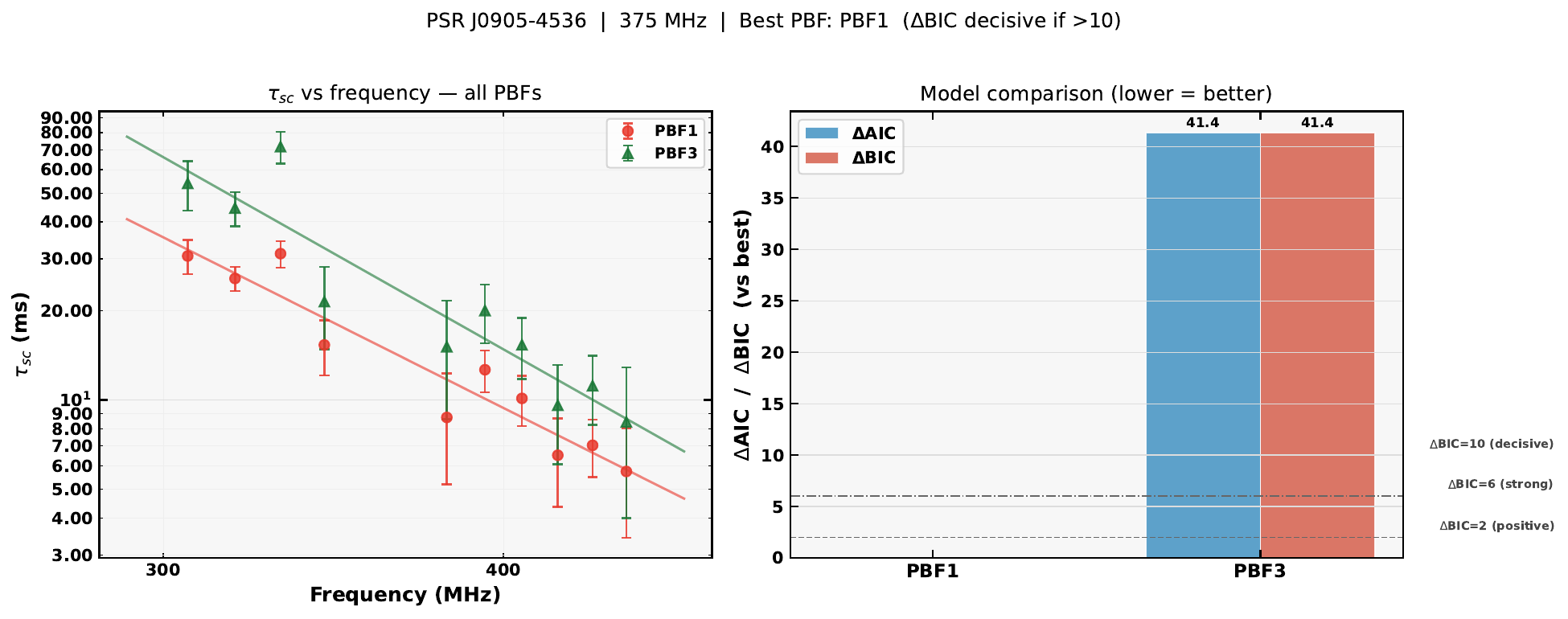}
\caption{PSR J0905$-$4536} 
\label{appfig27}
\end{figure}

\begin{figure}[H]
\centering
\includegraphics[scale=0.5]{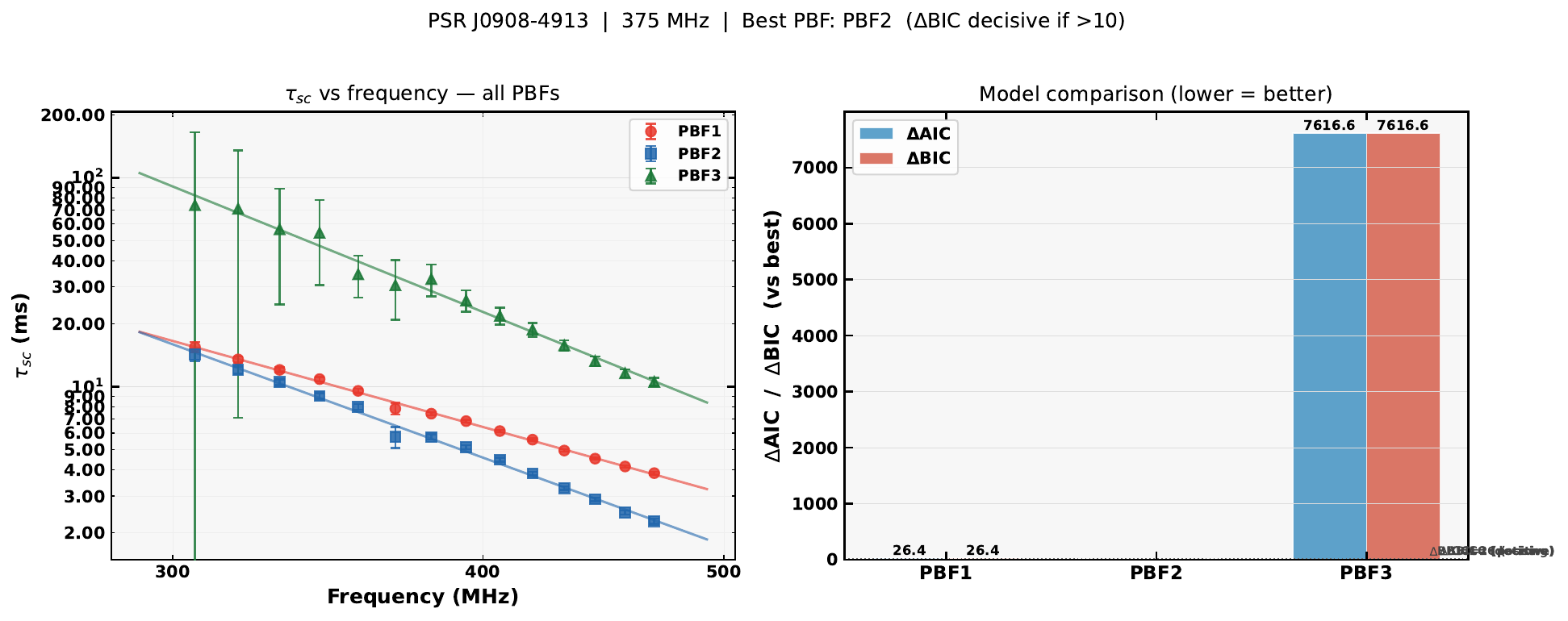}
\caption{PSR J0908$-$4913} 
\label{appfig28}
\end{figure}

\clearpage
\FloatBarrier
\section{PBF2 and PBF3 Results for all the pulsars}
\label{app3}
\renewcommand{\thefigure}{C\arabic{figure}}
\setcounter{figure}{0}
The results of PBF2 and PBF3 profile fit $\alpha$ estimate are given in this appendix. The plots are made in the same way as in ~\ref{app1} Figures \ref{appfig29} to \ref{appfig36} show the fit results of PBF2 and figures \ref{appfig37} to \ref{appfig50} show the fit results for PBF3.

\begin{figure}[H]
\centering
\includegraphics[scale=0.5]{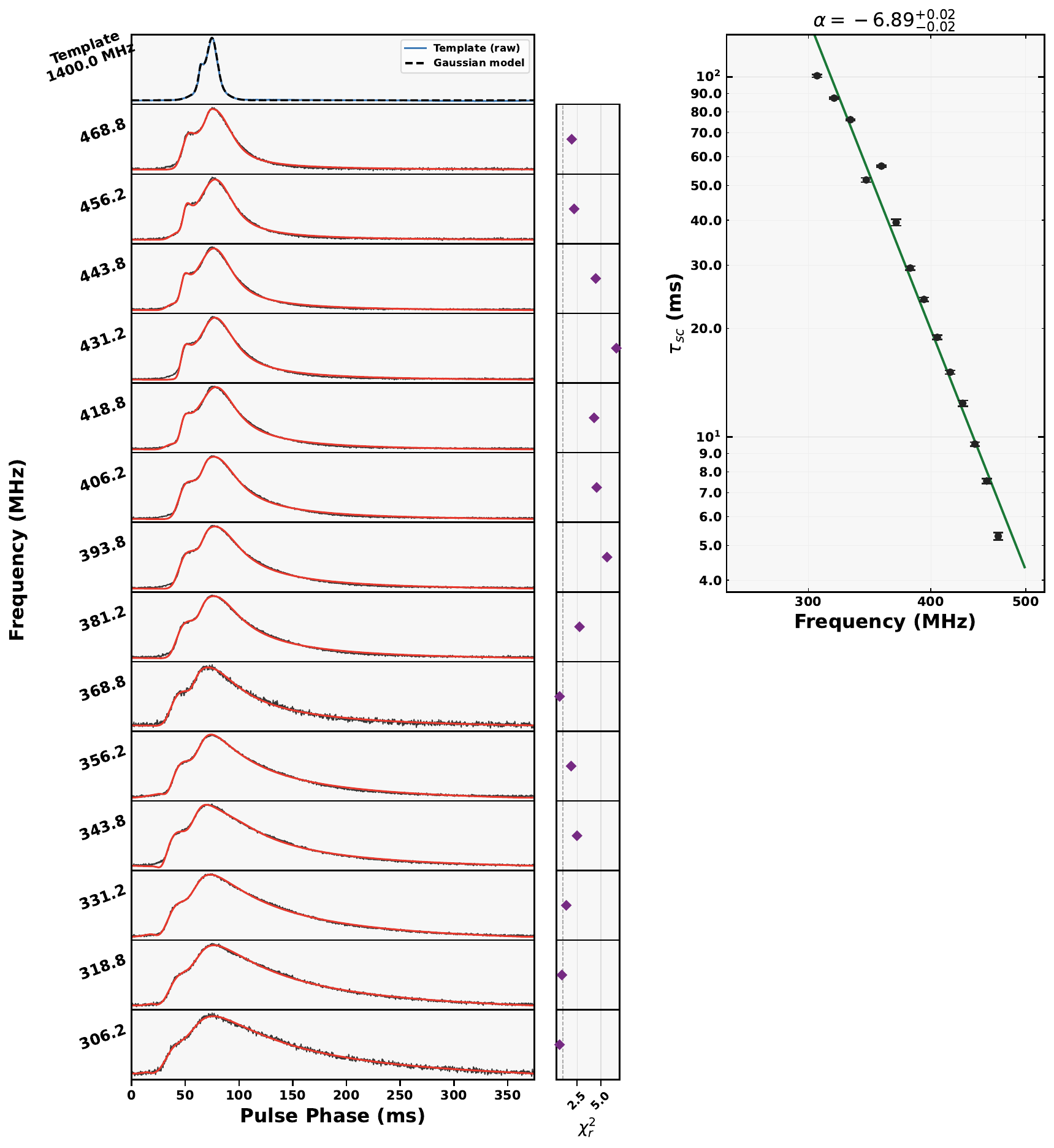}
\caption{PSR J0738$-$4042} 
\label{appfig29}
\end{figure}

\begin{figure}[H]
\centering
\includegraphics[scale=0.5]{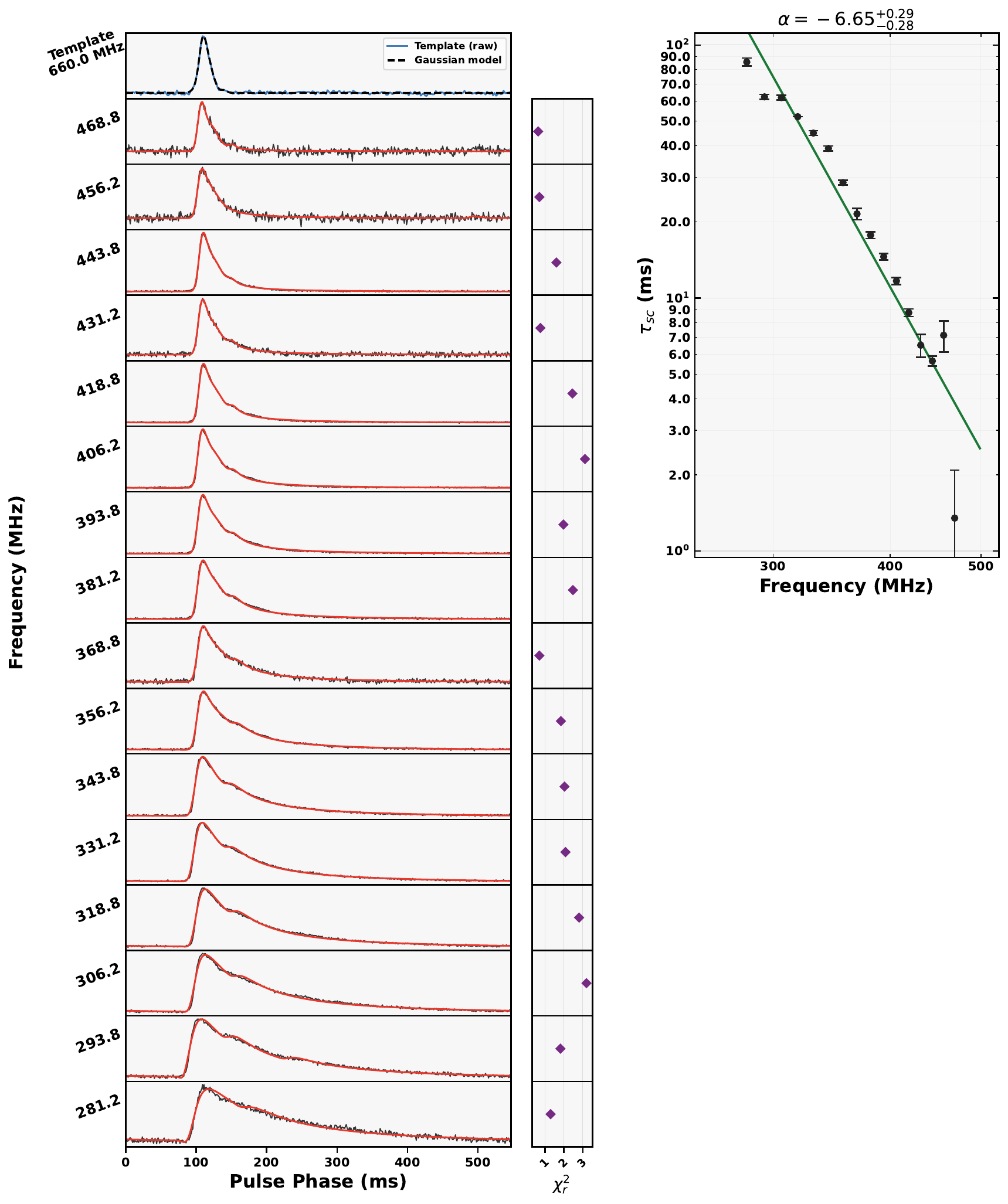}
\caption{PSR J0809$-$4753} 
\label{appfig30}
\end{figure}

\begin{figure}[H]
\centering
\includegraphics[scale=0.5]{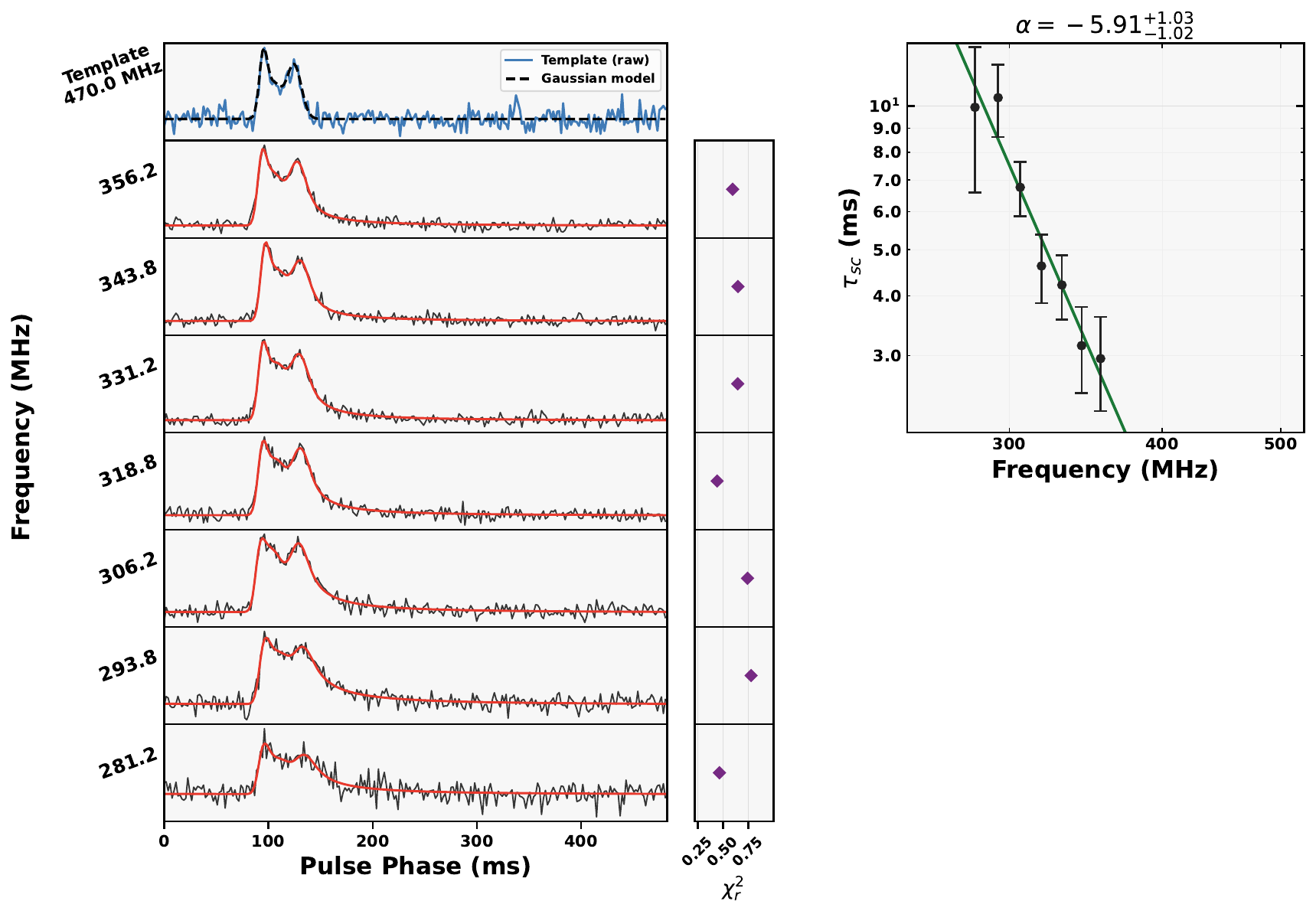}
\caption{PSR J0812$-$3905} 
\label{appfig31}
\end{figure}

\begin{figure}[H]
\centering
\includegraphics[scale=0.5]{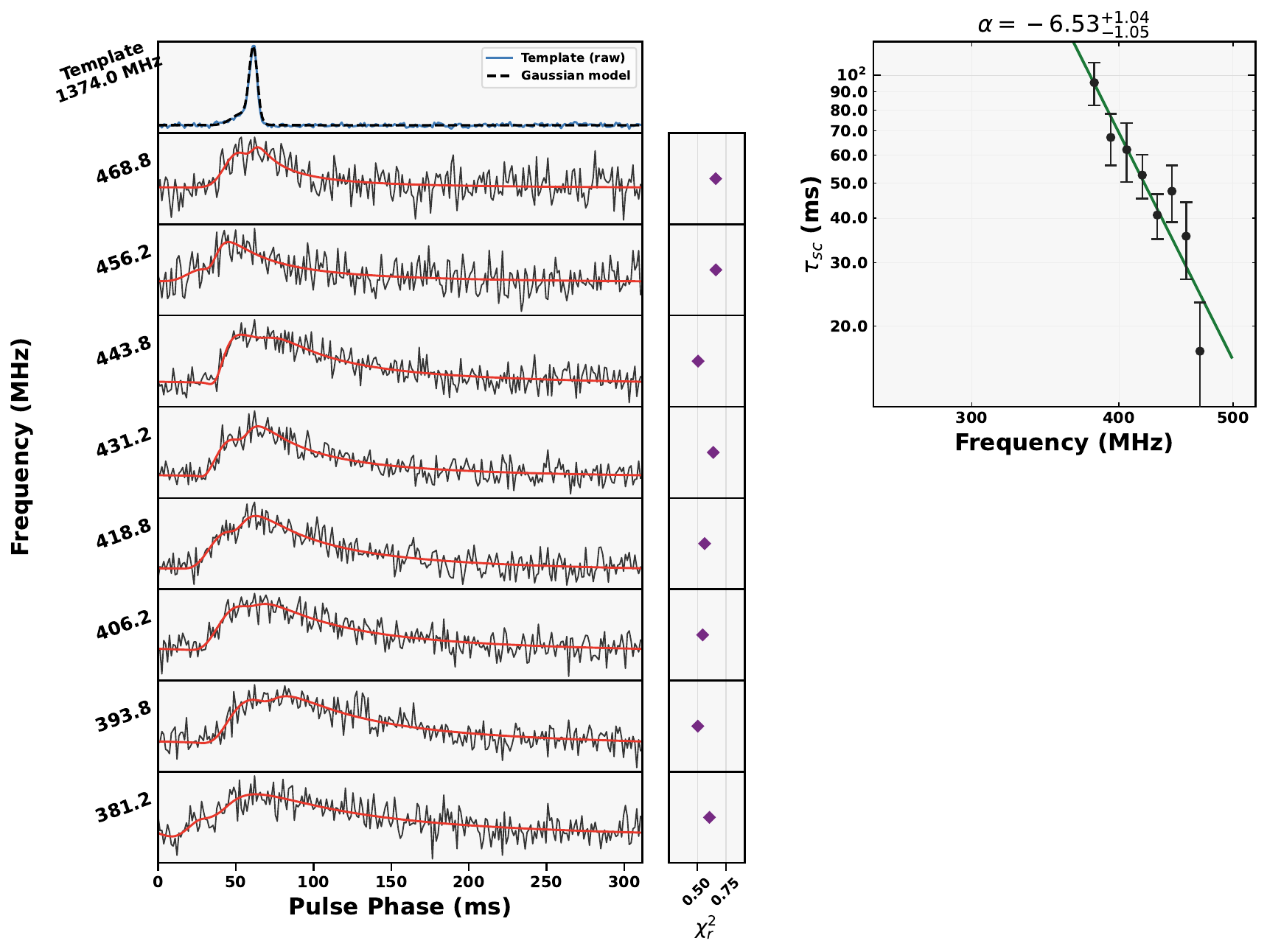}
\caption{PSR J0831$-$4406} 
\label{appfig32}
\end{figure}

\begin{figure}[H]
\centering
\includegraphics[scale=0.5]{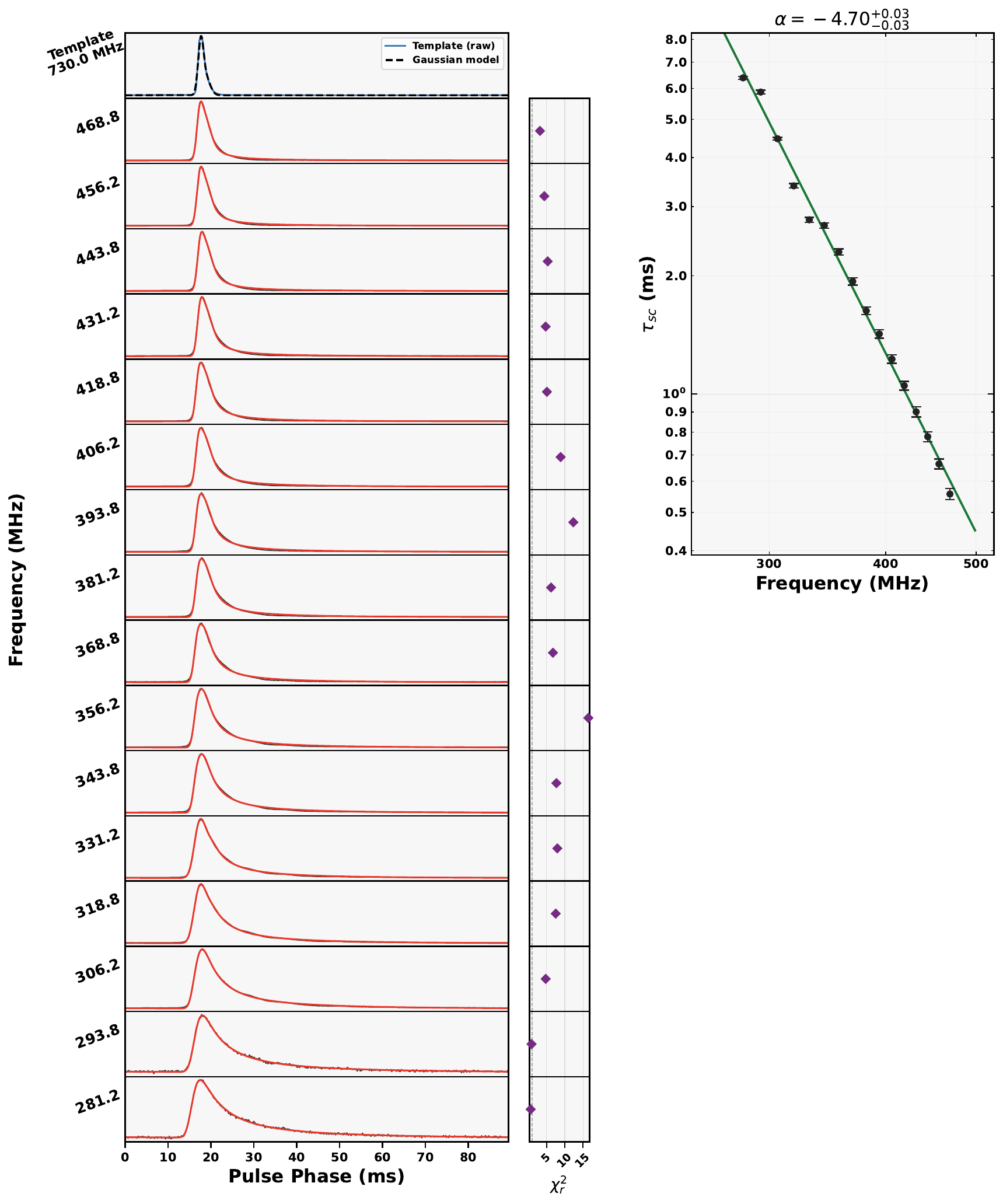}
\caption{PSR J0835$-$4510} 
\label{appfig33}
\end{figure}

\begin{figure}[H]
\centering
\includegraphics[scale=0.5]{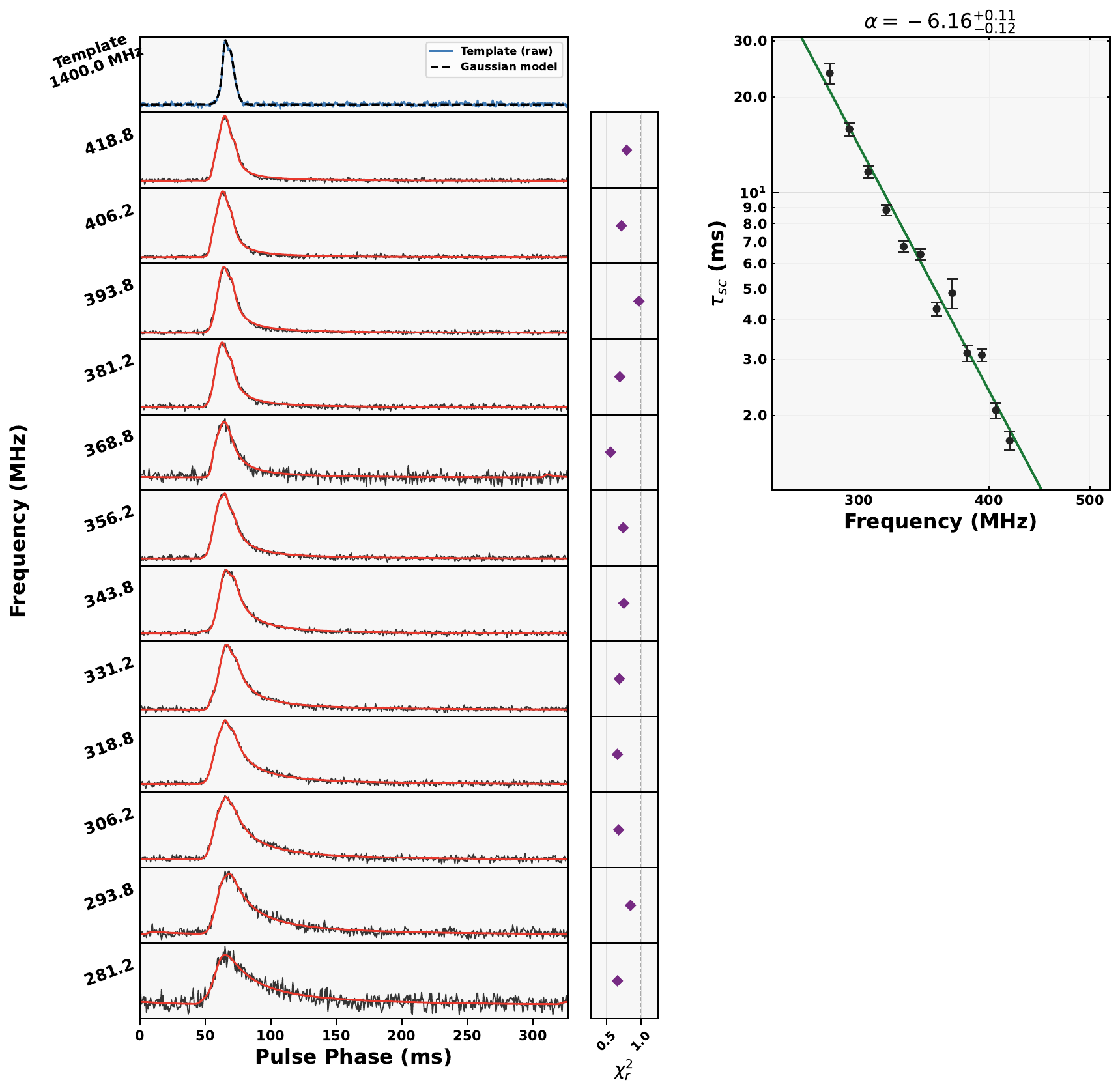}
\caption{PSR J0857$-$4424} 
\label{appfig34}
\end{figure}

\begin{figure}[H]
\centering
\includegraphics[scale=0.5]{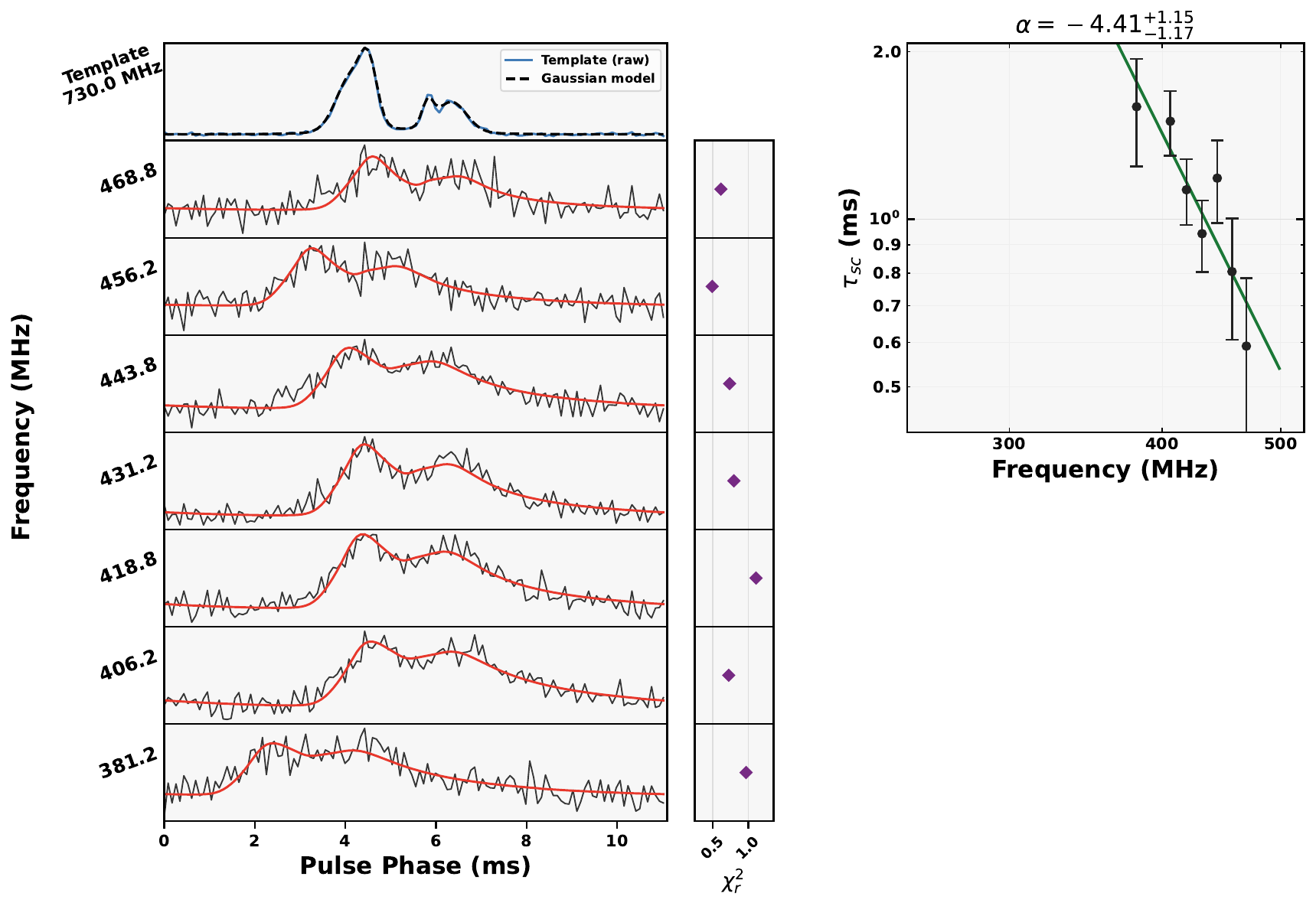}
\caption{PSR J0900$-$3144} 
\label{appfig35}
\end{figure}

\begin{figure}[H]
\centering
\includegraphics[scale=0.5]{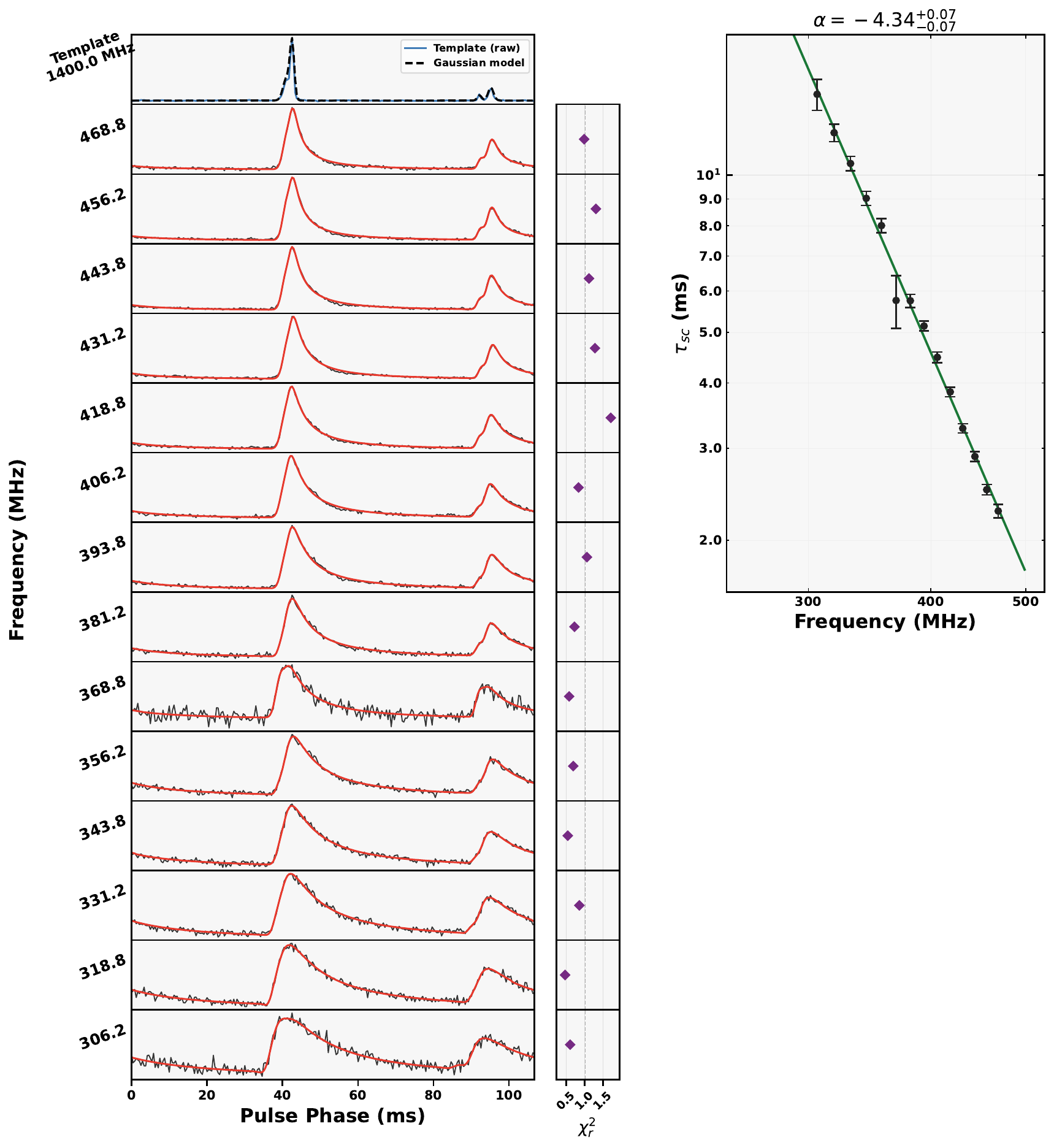}
\caption{PSR J0908$-$4913} 
\label{appfig36}
\end{figure}

\begin{figure}[H]
\centering
\includegraphics[scale=0.5]{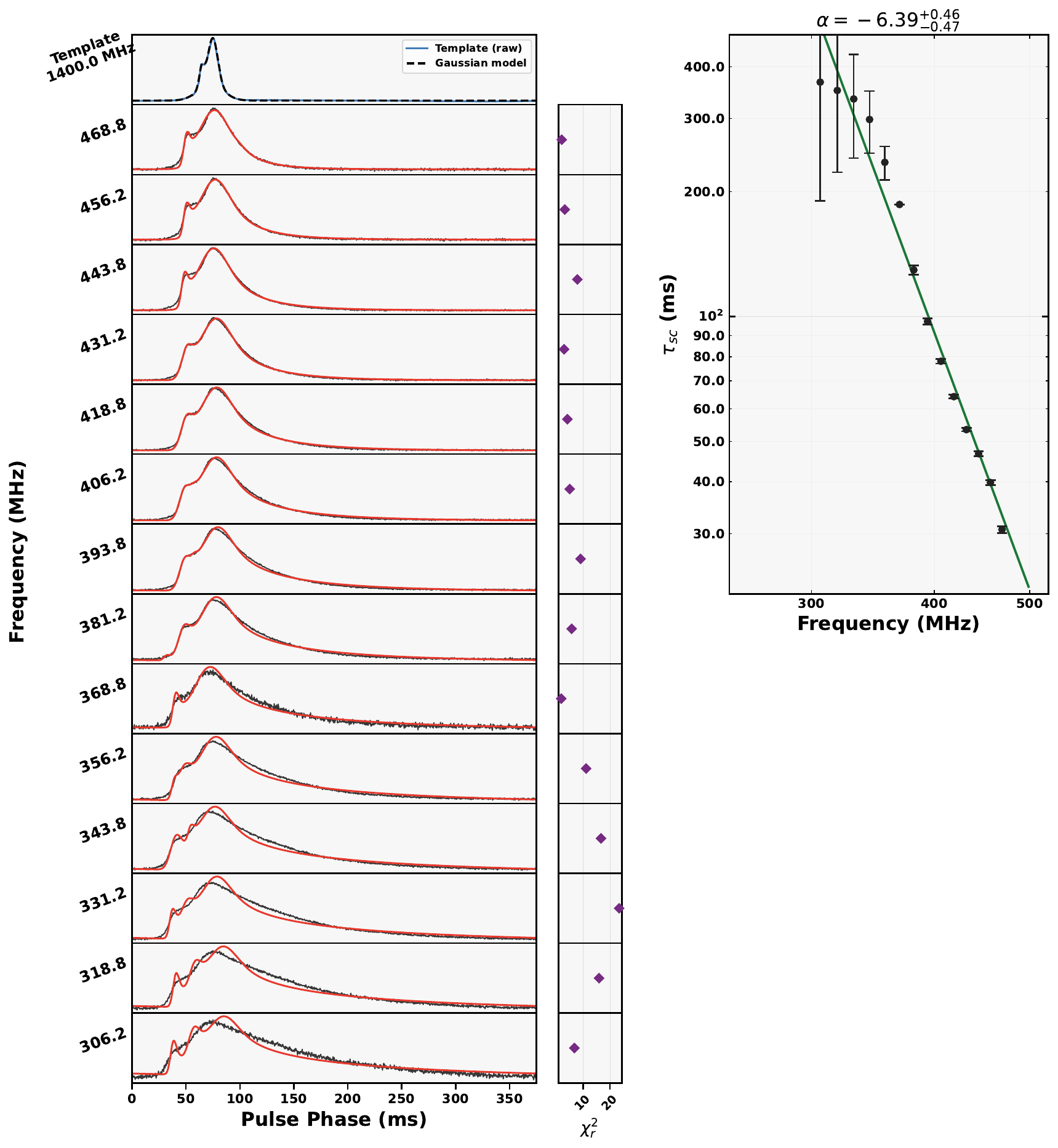}
\caption{PSR J0738$-$4042} 
\label{appfig37}
\end{figure}

\begin{figure}[H]
\centering
\includegraphics[scale=0.5]{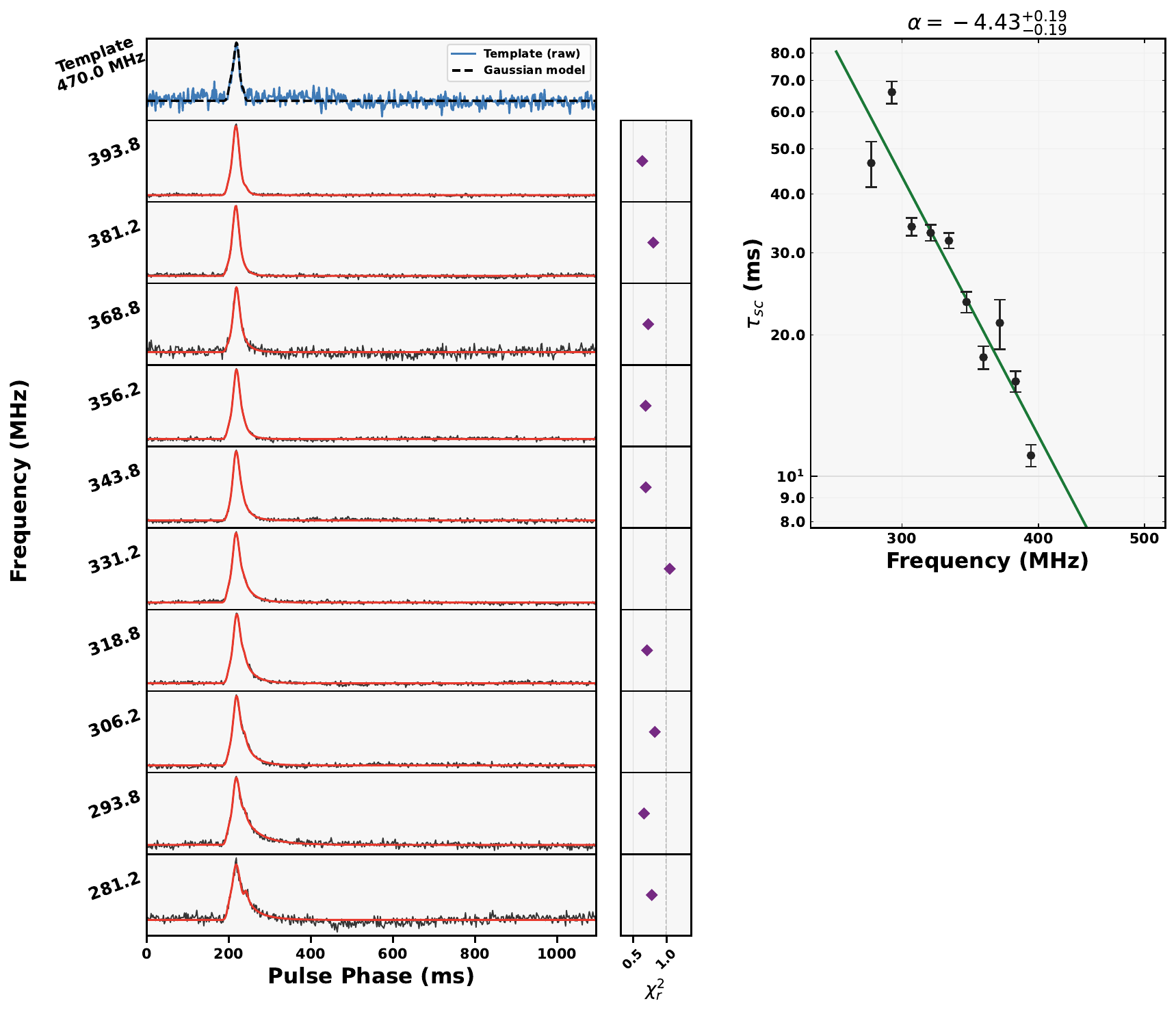}
\caption{PSR J0749$-$4247} 
\label{appfig38}
\end{figure}

\begin{figure}[H]
\centering
\includegraphics[scale=0.5]{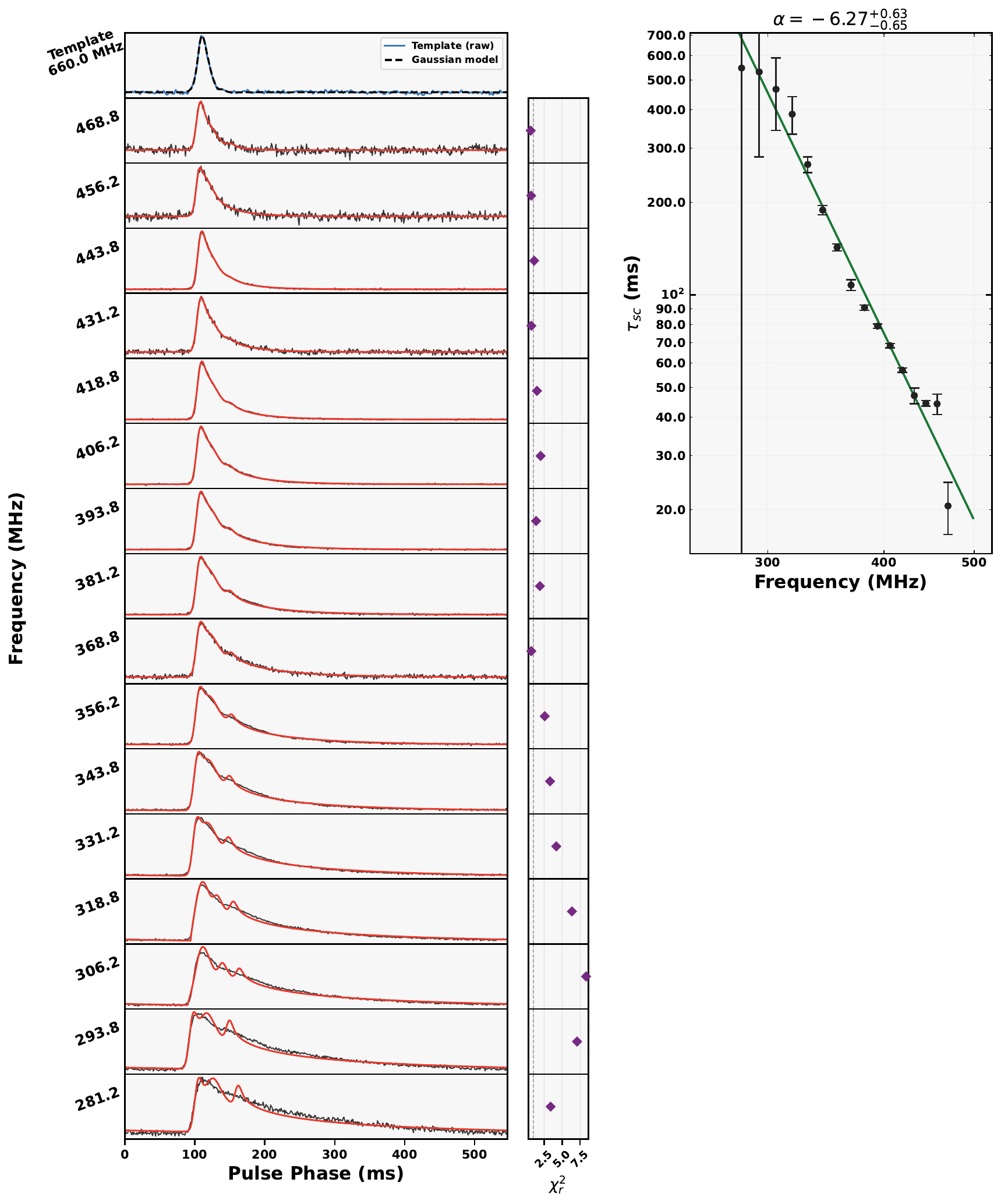}
\caption{PSR J0809$-$4753} 
\label{appfig39}
\end{figure}

\begin{figure}[H]
\centering
\includegraphics[scale=0.5]{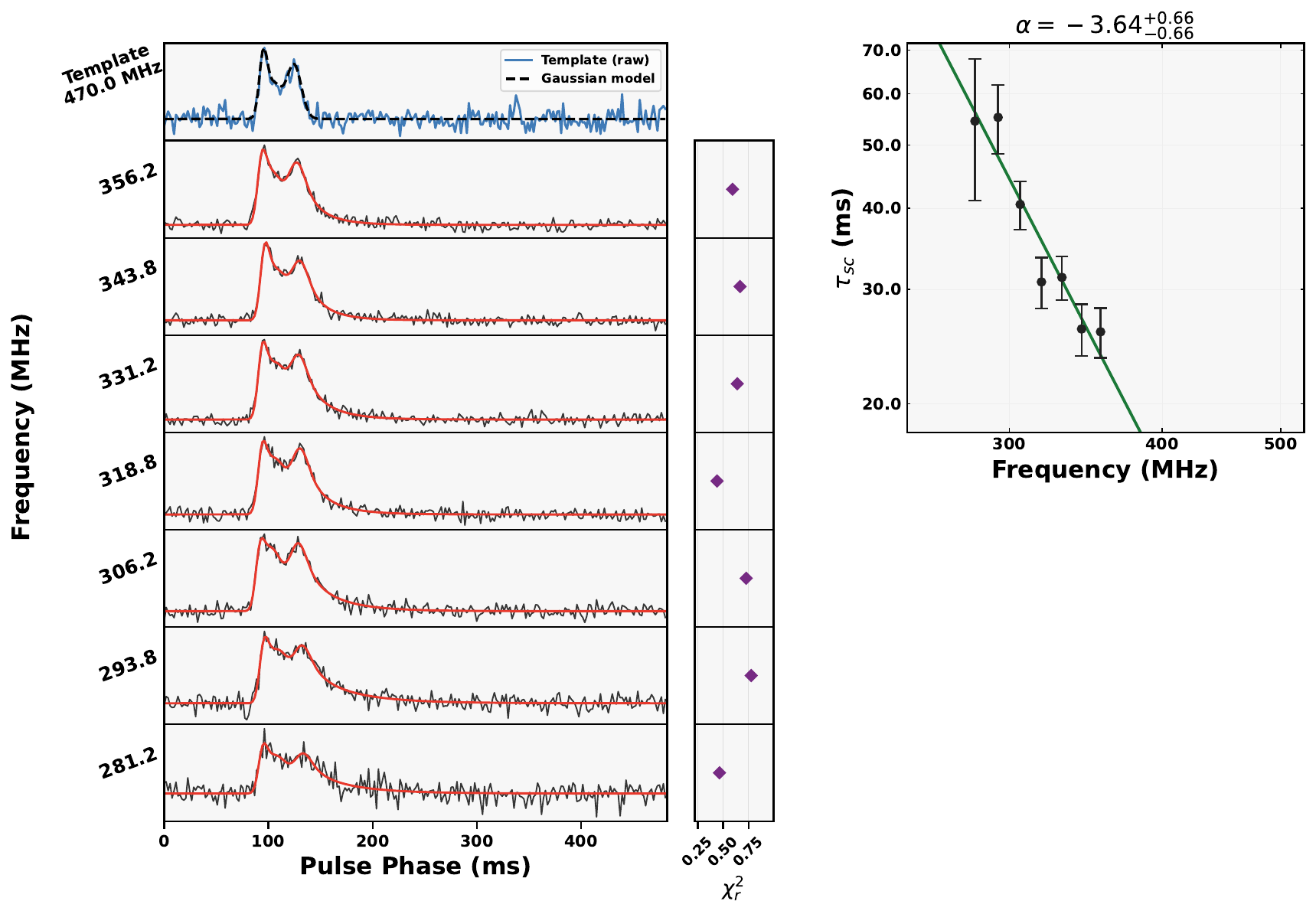}
\caption{PSR J0812$-$3905} 
\label{appfig40}
\end{figure}

\begin{figure}[H]
\centering
\includegraphics[scale=0.5]{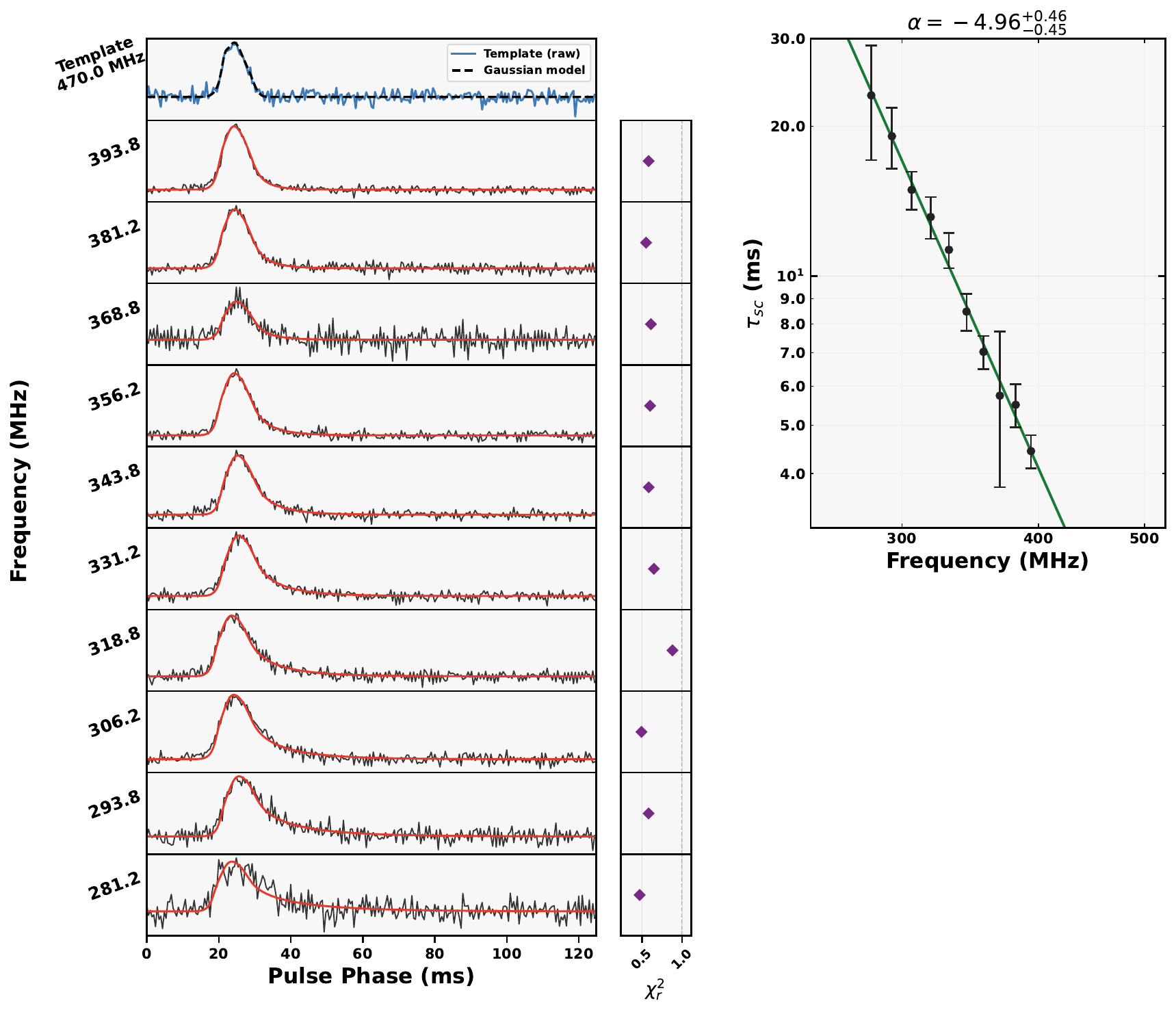}
\caption{PSR J0820$-$3826} 
\label{appfig41}
\end{figure}

\begin{figure}[H]
\centering
\includegraphics[scale=0.5]{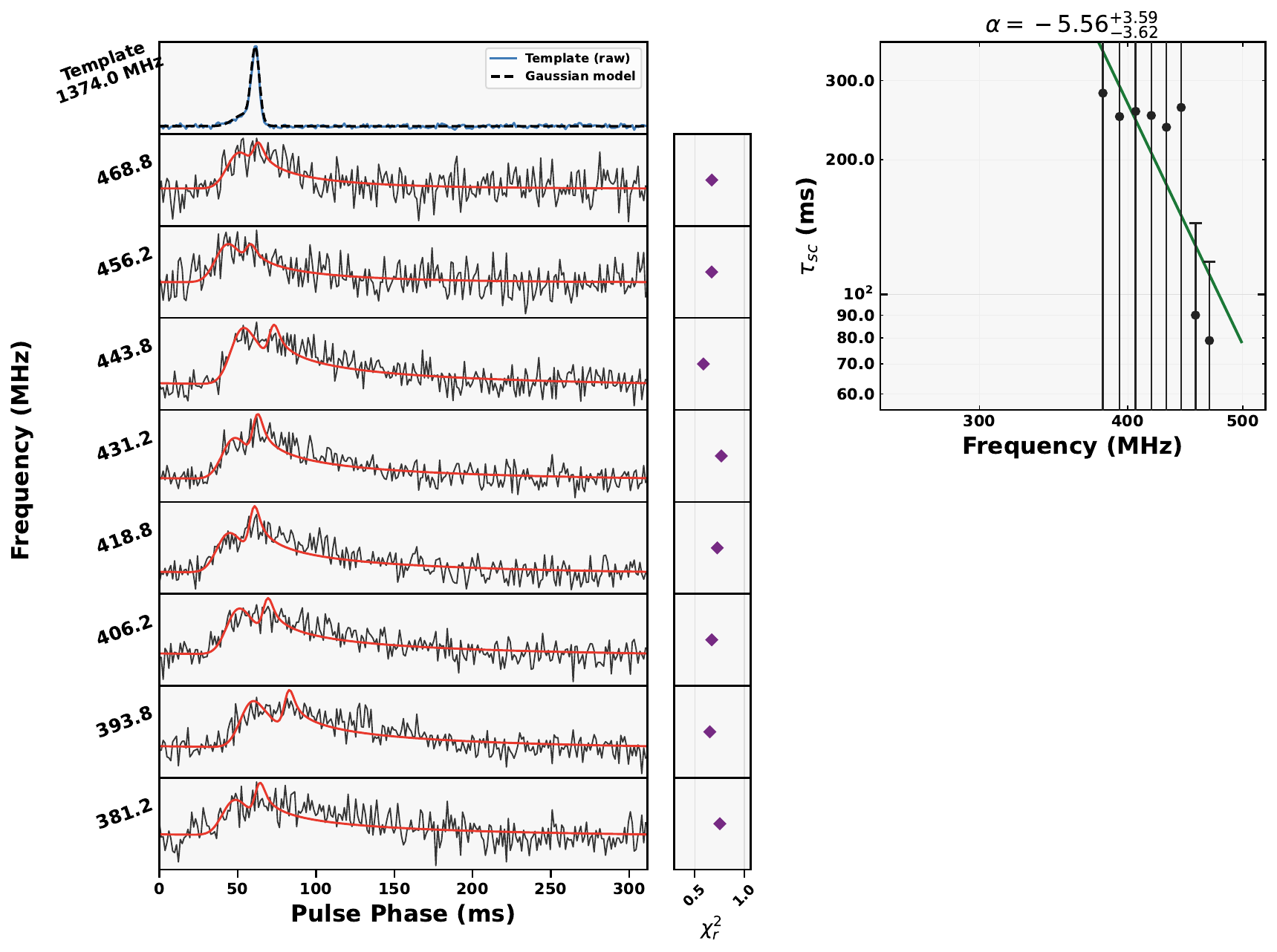}
\caption{PSR J0831$-$4406} 
\label{appfig42}
\end{figure}

\begin{figure}[H]
\centering
\includegraphics[scale=0.5]{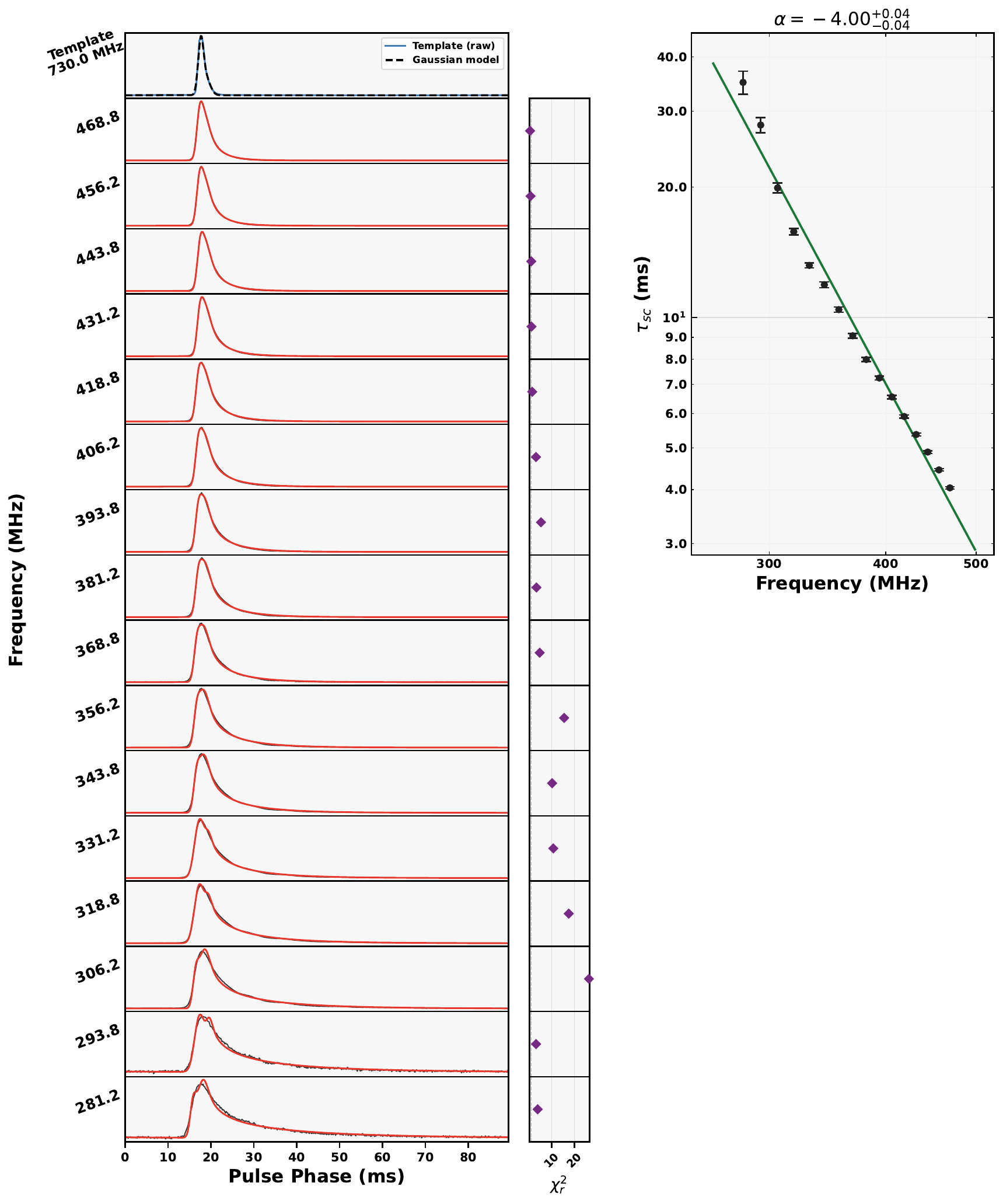}
\caption{PSR J0835$-$4510} 
\label{appfig43}
\end{figure}

\begin{figure}[H]
\centering
\includegraphics[scale=0.5]{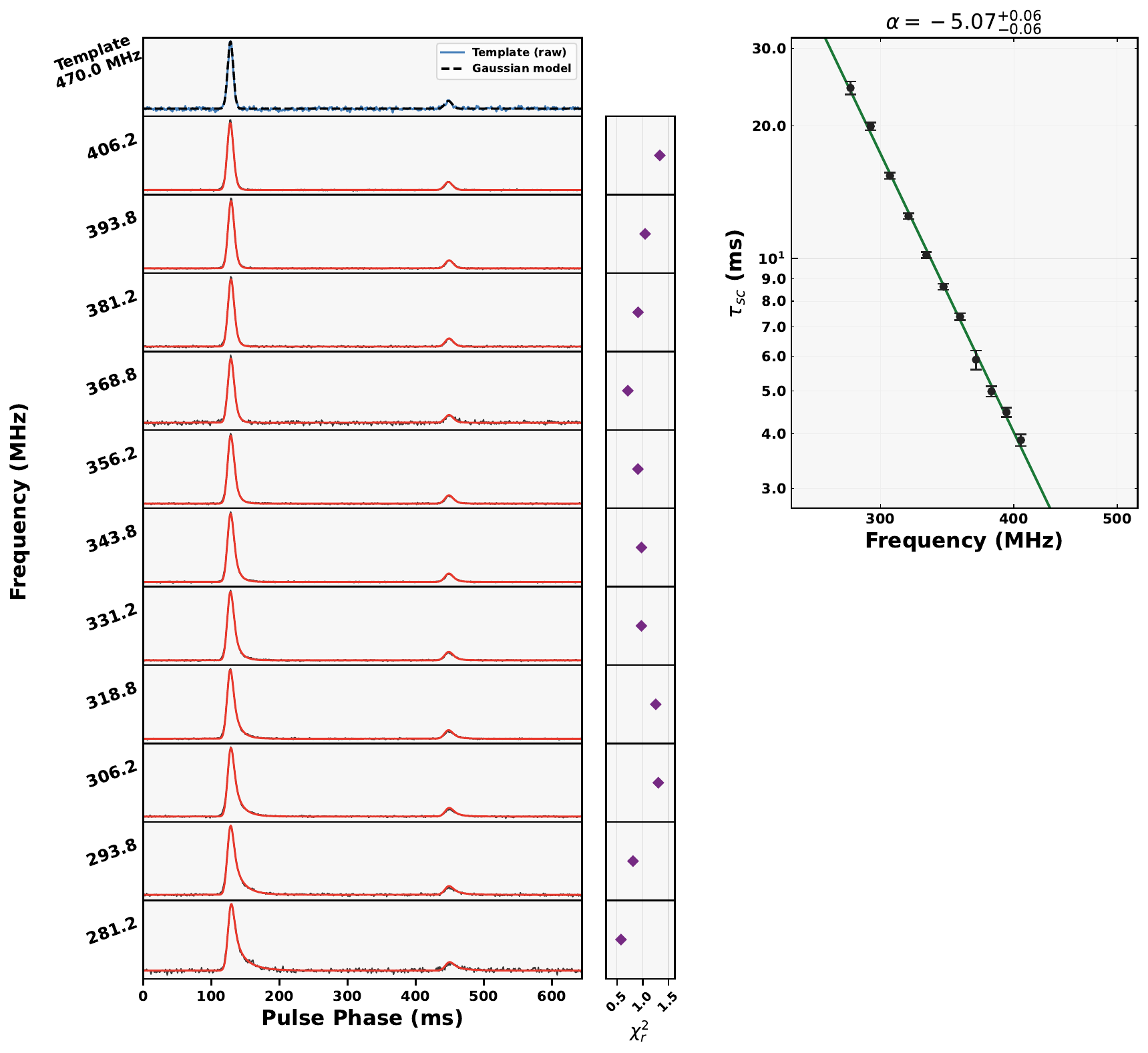}
\caption{PSR J0842$-$4851} 
\label{appfig44}
\end{figure}

\begin{figure}[H]
\centering
\includegraphics[scale=0.5]{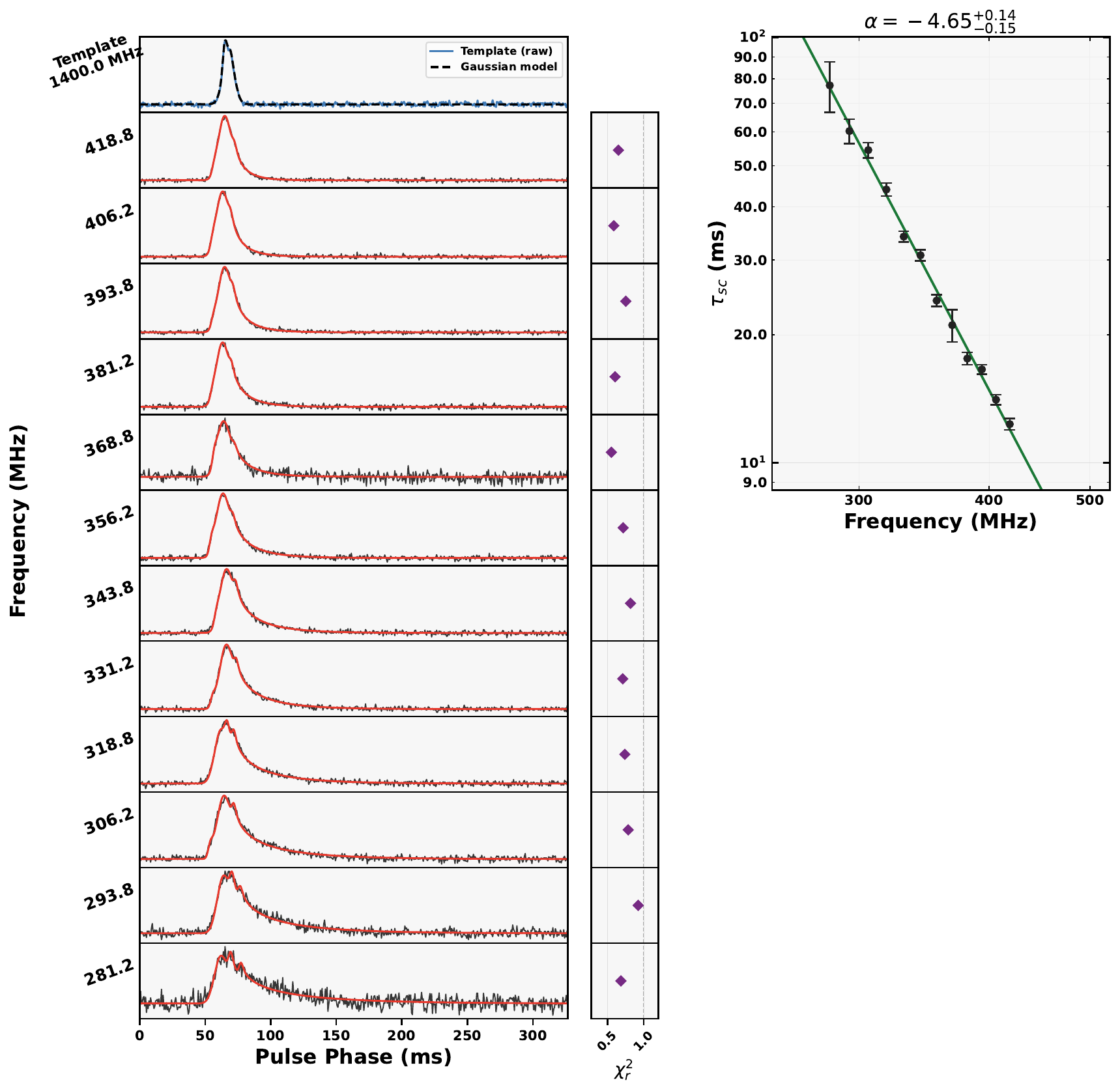}
\caption{PSR J0857$-$4424} 
\label{appfig45}
\end{figure}

\begin{figure}[H]
\centering
\includegraphics[scale=0.5]{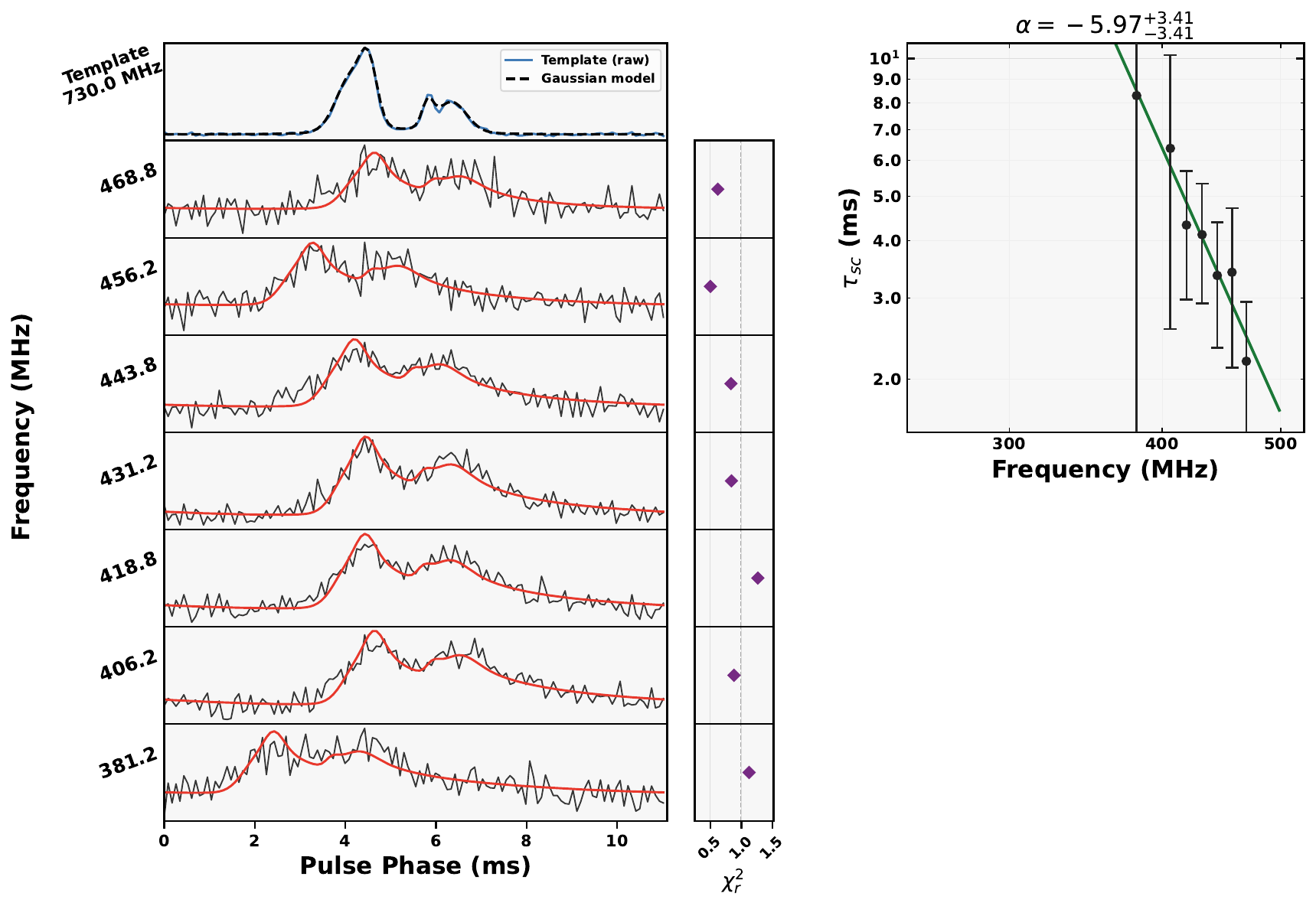}
\caption{PSR J0900$-$3144} 
\label{appfig46}
\end{figure}

\begin{figure}[H]
\centering
\includegraphics[scale=0.5]{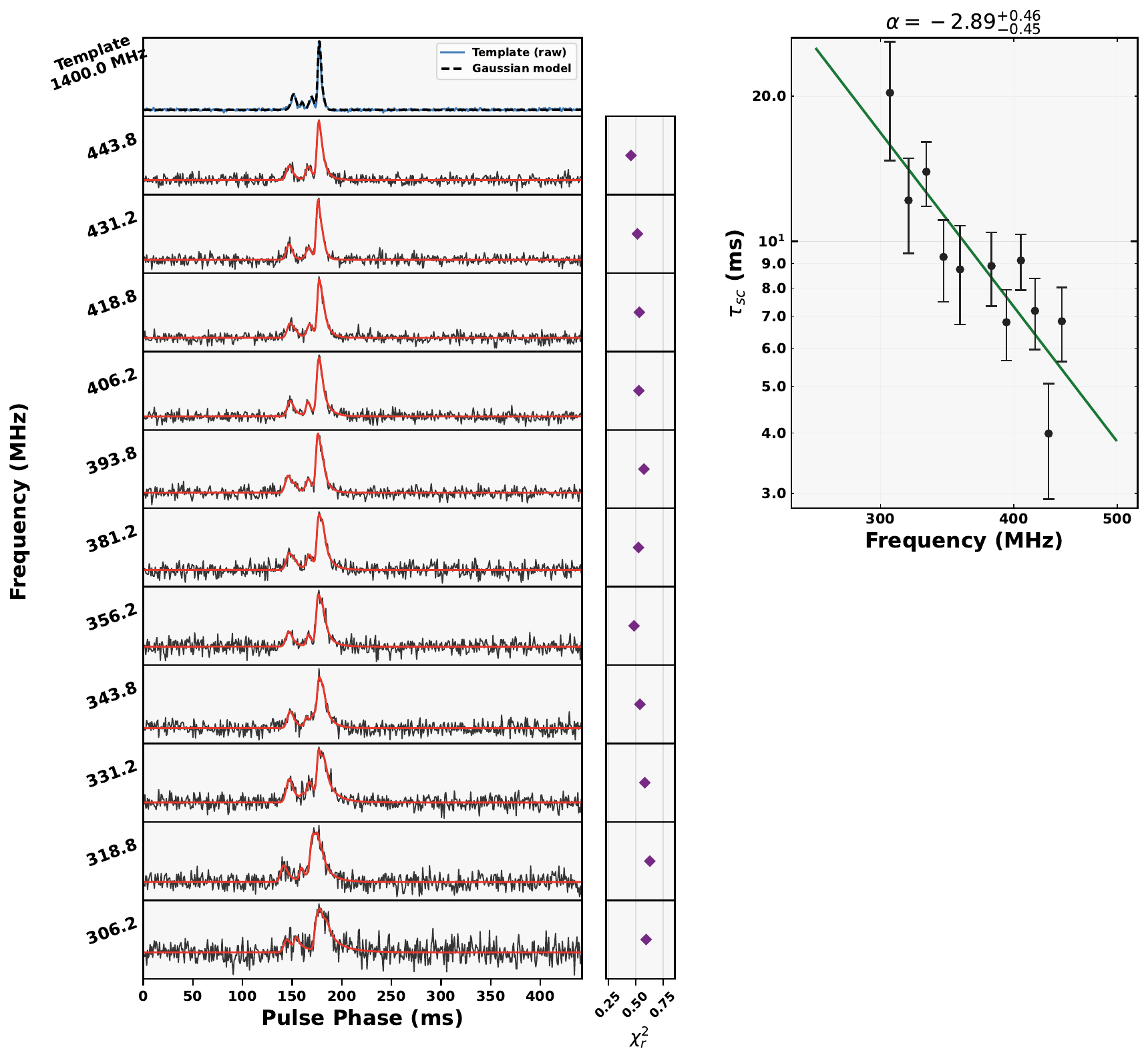}
\caption{PSR J0901$-$4624} 
\label{appfig47}
\end{figure}

\begin{figure}[H]
\centering
\includegraphics[scale=0.5]{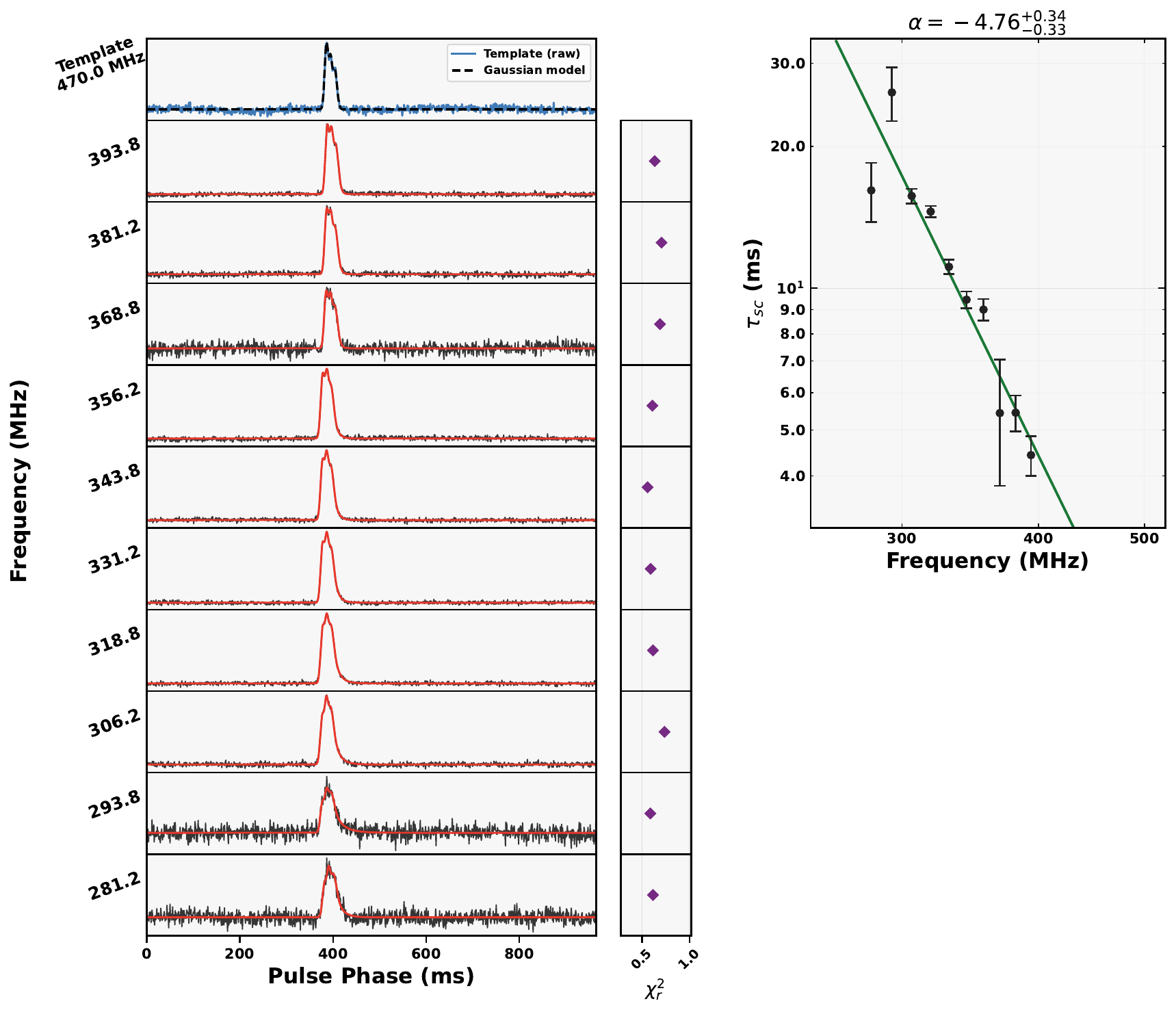}
\caption{PSR J0904$-$4246} 
\label{appfig48}
\end{figure}

\begin{figure}[H]
\centering
\includegraphics[scale=0.5]{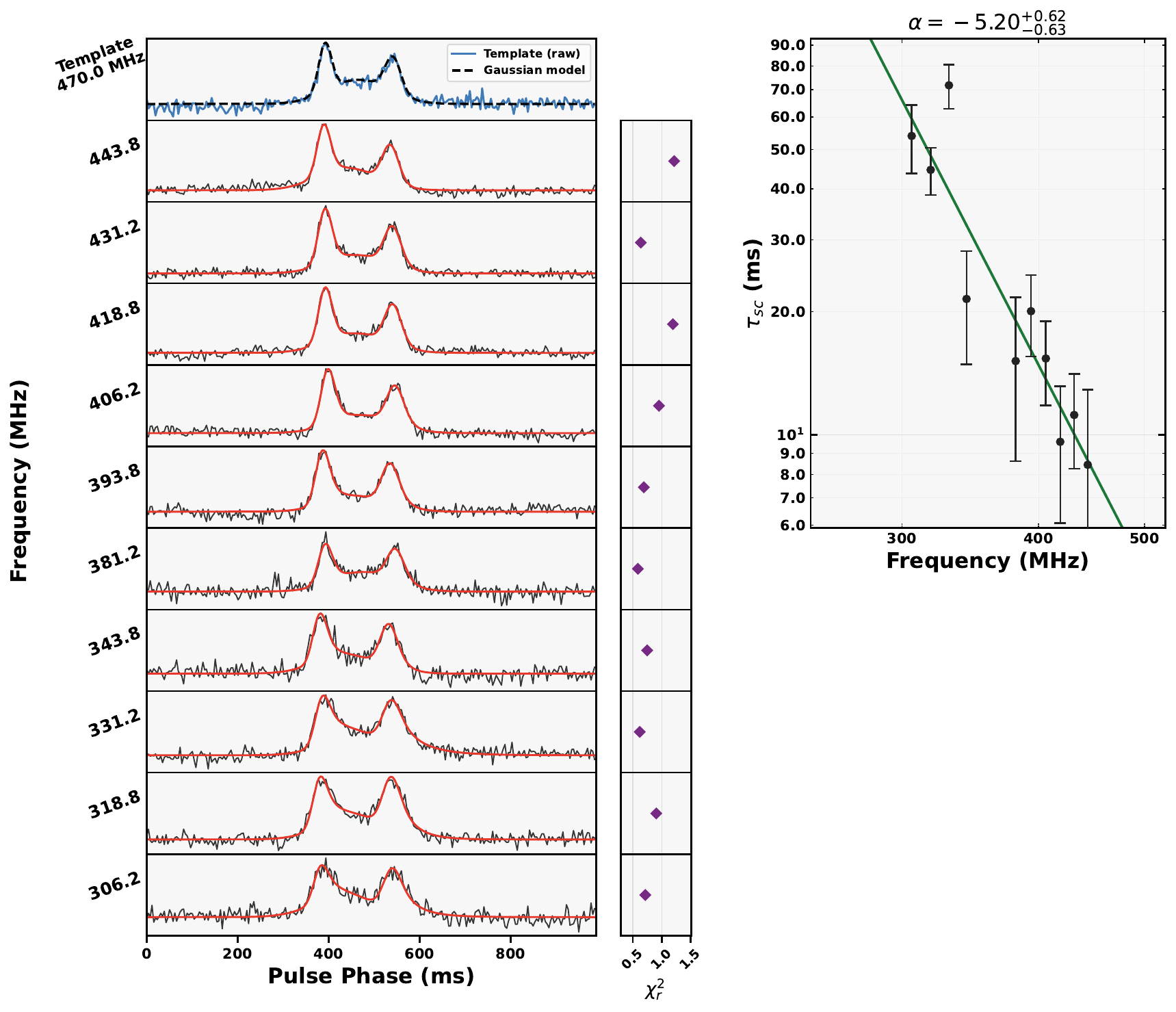}
\caption{PSR J0905$-$4536} 
\label{appfig49}
\end{figure}

\begin{figure}[H]
\centering
\includegraphics[scale=0.5]{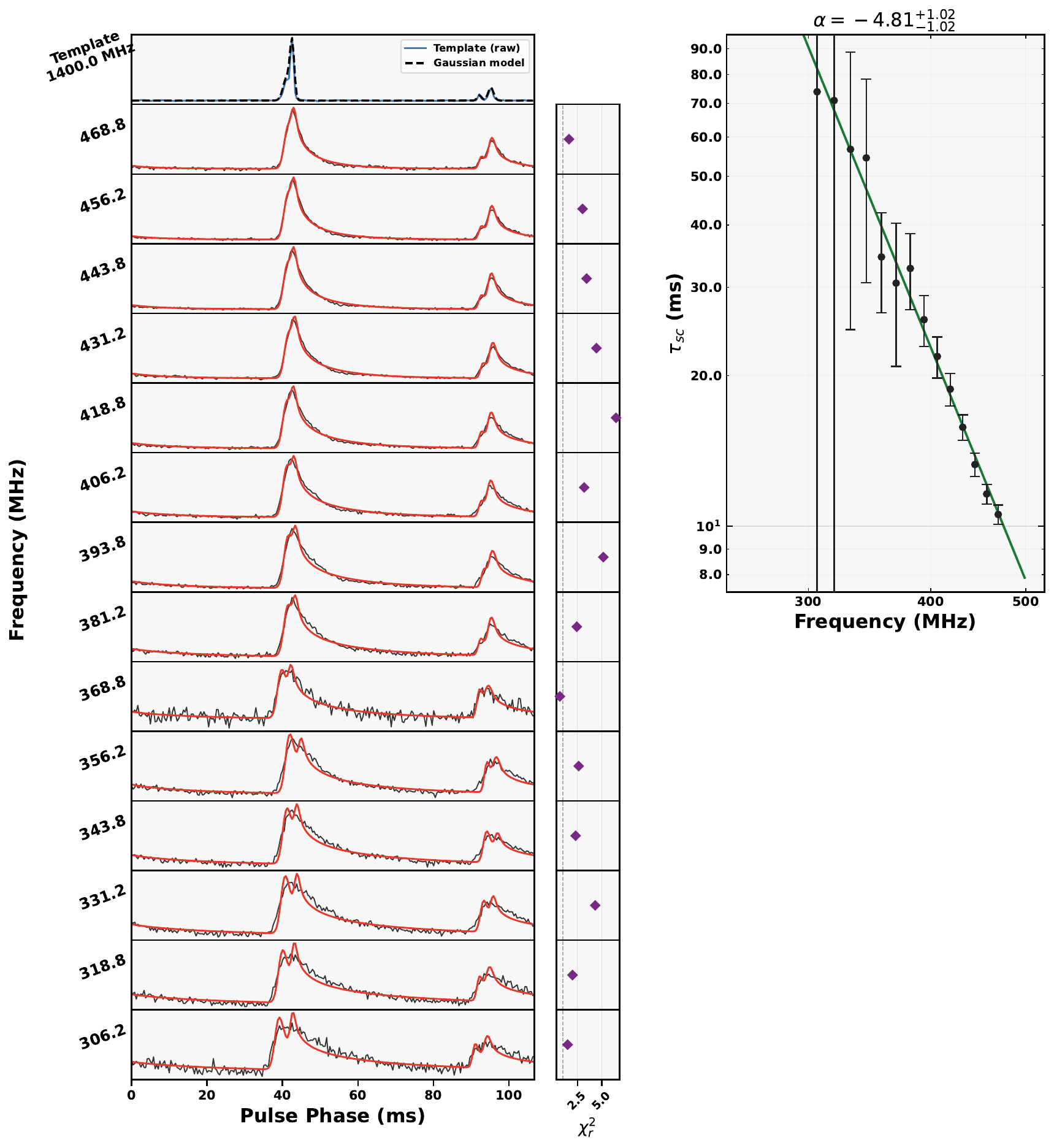}
\caption{PSR J0908$-$4913} 
\label{appfig50}
\end{figure}

\end{document}